\documentclass[
11pt, 
{portuguese,english}, 
onehalfspacing, 
liststotoc, 
headsepline, 
]{MastersDoctoralThesis} 

\usepackage[utf8]{inputenc} 
\usepackage[OT1]{fontenc} 

\usepackage[autostyle=true]{csquotes} 
\usepackage[noadjust]{cite}
\usepackage{amsmath}
\usepackage{graphicx}
\usepackage{subfig}
\usepackage{bm}
\usepackage[final]{pdfpages}
\usepackage{pbox}
\usepackage{float}
\usepackage{physics}
\usepackage{upgreek}
\usepackage[urlcolor=blue,colorlinks,breaklinks=true]{hyperref}

\usepackage{hyperref}
\usepackage{amsfonts}
\usepackage{amssymb}
\usepackage[mathscr]{euscript}
\usepackage{pzccal}
\DeclareFontFamily{OT1}{pzc}{}
\DeclareFontShape{OT1}{pzc}{m}{it}{<-> s * [1.10] pzcmi7t}{}
\DeclareMathAlphabet{\mathpzc}{OT1}{pzc}{m}{it}
\DeclareUnicodeCharacter{2212}{-}

\thesistitle{Cosmological Implications of Nonminimally-Coupled $f(R)$ Gravity and the Lagrangian of Cosmic Fluids} 
\thesistitlealt{Implicações Cosmológicas de Gravidade $f(R)$ Não-Minimamente Acoplada e o Lagrangiano de Fluidos Cósmicos}
\supervisor{Dr. Pedro Pina Avelino} 
\examiner{} 
\degree{Doctor of Philosophy in Physics} 
\degreealt{Doutoramento em Física}
\author{Rui Pedro Lopes de \textsc{Azevedo}} 
\addresses{} 
\subject{Physics} 
\university{\href{http://sigarra.up.pt/up/pt/web_base.gera_pagina?p_pagina=home}{Universidade do Porto}} 
\department{\href{http://dfa.fc.up.pt/}{Departamento de F\'isica e Astronomia}} 
\faculty{\href{http://sigarra.up.pt/fcup/pt/web_page.inicial}{Faculdade de Ciências}} 

\hypersetup{pdftitle=\ttitle} 
\hypersetup{pdfauthor=\authorname} 
\hypersetup{pdfkeywords=\keywordnames} 

\begin{document}

\frontmatter 

\pagestyle{plain} 




 
 
 

\includepdf{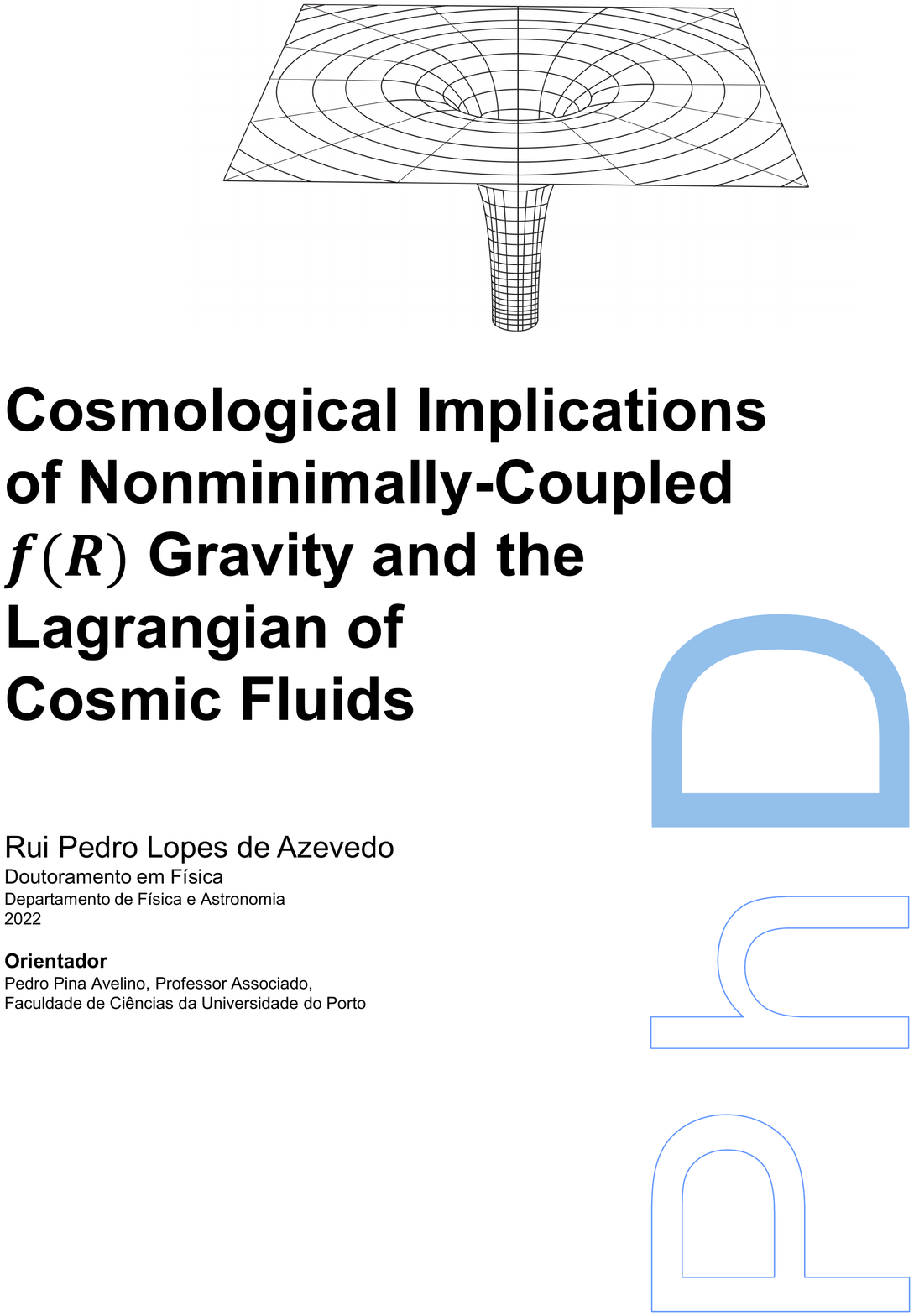}

\cleardoublepage


\cleardoublepage


\vspace*{0.2\textheight}



\noindent\enquote{\itshape This planet has - or rather had - a problem, which was this: most of the people living on it were unhappy for pretty much of the time. Many solutions were suggested for this problem, but most of these were largely concerned with the movement of small green pieces of paper, which was odd because on the whole it wasn't the small green pieces of paper that were unhappy.}\bigbreak

\hfill Douglas Adams, {\it The Hitchhiker's Guide to the Galaxy}


\begin{acknowledgements}
	\addchaptertocentry{\acknowledgementname} 
	
	\vfill
	
	I would like to thank my supervisor, Pedro Avelino. All that help and the fruitful discussions had during the last four years were invaluable to the completion of this thesis.
	
	I would like to thank my collaborator Vasco Ferreira, as well as the members of the IA Diversity \& Inclusion Group, with whom I enjoyed working and that were the source of many interesting discussions. I would also like to thank all the staff at the  physics and astronomy department at FCUP and CAUP for the friendliness and availability they showed throughout my time there.
	
	I would like to thank my therapists, for encouraging me learn more about myself and helping me through darker times.

	I would like to thank my family, for all the support they have shown along the duration of this degree, and for being there when I needed them.
	
	Finally, I am especially grateful to my friends, both at home and abroad, for helping me endure the cloudy skies, and enjoy the sunny days all that much more. I am certain I would not have been able to complete this work if not for their invaluable support. Their friendship and love are gifts that I will treasure for the rest of my life.
	
	\vfill
	
	This work was financially supported by the Portuguese funding agency, FCT --- Fundação
	para a Ciência e Tecnologia, through the Ph.D. research grant SFRH/BD/132546/2017.
	
	\vfill\vfill\vfill

\end{acknowledgements}


\begin{abstract}
\addchaptertocentry{\abstractname} 

In the standard model of cosmology, the background evolution of the Universe can in general be adequately described by general relativity and a uniform and isotropic metric minimally coupled with a collection of perfect fluids. These fluids are usually described by their energy-momentum tensor, which can be derived from the fluid's Lagrangian density. Under general relativity, the Lagrangian density is only relevant to the extent that it results in the correct energy-momentum tensor for a specific perfect fluid.  This is not the case in theories that feature a nonminimal coupling (NMC) between the matter fields and gravity. In such cases, the on-shell Lagrangian density of the matter fields appears explicitly in the equations of motion, in addition to their energy-momentum tensor. The determination of the correct on-shell Lagrangian density for a particular fluid is therefore of paramount importance in order to provide an accurate description of the corresponding cosmological implications. In essence, this is the problem tackled in this thesis.

We have aimed at addressing three key points. We covered some of the results in the literature regarding the Lagrangian density of cosmic fluids, and cleared up some misunderstandings regarding the freedom of choice (or lack thereof) of its on-shell form, both in general relativity and in theories featuring an NMC. In addition, we derived the correct Lagrangian density for fluids composed of solitonic particles with fixed rest mass and structure. Secondly, we studied the thermodynamic behaviour of perfect fluids of this type in the context of theories featuring an NMC between gravity and the matter fields. Finally, we used these results to derive novel cosmological constraints on specific NMC gravity models, using data from cosmic microwave background, big-bang nucleosynthesis, type Ia supernovae and baryon acoustic oscillations observations.
\end{abstract}

\begin{otherlanguage}{portuguese} 
\begin{altabstract}
\addchaptertocentry{\abstractname} 

No modelo padrão da cosmologia, a evolução de fundo do Universo pode em geral ser adequadamente descrita pela relatividade geral e uma métrica uniforme e isotrópica minimamente acoplada a um conjunto de fluidos perfeitos. Estes fluidos são geralmente descritos pelo seu tensor de energia-momento, que pode ser derivado da densidade Lagrangiana do fluido. Em relatividade geral, a densidade Lagrangiana só é relevante na medida em que resulta no tensor de energia-momento correto para um fluido perfeito específico. Este não é o caso em teorias que apresentem um acoplamento não-mínimo (NMC) entre os campos de matéria e a gravidade. Nestes casos, a densidade Lagrangiana on-shell dos campos de matéria aparece explicitamente nas equações de movimento, juntamente com o seu tensor de energia-momento. A determinação da densidade Lagrangiana on-shell correta para um fluido em particular é, portanto, de suma importância para permitir uma descrição correta das implicações cosmológicas correspondentes. Em essência, este é o problema abordado nesta tese.

Tivemos como objetivo abordar três pontos-chave. Cobrimos alguns dos resultados presentes na literatura sobre a densidade Lagrangiana de fluidos cósmicos, e esclarecemos alguns equívocos sobre a liberdade de escolha (ou falta dela) da sua forma on-shell, tanto em relatividade geral quanto em teorias que apresentam um NMC. Além disso, derivámos a densidade Lagrangiana correta para fluidos compostos de partículas solitónicas com massa e estrutura em repouso fixas. Em segundo lugar, estudámos o comportamento termodinâmico de fluidos perfeitos deste tipo no contexto de teorias que apresentam um NMC entre a gravidade e os campos de matéria. Finalmente, usámos esses resultados para derivar novas restrições cosmológicas em modelos de gravidade NMC específicos, usando dados da radiação cósmica de fundo, nucleossíntese do big bang, supernovas do tipo Ia e observações de oscilações acústicas de bariões.

\end{altabstract}
\end{otherlanguage}


\begin{publications}

This work originated eight papers published in peer-reviewed journals:
\begin{itemize}
	\item[\cite{Avelino2018}] Avelino, P. P. \& \textbf{Azevedo, R. P. L.} ``Perfect 
Fluid Lagrangian and its cosmological implications in theories of gravity with nonminimally coupled matter fields''. \textit{Phys. Rev. D} 97, 064018 (2018). 
	\item[\cite{Azevedo2018a}] \textbf{Azevedo, R. P. L.} \& Avelino, P. P. ``Big-bang nucleosynthesis and cosmic microwave background constraints on nonminimally coupled theories of gravity''. \textit{Phys. Rev. D} 98, 064045 (2018).	
	\item[\cite{Azevedo2019b}] \textbf{Azevedo, R. P. L.} \& Avelino, P. P. ``Particle creation and decay in nonminimally coupled models of gravity''. \textit{Phys. Rev. D} 99, 064027 (2019).	
	\item[\cite{Ferreira2020}] Ferreira, V. M. C., Avelino, P. P. \& \textbf{Azevedo, R. P. L.} ``Lagrangian description of cosmic fluids: Mapping dark energy into unified dark	energy''. \textit{Phys. Rev. D} 102, 063525 (2020).
	\item[\cite{Avelino2020}] Avelino, P. P. \& \textbf{Azevedo, R. P. L.} ``Boltzmann's H-theorem, entropy and the strength of gravity in theories with a nonminimal coupling between matter and geometry''. \textit{Phys. Lett. B} 808, 135641 (2020). 
	\item[\cite{Azevedo2020}] \textbf{Azevedo, R. P. L.} \& Avelino, P. P. ``Second law of thermodynamics in nonminimally coupled gravity''. \textit{Europhys. Lett.}	132, 30005 (2020).
	\item[\cite{Azevedo2021}] \textbf{Azevedo, R. P. L.} \& Avelino, P. P. ``Distance-duality in theories with a nonminimal coupling to gravity''. \textit{Phys. Rev. D} 104, 084079 (2021).
	\item[\cite{Avelino2022}] Avelino, P. P. \& \textbf{Azevedo, R. P. L.} ``On-shell Lagrangian of an ideal gas''. \textit{Phys. Rev. D} 105, 104005 (2022).
\end{itemize} 
\end{publications}


\tableofcontents 

\listoffigures 

\listoftables 


\begin{abbreviations}{ll} 

\textbf{BAO} & \textbf{B}aryon \textbf{A}ccoustic \textbf{O}scillations\\
\textbf{BBN} & \textbf{B}ig-\textbf{B}ang \textbf{N}ucleosynthesis\\
\textbf{CDM} & \textbf{C}old \textbf{D}ark \textbf{M}atter\\
\textbf{CMB} & \textbf{C}osmic \textbf{M}icrowave \textbf{B}ackground\\
\textbf{COBE} & \textbf{CO}smic \textbf{B}ackground \textbf{E}xplorer\\
\textbf{DDR} & \textbf{D}istance \textbf{D}uality \textbf{R}elation\\
\textbf{EMT} & \textbf{E}nergy-\textbf{M}omentum \textbf{T}ensor\\
\textbf{EOM} & \textbf{E}quations \textbf{O}f \textbf{M}otion\\
\textbf{EOS} & \textbf{E}quation \textbf{O}f \textbf{S}tate\\
\textbf{FIRAS} & \textbf{F}ar\textbf{I}nfra\textbf{R}ed \textbf{A}bsolute \textbf{S}pectrophotometer\\
\textbf{FLRW} & \textbf{F}riedmann-\textbf{L}ema\^itre-\textbf{R}obertson-\textbf{W}alker\\
\textbf{GR} & \textbf{G}eneral \textbf{R}elativity\\
\textbf{JBD} & \textbf{J}ordan-\textbf{B}rans-\textbf{D}icke\\
\textbf{MCMC} & \textbf{M}arkov \textbf{C}hain \textbf{M}onte \textbf{C}arlo\\
\textbf{MFE} & \textbf{M}odified \textbf{F}riedmann \textbf{E}quation\\
\textbf{MRE} & \textbf{M}odified \textbf{R}aychaudhuri \textbf{E}quation\\
\textbf{NMC} & \textbf{N}on\textbf{M}inimal \textbf{C}oupling\\
\textbf{SnIa} & Type \textbf{Ia} \textbf{S}uper\textbf{n}ova\\
\textbf{UDE} & \textbf{U}nified \textbf{D}ark \textbf{E}nergy\\
\textbf{WMAP} & \textbf{W}ilkinson \textbf{M}icrowave \textbf{A}nisotropy \textbf{P}robe\\

\end{abbreviations}




\mainmatter 

\pagestyle{thesis} 



\chapter{The Standard Model of Cosmology} 

\label{chapter_intro} 

Albert Einstein's general relativity (GR) remains the most successful theory of gravitation, boasting an enormous body of experimental evidence \cite{Will2014}, from the accurate prediction of Mercury's orbit to the more recent detection of gravitational waves \cite{Abbott2016}. 

Despite this backing, when coupled only with baryonic matter GR still fails to account for the rotational speeds of galaxies and offers little explanation for the accelerated expansion of the Universe. Namely, the rotational speed and mass of galaxies as predicted by GR do not match observations, indicating the presence of ``dark'' matter that does not interact electromagnetically \cite{Clowe2006,Bertone2005}. Moreover, in the past two decades, observations of supernovae have signalled that the Universe is presently expanding at an accelerating rate \cite{Carroll2001}. Although an accelerated expansion via a cosmological constant is a possible scenario in GR, the origin of this acceleration is still unknown. For example, if one attempts to connect the small energy density required to explain such acceleration with the much larger vacuum energy predicted by quantum field theory one faces a discrepancy of about 120 orders of magnitude. 

The $\Lambda$CDM model was formulated to model these galactic and universal dynamics, consisting of a Universe evolving under GR and with the addition of uniform dark energy with a negative equation of state (EOS) parameter, represented by a cosmological constant $\Lambda$, responsible for this accelerated expansion, and cold dark matter (CDM), a non-baryonic type of matter that has at best a vanishingly small nonminimal interaction with the other matter fields, responsible for the missing mass in galaxies. This model is often supplemented by an inflationary scenario in the early Universe that attempts to provide a solution to the horizon, flatness and relic problems.

In this chapter, we will give an overview of the standard model of cosmology and cover the most relevant observational constraints to this thesis. Unless otherwise stated, in the remainder of this work we will use units such that $c=\hbar=k_\text{B}=1$, where $c$ is the speed of light, $\hbar$ is the reduced Planck constant and $k_\text{B}$ is Boltzmann's constant.

\section{General relativity}\label{subsec:GR}
In 1915 Einstein arrived at his theory of general relativity, which can be derived from the Einstein-Hilbert action
\begin{equation}
	\label{eq:actionGR}
	S = \int d^4x \sqrt{-g} \left[\kappa(R-2\Lambda)+ \mathcal{L}_\text{m} \right]\,,
\end{equation}
where $g$ is the determinant of the metric tensor, with components $g_{\mu\nu}$, $R$ is the Ricci scalar, $\mathcal{L}_{\rm m}$ is the Lagrangian density of the matter fields, $\Lambda$ is the cosmological constant, and $\kappa = c^4/(16\pi G)$, with $c$ the speed of light in vacuum and $G$ Newton’s gravitational constant. For the remainder of this work, we will refer to Lagrangian densities $\mathcal{L}$ as simply ``Lagrangians''.  Assuming a Levi-Civita connection, the Einstein field equations can be derived by requiring that the variation of the action  with respect to the components of the inverse metric tensor $g^{\mu\nu}$ vanishes, which returns
\begin{equation}
	\label{eq:fieldGR}
	G_{\mu\nu} +\Lambda g_{\mu\nu}= \frac{1}{2\kappa}T_{\mu\nu}\,,
\end{equation}
where $G_{\mu\nu}\equiv R_{\mu\nu}-\frac{1}{2}Rg_{\mu\nu}$ is the Einstein tensor, $R_{\mu\nu}$ is the Ricci tensor and $T_{\mu\nu}$ are the components of the matter energy-momentum tensor (EMT), given by
\begin{equation}
	\label{eq:energy-mom}
	T_{\mu\nu} = - \frac{2}{\sqrt{-g}}\frac{\delta\left(\sqrt{-g}\mathcal{L}_\text{m}\right)}{\delta g^{\mu\nu}}\,.
\end{equation}
If one takes the trace of the Einstein field equations we obtain
\begin{equation}
	\label{eq:EFE-trace}
	R = 4\Lambda -\frac{1}{2\kappa}T \,,
\end{equation}
where $T=T_{\mu\nu}g^{\mu\nu}$ is the trace of the EMT. One can write the field equations in an equivalent alternative form
\begin{equation}
	\label{eq:fieldGR-alt}
	R_{\mu\nu}-\Lambda g_{\mu\nu} = \frac{1}{2\kappa}\left(T_{\mu\nu}-\frac{1}{2}Tg_{\mu\nu}\right) \,.
\end{equation}
Taking the covariant derivative of Eq. \eqref{eq:fieldGR} and using the second Bianchi identity
\begin{equation}
	\label{eq:Bianchi}
	R_{\alpha\beta\mu\nu;\sigma}+	R_{\alpha\beta\sigma\mu;\nu}+R_{\alpha\beta\nu\sigma;\mu}=0\,,
\end{equation}
one can obtain a conservation law for the energy-momentum tensor
\begin{equation}
	\label{eq:conservGR}
	\nabla^\mu T_{\mu\nu} = 0\,.
\end{equation}

The standard model of cosmology is built around the assumption that the Universe is homogeneous and isotropic on cosmological scales, \textit{i.e.} the Cosmological Principle. With this consideration in mind, it can be  described by the Friedmann-Lemaître-Robertson-Walker (FLRW) metric, represented by the line element
\begin{equation}
	\label{eq:line}
	ds^2=-dt^2+a^2(t)\left[\frac{dr^2}{1-kr^2} +r^2 d\theta^2 +r^2\sin^2\theta d\phi^2\right]\,,
\end{equation}
where $a(t)$ is the scale factor, $k$ is the spatial curvature of the Universe, $t$ is the cosmic time, and $r$, $\theta$ and $\phi$ are polar comoving coordinates. Current observational data strongly suggest that the Universe is spatially flat, \textit{i.e.} $k=0$, or very nearly so, and this is usually taken as an assumption when spatial curvature is not important.  As a source term for the Einstein field equations, it is often considered that the matter content of the Universe on sufficiently large scales is well described as a collection of perfect fluids, with EMT of the form
\begin{equation}\label{eq:pf_emt}
	T^{\mu\nu}=(\rho+p)U^\mu U^\nu + p g^{\mu\nu}\,,
\end{equation}
where $\rho$ and $p$ are respectively the density and pressure of the fluid, and $U^\mu$ are the components of the 4-velocity of a fluid element, satisfying $U_\mu U^\mu = -1$. The pressure and density of the fluid are related by the equation of state (EOS), $p=w\rho$, where $w$ is the EOS parameter. For non-relativistic matter, one has $p_\text{m}=0$ and $w_\text{m}= 0$, while for radiation $p_\text{r}=\rho_\text{r}/3$ and $ w_\text{r}= 1/3$. In the case of a constant density, such as the cosmological constant, one would require $w_\Lambda = -1$, so that $p_\Lambda=-\rho_{\Lambda}$.

Using Eqs. \eqref{eq:conservGR} and \eqref{eq:pf_emt} one can obtain the continuity equation for the perfect fluid
\begin{equation}
	\label{eq:contin}
	\dot{\rho}+3H(\rho+p)=0,
\end{equation}	
where $H\equiv\dot{a}/a$ is the Hubble parameter. If $w=\text{const.}$, one can find the general solution for $\rho$ via direct integration 
\begin{equation}
	\label{eq:density}
	\rho(t)=\rho_0 \, a(t)^{-3(1+w)}\,,
\end{equation}
where $\rho_0$ is the value of the energy density at time $t_0$ and $a(t_0)\equiv 1$. We will therefore consider three different components for the energy density in the Universe: a matter term $\rho_{\rm m}$ composed of both baryonic and cold dark matter, a radiation term $\rho_{\rm r}$ composed essentially of photons, and a dark energy term $\rho_\Lambda$ which will be expressed by a cosmological constant. Using Eq. \eqref{eq:density}, these components evolve as, respectively,
\begin{align}
	\label{eq:densities}
	\rho_{\rm m}(t)&=\rho_{{\rm m},0} \, a(t)^{-3}\,, \nonumber \\
	\rho_{\rm r}(t)&=\rho_{{\rm r},0} \, a(t)^{-4}\,, \nonumber \\
	\rho_\Lambda(t)&=\rho_{\Lambda,0} = 2\kappa\Lambda\,.
\end{align} 

Having both defined a metric and a source term for the field equation, one can now write the explicit forms of its $tt$ and $ii$ components as the well-known Friedmann and Raychaudhuri equations
\begin{equation}
	\label{eq:friedGR}
	H^2=\frac{1}{ 6\kappa}(\rho_{\rm m}+\rho_{\rm r})-\frac{k}{a^2}+\frac{\Lambda }{3}\,,
\end{equation}
\begin{equation}
	\label{eq:rayGR}
	2\dot{H}+3H^2=-\frac{1}{ 2 \kappa}(\rho_{\rm m}+\rho_{\rm r}+3p_{\rm r})+\Lambda\,,
\end{equation}
respectively, and the Ricci scalar yields
\begin{equation}
	\label{eq:Ricci_flat}
	R=6\left(2H^2+\dot{H}\right)\,.
\end{equation}

It is sometimes useful to recast the scale factor $a(t)$ as a function of the redshift $z$ that photons experience when propagating through spacetime. Light travels along null geodesics
\begin{equation}
	\label{eq:null geodesic}
	ds^2 = - dt^2 + a(t)^2\frac{dr^2 }{1-kr^2} = 0 \,,
\end{equation}
therefore light that leaves a source at comoving distance $r_\text{em}$ at time $t_\text{em}$ will arrive at the observer $r_\text{obs}=0$ at a time $t_\text{obs}$, given by
\begin{equation}
	\label{eq:null geodesic2}
	\int_{t_\text{em}}^{t_\text{obs}} \frac{dt}{a(t)}= \int_0^{r_\text{em}} \frac{dr}{\sqrt{1-kr^2}} \,.
\end{equation}
Suppose that we observe a signal that is emitted over a  time $\delta t_\text{em}$, and received over a time $\delta t_\text{obs}$. The r.h.s. of Eq. \eqref{eq:null geodesic2} remains the same regardless of the limits of the integral on the l.h.s., so we can write
\begin{align}
	\label{eq:null geodesic3}
	&\int_{t_\text{em}+\delta t_\text{em}}^{t_\text{obs}+\delta t_\text{obs}} \frac{dt}{a(t)} = \int_{t_\text{em}}^{t_\text{obs}} \frac{dt}{a(t)} \nonumber\\
	\Rightarrow&\int_{t_\text{em}+\delta t_\text{em}}^{t_\text{obs}+\delta t_\text{obs}} \frac{dt}{a(t)} = \int_{t_\text{em}}^{t_\text{em}+\delta t_\text{em}} \frac{dt}{a(t)} +\int_{t_\text{em}+\delta t_\text{em}}^{t_\text{obs}} \frac{dt}{a(t)} \nonumber	\\
	\Rightarrow&\int_{t_\text{em}+\delta t_\text{em}}^{t_\text{obs}+\delta t_\text{obs}} \frac{dt}{a(t)} +\int^{t_\text{em}+\delta t_\text{em}}_{t_\text{obs}} \frac{dt}{a(t)}= \int_{t_\text{em}}^{t_\text{em}+\delta t_\text{em}} \frac{dt}{a(t)} \nonumber \\
	\Rightarrow&\int_{t_\text{obs}}^{t_\text{obs}+\delta t_\text{obs}} \frac{dt}{a(t)} = \int_{t_\text{em}}^{t_\text{em}+\delta t_\text{em}} \frac{dt}{a(t)} \, .
\end{align}	
If we assume that the intervals $\delta t_\text{obs/em}$ are much smaller than the timescale of the variation of $a(t)$, then $a(t_\text{em})= a(t_\text{em}+\delta t_\text{em})$, and likewise for $a(t_\text{obs})$. Therefore we can rewrite Eq. \eqref{eq:null geodesic3} as
\begin{align}
	\label{eq:null geodesic4}
	&\frac{1}{a(t_\text{obs})}\int_{t_\text{obs}}^{t_\text{obs}+\delta t_\text{obs}} dt = \frac{1}{a(t_\text{em})}\int_{t_\text{em}}^{t_\text{em}+\delta t_\text{em}} dt \nonumber \\
	\Rightarrow&\frac{\delta t_\text{obs}}{a(t_\text{obs})}= \frac{\delta t_\text{em}}{a(t_\text{em})} \,.
\end{align}
If the intervals $\delta t_\text{em}$ and $\delta t_\text{obs}$ are the time it takes to emit and receive one full wavelength $\lambda$ of a light ray, then 
\begin{equation}
	\label{eq:wavelenght_dif}
	\frac{\lambda_\text{obs}}{a(t_\text{obs})}= \frac{\lambda_\text{em}}{a(t_\text{em})} \,.
\end{equation}
The redshift $z$ is defined as
\begin{equation}
	\label{eq:redshift_def}
	z= \frac{\lambda_\text{obs}-\lambda_\text{em}} {\lambda_\text{em}} \,,
\end{equation}
and using Eq. \eqref{eq:wavelenght_dif} we obtain
\begin{align}
	\label{eq:redshift_scale}
	&z = \frac{a(t_\text{obs})}{a(t_\text{em})}-1 \nonumber\\
	\Rightarrow &1+z = \frac{1}{a(t)} \,,
\end{align}
where we have taken $t\equiv t_\text{em}$ and fixed the scale factor at the present time at unity $a(t_\text{obs})\equiv a_0=1$.  

The density of a flat universe is known as the critical density $\rho_\text{c}=6\kappa H^2$. One can therefore write the various contributions to the right hand side of Eq. \eqref{eq:friedGR} as a function of the corresponding density parameters $\Omega\equiv\rho/\rho_\text{c}$. If one also includes the explicit dependence of the density on the scale factor $a$, one can rewrite the Friedmann equation as
\begin{equation}
	\label{eq:friedGR_dens}
	H^2=H^2_0\left(\Omega_{\text{r},0}a^{-4}+\Omega_{\text{m},0}a^{-3}+\Omega_{\text{k},0}a^{-2}+\Omega_{\Lambda,0} \right) \,,
\end{equation}
where $H_0$ is the Hubble constant, $\Omega_{\text{r},0}$ and $\Omega_{\text{m},0}$ are the radiation and matter density parameters, respectively, 
\begin{equation}
	\Omega_{\text{k},0}=1-\Omega_0=1-6\kappa H^2_0 \,,
\end{equation}
is the spatial curvature density parameter, and \begin{equation}
	\Omega_{\Lambda,0}=\Lambda/3H_0^2 \,,
\end{equation}
is the cosmological constant density parameter. The subscript $0$ denotes that the quantities are measured at the present time.

Current observational evidence allows one to roughly split the background evolution of the Universe into four stages. Soon after the Big Bang, it is hypothesized that the Universe went through a rapid inflationary period. While the telltale signs of inflation have evaded detection so far, it provides a neat solution to the horizon, flatness and relic problems, and provides a mechanism that explains the origin of the density fluctuations which gave rise to the large-scale structure observed at the present day \cite{Liddle2000}. There are many models of inflation, most of them relying on the addition of a scalar field to the Einstein-Hilbert action, but they are not an essential part of the $\Lambda$CDM model, so we will not cover them here.

After inflation the Universe became dominated by radiation, with $H^2\simeq H^2_0\Omega_{\text{r},0} a^{-4}$ and $w_\text{r}=p_\text{r}/\rho_\text{r}=1/3$. Via the continuity \eqref{eq:contin} and Friedmann \eqref{eq:friedGR_dens} equations we obtain $a(t)\propto t^{1/2}$ and $\rho_\text{r}\propto a^{-4}\propto t^{-2}$ \cite{Weinberg2008}. It is during this epoch that the majority of the light elements (mainly stable isotopes of hydrogen, helium and lithium) were formed via a process called big-bang nucleosynthesis (BBN).

Following BBN the Universe remained so hot that nucleons and electrons were not able to come together into atoms, and it remained opaque to radiation. As it expanded and cooled, the temperature dropped enough for electrons to finally be captured into (mostly) hydrogen atoms, in a process known as recombination. It is during this epoch that photons decoupled from matter, and the cosmic microwave background (CMB) was emitted.

Following recombination came an era of matter domination. It can be described by $H^2 \simeq H^2_0\Omega_{\text{m},0}a^{-3}$, which leads to the solutions $a(t)\propto t^{2/3}$ and $\rho_\text{m}\propto a^{-3}\propto t^{-2}$.

We are currently experiencing a transition period where the expansion of the Universe is accelerating. In the future, the Universe will most likely be dominated by dark energy, and in the case of a cosmological constant one obtains $H^2 \simeq H^2_0\Omega_{\Lambda,0}$, leading to an exponential expansion $a(t)\propto \exp(H_0 t)$.


\section{Distance measures in general relativity}

Measuring distances in astronomy can be done in several different ways, all of which necessarily take into account the curved nature of spacetime. Therefore, the measurement of these distances will depend on the spacetime metric and how it changes as light propagates through it. Two very useful distance definitions are the angular diameter distance, $d_\text{A}$, and the luminosity distance, $d_\text{L}$.

In Euclidean space, an object with diameter $s$ at some distance $d$ away would extend across an angular diameter $\theta$, i.e.
\begin{equation}
	\label{eq:euc_dA}
	\theta = \frac{s}{d} \,.
\end{equation}
In curved spacetime, however, the proper distance to a distant object such as a star or a galaxy cannot be so easily measured. Nevertheless, one can define an angular diameter distance $d_\text{A}$ as a recasting of $d$ such that the relation between the object's size and angular diameter would maintain the Euclidean-space relation \eqref{eq:euc_dA}, \textit{i.e.}
\begin{equation}
	\label{eq:angle}
	\theta=\frac{s}{d_\text{A}}\,.
\end{equation}
In a FLRW universe, the proper distance $s$ corresponding to an angle $\theta$ is simply
\begin{equation}
	\label{eq:prop_dist}
	s= a(t_\text{em})r\theta= \frac{r\theta}{1+z}\,,
\end{equation}
where $r$ is the coordinate distance to the object.
Assuming a spatially flat Universe, $k=0$, the distance $r$ can be calculated by just integrating over a null geodesic, that is
\begin{align}
	\label{eq:null geodesic_distance}
	ds^2 &= - dt^2 + a(t)^2dr^2 = 0\nonumber \\
	\Rightarrow dr &= -\frac{dt}{a(t)} \nonumber \\
	\Rightarrow  r &= \int_{t_\text{em}}^{t_\text{obs}} \frac{dt}{a(t)}= \int_0^z \frac{dz'}{H(z')} \,,
\end{align}
so the angular diameter distance to an object at redshift $z$ is simply
\begin{equation}
	\label{eq:ang_dist}
	d_\text{A}=\frac{1}{1+z}\int_0^z \frac{dz'}{H(z')}\,.
\end{equation}

On the other hand, the luminosity distance $d_\text{L}$ of an astronomical object relates its absolute luminosity $L$, \textit{i.e.} its radiated energy per unit time, and its energy flux at the detector $l$, so that they maintain the usual Euclidean-space relation
\begin{equation}
	\label{eq:apparent luminosity}
	l = \frac{L}{4\pi d_\text{L}^2}\,,
\end{equation}
or, rearranging,
\begin{equation}
	\label{eq:lum_dist_1}
	d_\text{L} = \sqrt{\frac{L}{4\pi l}}\,.
\end{equation}
Over a small emission time $\Delta t_\text{em}$ the absolute luminosity can be written as
\begin{equation}
	\label{eq:abs_lum_1}
	L = \frac{N_{\gamma,\text{em}} E_\text{em}}{\Delta t_\text{em}}\,,
\end{equation}
where $N_{\gamma,\text{em}}$ is the number of emitted photons and $E_\text{em}$ is the average photon energy. An observer at a coordinate distance $r$ from the source will on the other hand observe an energy flux given by
\begin{equation}
	\label{eq:app_lum}
	l = \frac{N_{\gamma,\text{obs}}E_\text{obs}}{\Delta t_\text{obs} 4\pi r^2}
\end{equation}
where $N_{\gamma,\text{obs}}$ is the number of observed photons and $E_\text{obs}$ is their average energy.

Note that while the number of photons is generally assumed to be conserved, $N_{\gamma,\text{obs}} = N_{\gamma,\text{em}}$, the time that it takes to receive the photons is increased by a factor of $1+z$, $\delta t_\text{obs}= (1+z)\delta t_\text{em}$ as shown in Eq. \eqref{eq:null geodesic4}, and their energy is reduced by the same factor 
\begin{equation}
	\label{eq:energy_obs}
	E_\text{obs} = \frac{E_\text{em}}{1+z} \,.
\end{equation}
Using Eqs. \eqref{eq:abs_lum_1}, \eqref{eq:app_lum}, \eqref{eq:energy_obs} and \eqref{eq:null geodesic} in Eq. \eqref{eq:lum_dist_1}, we finally obtain
\begin{equation}
	\label{eq:lum_dist}
	d_\text{L} = (1+z)\int_0^z \frac{dz'}{H(z')} \, .
\end{equation}

Eqs. \eqref{eq:ang_dist} and \eqref{eq:lum_dist} differ only in a factor of $(1+z)^2$,
\begin{equation}
	\label{eq:DDR}
	\frac{d_\text{L}}{d_\text{A}}=(1+z)^2\,.
\end{equation}
Eq. \eqref{eq:DDR} is called Etherington's distance-duality relation (DDR), and it can be a useful tool for probing theories of modified gravity or theories that do not conserve the photon number \cite{Bassett2004,Ruan2018}.


\section{The cosmic microwave background}

In 1965, Arno Penzias and Robert Wilson encountered a source of isotropic  ``noise'' in the microwave part of the radiation spectrum as they were experimenting with the Holmdel Horn Antenna. At the same time, Robert Dicke, Jim Peebles and David Wilkinson were preparing a search for primordial radiation emitted soon after recombination. The concurrence of these two events allowed these physicists to quickly identify Penzias and Wilson's accidental discovery as the Cosmic Microwave Background (CMB) radiation \cite{Penzias1965,Dicke1965}, one of the most important astronomical discoveries ever made, and one that granted Penzias, Wilson and Peebles the Nobel prize. Despite the nature of its discovery, the CMB had been a prediction of cosmology for some time, and it is one that has continued to be studied to this day using sophisticated ground- and balloon-based telescopes, and several space-based observatories such as the Cosmic Microwave Background Explorer (COBE) \cite{Bennett1996}, the Wilkinson Microwave Anisotropy Probe (WMAP) \cite{Bennett2013}, and more recently the \textit{Planck} satellite \cite{Aghanim2020a}.

Before recombination, radiation was kept in thermal equilibrium with baryonic matter via frequent collisions between photons and free electrons, and thus the photons maintained a black body-spectrum. For a given temperature $\mathcal{T}$ the spectral and number densities of photons at frequency $\nu$ are given by, respectively,
\begin{equation}
	\label{eq:black_spectral}
	u(\nu)=\frac{8 \pi h \nu^3}{e^{h \nu/(k_B \mathcal{T})}-1} \,,
\end{equation}
and
\begin{equation}
	\label{eq:black_number}
	n(\nu) = \frac{u(\nu)}{h \nu} = \frac{8 \pi \nu^2}{e^{h \nu/(k_B \mathcal{T})}-1} \,,
\end{equation}
where $h$ is the Planck constant.

As spacetime expanded, matter cooled enough for the free electrons to be captured by the positive nuclei, and radiation began propagating freely across space. While this radiation is isotropic to a large degree, it does present small anisotropies. These can be primordial in origin, but can also be due to effects along the path photons travelled until detection. Some of these effects are temperature fluctuations at the last-scattering surface, Doppler effects due to velocity variations in the plasma, the peculiar velocity of the Earth relative to the CMB, gravitational redshift (Sachs-Wolfe and integrated Sachs-Wolfe effects), and photon scattering with electrons in the intergalactic medium (Sunyaev–Zel’dovich effect). Despite this, the CMB remains the best black body ever measured. The detailed description of these anisotropies is not particularly relevant to this thesis. Nevertheless, they are central to the determination of constraints on the cosmological parameters \cite{Aghanim2020}.

As spacetime continued to expand after photon decoupling, the frequency of the photons emitted at the so-called last scattering surface becomes redshifted over time. That is to say, a photon emitted at the time of last scattering $t_\text{L}$ with frequency $\nu_\text{L}$ would be detected at time $t$ with a frequency $\nu = \nu_\text{L} a(t_\text{L})/a(t)$ and therefore the spectral and number densities measured at time $t$ would be
\begin{equation}
	\label{eq:black_spectral_red}
	u(\nu)=\frac{8 \pi h \nu^3}{e^{h \nu/(k_B \mathcal{T}(t))}-1} \,,
\end{equation}
and
\begin{equation}
	\label{eq:black_number_red}
	n(\nu) = \frac{8 \pi \nu^2}{e^{h \nu/(k_B \mathcal{T}(t))}-1} \,,
\end{equation}
where
\begin{equation}
	\label{eq:black_temp_red}
	\mathcal{T}(t)= \mathcal{T}(t_\text{L})\frac{a(t_\text{L})}{a(t)} \,,
\end{equation}
is the temperature of the black-body spectrum. Alternatively, we can write this temperature as,
\begin{equation}
	\label{eq:redshift_cmb_temp}
	\mathcal{T}(z)= \mathcal{T}(z_\text{L})\frac{1+z}{1+z_\text{L}} \,,	
\end{equation}
where $z_\text{L}$ is the redshift at the time of last scattering. In essence, the black-body spectrum is maintained, but with a ``redshifted'' temperature.


\section{Big-bang nucleosynthesis}

One of the great successes of modern cosmology is the prediction of the abundances of light elements formed in the early Universe (mainly $^2$H, $^3$He, $^4$He and $^7$Li), which are impossible to achieve in such quantities via stellar processes alone, through a process called big-bang nucleosynthesis (BBN). Current astronomy is unable to observe past the CMB radiation emitted at the last scattering surface ($z\simeq1090$), but it is possible to infer from the high degree of isotropy of the CMB itself that matter and radiation were in thermal equilibrium prior to decoupling. We can therefore use statistical physics and thermodynamics to predict the evolution of the early Universe when the temperature and density were much higher than the present day and radiation dominated the expansion of the universe.

The primordial soup was originally composed of all known species of elementary particles, ranging from the heaviest particles like the top quark down to neutrinos and photons. Then, as the Universe progressively expanded and cooled, the heavier species slowly became non-relativistic, and particle-antiparticle pairs annihilated themselves and unstable particles decayed. After the electroweak transition at around $\mathcal{T}_\text{EW}\sim100\text{ TeV}$, the first particles to disappear were the top quark, followed soon after by the Higgs boson and the $W^\pm$ and $Z^0$ gauge bosons. As the temperature dropped below $\mathcal{T}\sim10\text{ GeV}$, the bottom and charm quarks, as well as the tau lepton, were also annihilated. 

At $\mathcal{T}_\text{QCD}\sim150 \text{ MeV}$, the temperature was low enough that quark-gluon interactions started binding quarks and antiquarks together into hadrons, composite particles with integer charge. These can be baryons, three-quark particles with half-integer spin (\textit{i.e.} fermions) such as the protons and neutrons, or mesons, quark-antiquark pairs with integer spin (\textit{i.e.} bosons) such as pions. This process is known as the quantum chromodynamics (QCD) phase transition. Soon after at $\mathcal{T}<100 \text{ MeV}$ the pions and muons also annihilated, and after this process finished the only particles left in large quantities were neutrons, protons, electrons, positrons, neutrinos and photons.

As the Universe continued to cool down, the weak interaction rates eventually dropped below the expansion rate $H$, with neutrinos departing from thermodynamic equilibrium with the remaining plasma. In the context of this thesis, the most relevant consequence of this phenomenon is the breaking of the neutron-proton chemical equilibrium at $\mathcal{T}_\text{D}\sim 0.7 \text{ MeV}$, which leads to the freeze-out of the neutron-proton number density ratio at $n_n/n_p=\exp(-\Delta m/\mathcal{T}_\text{D}) \sim 1/7$, where $\Delta m = 1.29 \text{ MeV}$ is the neutron-proton mass difference (the ratio $n_n/n_p$ is then  slightly reduced by subsequent neutrons decays). Soon after, at $\mathcal{T}_\text{N}\sim 100 \text{ keV}$, the extremely high photon energy density has been diluted enough to allow for the formation of the first stable $^2$H deuterium nuclei.

Once $^2$H starts forming, an entire nuclear process network is set in motion, leading to the production of light-element isotopes and leaving all the decayed neutrons bound into them, the vast majority in helium-4 $^4$He nuclei, but also in deuterium $^2$H, helium-3 $^3$He and lithium-7 $^7$Li. Unstable tritium $^3$H and beryllium-7 $^7$Be nuclei were also formed, but quickly decayed into $^3$He and  $^7$Li, respectively. Primordial nucleosynthesis may be described by the evolution of a set of differential equations \cite{Wagoner1966,Wagoner1967,Wagoner1969,Wagoner1973,Esposito2000a,Esposito2000}, namely the Friedmann equation \eqref{eq:friedGR}, the continuity equation \eqref{eq:contin}, the equation for baryon number conservation
\begin{equation}\label{eq:bary_cons}
	\dot{n}_\text{B}+3Hn_\text{B}=0 \,,
\end{equation}
the Boltzmann equations describing the evolution of the average density of each species
\begin{equation}
	\label{eq:nuclide_evo}
	\dot{X}_i = \sum_{j,k,l}N_i\left(\Gamma_{kl\rightarrow ij}\frac{X_l^{N_l}X_k^{N_k}}{N_l!N_k!} -\Gamma_{ij\rightarrow kl}\frac{X_i^{N_i}X_j^{N_j}}{N_i!N_j!}\right)\equiv \Gamma_i \,,
\end{equation}
and the equation for the conservation of charge
\begin{equation}
	\label{eq:charge_cons}
	n_\text{B}\sum_j Z_j X_j = n_{e^-}-n_{e^+} \,,
\end{equation}
where $i$, $j$, $k$, $l$ denote nuclear species ($n$, $p$, $^2$H, ...), $X_i=n_i/n_\text{B}$, $N_i$ is the number of nuclides of type $i$ entering a given reaction, the $\Gamma$s represent the reaction rates, $Z_i$ is the charge number of the $i$-th nuclide, and $n_{e^\pm}$ is the number density of electrons ($-$) and positrons ($+$).

As one could expect, the accurate computation of element abundances cannot be done without resorting to numerical algorithms \cite{Wagoner1973,Kawano1992,Smith:1992yy,Pisanti2008,Consiglio2017}. It is worthy of note at this stage that these codes require one particular parameter to be set a priori, the baryon-to-photon ratio $\eta$, whose value is fixed shortly after BBN and in the standard model of cosmology is expected to remain constant after that.

\section{Constraints on the cosmological parameters}\label{subsec:cosmo_cons}

In the first decades following the formulation of GR, cosmology remained a largely theoretical pursuit, with the discovery of several useful metrics (such as de Sitter and FLRW), the formulation of the cosmological principle and the early predictions of the CMB and the big bang. Concurrently there were significant advancements in astronomy, but the data collected was not yet enough to derive precise constraints on cosmological parameters, nor to detect the accelerated expansion of the Universe.

Nevertheless, several key observations were made in the decades that followed. In 1929 Edwin Hubble shows that the Universe is expanding by measuring the redshift-distance relation to a linear degree \cite{Hubble1931}; in 1965 Penzias and Wilson make the first observations of the CMB \cite{Penzias1965,Dicke1965}; and in 1970 Vera Rubin presents evidence for the existence of a substantial amount of dark matter by measuring the rotation curve of galaxies \cite{Rubin1970}. Several key developments were made during the 1980s, such as the independent proposals and work on inflation by Alan Guth \cite{Guth1981}, Alexei Starobinsky \cite{Starobinsky1980}, Andrei Linde \cite{Linde1982}, and Andreas Albrecht and Paul Steinhardt \cite{Albrecht1982}, the increasing evidence pointing to a Universe currently dominated by cold dark matter, and culminating with the 1989 launch of COBE \cite{Bennett1996} and the detection of CMB anisotropies.

COBE's launch marked the beginning of the \textit{golden age of cosmology}. The exponential increase in the establishment of space- and ground-based observatories like the Hubble Space Telescope (HST), the Wilkinson Microwave Anisotropy Probe (WMAP), the \textit{Planck} satellite, and the Very Large Telescope (VLT) has progressed to today and shows no signs of slowing. As of this writing, four significant projects fully underway are the construction of the largest optical/near-infrared telescope in the form of the Extremely Large Telescope (ELT), the calibration of HST's successor \textit{i.e.} the James Webb Space Telescope (JWST), the planning of the largest radio observatory ever built in the Square Kilometer Array (SKA), and the planning of the first space-based gravitational wave detector in the Laser Interferometer Space Antenna (LISA). Also worthy of mention are the many ongoing surveys like the Dark Energy Survey (DES), the Sloan Digital Sky Survey (SDSS) and the Galaxy and Mass Assembly (GAMA).

As such, the constraints on the cosmological model ruling our Universe have greatly improved in the last two decades. In what follows we will present current constraints on the cosmological parameters relevant to this work, most of which can be seen in Table \ref{tab:cosmo_constraints}.

\subsection{Hubble constant $H_0$}

The most precise constraints available today for $H_0$ come from CMB data collected by the aforementioned \textit{Planck} mission. Unfortunately, both $H_0$ and $\Omega_{\text{m},0}$ are derived parameters from that analysis, and suffer from degeneracy with each other. Nevertheless, with the inclusion of lensing considerations, the Planck collaboration was able to constrain the Hubble constant for a flat Universe to \cite{Aghanim2020}
\begin{equation}
	\label{eq:H0_planck}
	H_0=(67.36\pm0.54)\text{km s}^{-1}\text{Mpc}^{-1} \,,
\end{equation}
at the 68\% credibility level (CL). This result can be further supplemented with data from baryon acoustic oscillations (BAO), for a slightly tighter constraint of (68\% CL)
\begin{equation}
	\label{eq:H0_planck_bao}
	H_0=(67.66\pm0.42)\text{km s}^{-1}\text{Mpc}^{-1} \,.
\end{equation}

There is, however, some tension in the measurement of $H_0$. The measurement of the cosmic distance ladder, and in particular the relation between an object's geometric distance and its redshift, can also be used to determine cosmological parameters. The recent Supernova H0 for the Equation of State (SH0ES) project used the HST to make an independent measurement of the Hubble constant (68\% CL) \cite{Riess2021}
\begin{equation}
	\label{eq:H0_shoes}
	H_0=(73.04\pm1.04)\text{km s}^{-1}\text{Mpc}^{-1} \,.
\end{equation}
While this value is less precise, the central values of both determinations have barely changed throughout the years while the errors are continuously decreasing, leading to the discrepancy of $2.5\sigma$ in 2013 (the time of the first Planck data release) growing to over $5\sigma$ today. Numerous investigations have been conducted, trying to explain this tension either with new physics or unaccounted-for systematic errors \cite{Divalentino2021} but, as of this writing, there is no definite answer.

\subsection{Density parameters}

As mentioned above, \textit{Planck} data is able to determine the matter density parameter $\Omega_{\text{m},0}$ along with $H_0$. With the inclusion of lensing, the 68\% constraint from the Planck collaboration for a flat Universe is
\begin{equation}
	\label{eq:omegaM_planck}
	\Omega_{\text{m},0}=0.3153\pm0.0073 \,,
\end{equation}
which can once again be constrained further with the inclusion of BAO data
\begin{equation}
	\label{eq:omegaM_planck_BAO}
	\Omega_{\text{m},0}=0.3111\pm0.0056 \,.
\end{equation}
These values naturally lead to a determination of the cosmological constant density parameter from CMB data 
\begin{equation}
	\label{eq:omegaL_planck}
	\Omega_{\Lambda,0}=0.6847\pm0.0073 \,.
\end{equation}
and with the addition of BAO
\begin{equation}
	\label{eq:omegaL_planck_BAO}
	\Omega_{\Lambda,0}=0.6889\pm0.0056 \,.
\end{equation}

In addition, \textit{Planck} is able to constrain values of $\Omega_{\text{b},0}h^2$ and $\Omega_{\text{c},0}h^2$, where $\Omega_{\text{b},0}$ and $\Omega_{\text{c},0}$ are the density parameters for baryonic matter and cold dark matter, respectively, and $h=H_0/(100\text{km s}^{-1}\text{Mpc}^{-1})$ is the dimensionless Hubble constant. The base estimates for these parameters from CMB data are
\begin{equation}
	\label{eq:omegab_planck}
	\Omega_{\text{b},0}h^2= 0.02237\pm0.00015 \,,
\end{equation}
\begin{equation}
	\label{eq:omegac_planck}
	\Omega_{\text{c},0}h^2= 0.1200\pm0.0012 \,,
\end{equation}
and with the addition of BAO data
\begin{equation}
	\label{eq:omegab_planck_BAO}
	\Omega_{\text{b},0}h^2= 0.02242\pm0.00014 \,,
\end{equation}
\begin{equation}
	\label{eq:omegac_planck_BAO}
	\Omega_{\text{c},0}h^2= 0.11933\pm0.00091 \,.
\end{equation}

\subsection{Cosmic microwave background temperature}

\begin{figure}[!ht]
	\centering
	\includegraphics[width=0.85\textwidth]{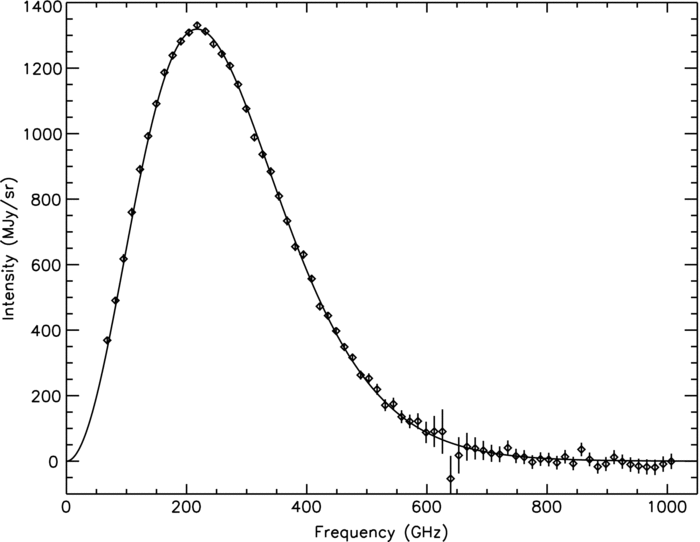}
	\caption[CMB spectrum from the COBE-FIRAS instrument]{The mean spectrum associated with the velocity of the solar system with respect to the CMB. The line is the a priori prediction	based on the WMAP velocity and  previous FIRAS calibration. The uncertainties are the noise from the FIRAS measurements. Figure taken from \cite{Fixsen2009}.\label{fig:cmb_firas_wmap}}
\end{figure}

The accurate characterization of the CMB spectrum has been the goal of many observational experiments so far, but as of this writing the best measurements of the black body spectrum still come from the FarInfraRed Absolute Spectrophotometer (FIRAS) instrument on board COBE \cite{Fixsen1996}, which has more recently been calibrated using WMAP data \cite{Fixsen2009}, and resulted in a temperature measurement of
\begin{equation}
	\mathcal{T}=2.760\pm0.0013\,\text{K} \,,
\end{equation}
as can be seen in Fig. \ref{fig:cmb_firas_wmap}.

\subsection{Baryon-to-photon ratio}

There are two main ways of estimating the value of the baryon-to-photon ratio $\eta$ at different stages of cosmological evolution. On one hand, one may combine the observational constraints on the light-element abundances with numerical simulations of primordial BBN to infer the allowed range of $\eta$. This is the method used in \cite{Iocco2009}, among others, leading to
\begin{equation}
	\label{eq:etabbn}
	\eta_\text{BBN}=(5.7\pm0.3)\times 10^{-10} \,,
\end{equation}
at 68\% CL just after nucleosynthesis (at a redshift $z_\text{BBN} \sim10^9$). More recently, an updated version of the program {\fontfamily{qcr}\selectfont PArthENoPE} ({\fontfamily{qcr}\selectfont PArthENoPE} 2.0), which computes the abundances of light elements produced during BBN, was used to obtain new limits on the baryon-to-photon ratio, at $1\sigma$ \cite{Consiglio2017}
\begin{equation}
	\label{eq:etabbn2}
	\eta_\text{BBN}=(6.23^{+0.12}_{-0.14})\times 10^{-10}\,.
\end{equation}

There is actually some variation of $\eta$ during nucleosynthesis due to the entropy transfer to photons associated with the $e^\pm$ annihilation. The ratio between the values of $\eta$ at the beginning and at the end of BBN is given approximately by a factor of $2.73$ \cite{Serpico2004}.

The baryon-to-photon ratio also affects the acoustic peaks observed in the CMB, generated at a redshift $z_\text{CMB} \sim10^3$. The 2017 Planck analysis \cite{Ade2016} constrains the baryon density $\omega_\text{b} = \Omega_\text{b} h^2$ from baryon acoustic oscillations, at 95\% CL,
\begin{equation}
	\label{eq:omegacmb}
	\omega_\text{b}=0.02229^{+0.00029}_{-0.00027}\, .
\end{equation}
This quantity is related to the baryon-to-photon ratio via $\eta = 273.7\times10^{-10} \omega_\text{b}$, leading to
\begin{equation}
	\label{eq:etacmb}
	\eta_\text{CMB}= 6.101^{+0.079}_{-0.074}\times 10^{-10}\,.
\end{equation}

\begin{table}[]
	\centering
	\footnotesize
	\caption[Current constraints on cosmological parameters]{Mean values and marginalized 68\% CI limits on the cosmological parameters, according to recent observational data. $H_0$ has units of km s$^{-1}$ Mpc$^{-1}$. \label{tab:cosmo_constraints}}
	\tabcolsep=0.11cm 
	\begin{tabular}{l|ccccc}
		\hline\hline
		& \textit{Planck} & \textit{Planck}+BAO & SH0ES          & BBN \cite{Iocco2009}  & BBN \cite{Consiglio2017}              \\ \hline
		{\boldmath$H_0$}                    & $67.36\pm0.54$       & $67.66\pm0.42$            & $73.52\pm1.62$ & \----                        & \----                                        \\
		{\boldmath$\Omega_{\text{m},0}$}    & $0.3153\pm0.0073$    & $0.3111\pm0.0056$         & \----          & \----                        & \----                                        \\
		{\boldmath$\Omega_{\Lambda,0}$}     & $0.6847\pm0.0073$    & $0.6889\pm0.0056$         & \----          & \----                        & \----                                        \\
		{\boldmath$\Omega_{\text{b},0}h^2$} & $0.02237\pm0.00015$  & $0.02242\pm0.00014$       & \----          & \----                        & \----                                        \\
		{\boldmath$\Omega_{\text{c},0}h^2$} & $0.1200\pm0.0012$    & $0.11933\pm0.00091$       & \----          & \----                        & \----                                        \\
		{\boldmath$\eta\times10^{10}$}      & $6.123\pm0.041$      & $6.136\pm0.038$           & \----          & $5.7\pm0.3$                  & $6.23^{+0.12}_{-0.14}$                        \\ \hline\hline
	\end{tabular}
\end{table}


\chapter{$f(R)$ Theories} 

\label{chapter_modgrav} 


Even though the standard $\Lambda$CDM model has great explanatory power and is very tightly constrained, it still has two core problems: the necessity to include dark energy and dark matter, both of which have managed to avoid direct detection. Thus, many physicists have proposed alternative theories to GR that do not require the addition of at least one of these dark components to the energy budget to explain cosmological phenomena.

The standard of proof required of these new theories is very high, as they have to satisfy stringent cosmological and astrophysical constraints. Another hurdle faced by these modified gravity theories is their origin: are they fundamental theories, or merely the low-energy manifestation of a grander theory of everything? As of this writing, GR remains the most successful theory, but it can nevertheless be fruitful to take a deeper look at modified gravity since it can often shine a light on other interesting problems and phenomena.

Many extensions of GR have been proposed in the literature. Among these are theories with additional fields, such as scalar-tensor theories (\textit{e.g.} Jordan-Brans-Dicke), Einstein-\ae{}ther theories (\textit{e.g.} modified Newtonian dynamics aka MoND) and bimetric theories, theories with more complex geometric terms such as $f(R)$  and $f(R, R_{\mu\nu}, R_{\mu\nu\alpha\beta})$ theories \cite{DeFelice2010,Sotiriou2010,Capozziello2009,Capozziello2007,Capozziello2005,Capozziello2003,Allemandi2004,Chiba2007}, and theories featuring a non-minimal coupling (NMC) between geometry and matter, such as $f(R,\mathcal{L}_m)$ theories \cite{Allemandi2005,Nojiri2004,Bertolami2007,Sotiriou2008,Harko2010,Harko2011,Nesseris2009,Thakur2013,Bertolami2008a,Harko2013,Harko2015}. An excellent review and introduction to these and other models and their cosmological impact can be found in \cite{Clifton2012}.

In this chapter, we take a closer look at the case of the aforementioned $f(R)$ theories, which will serve as a stepping stone for the NMC models we will present later. $f(R)$ theories need not be looked at as fundamental, but rather as phenomenological gravity models that may describe the low-energy gravitational dynamics of a grander ``theory of everything''. For a good review and deeper dive on $f(R)$ theories beyond what is covered in this chapter, see \cite{DeFelice2010,Sotiriou2010} and references therein.

\section{Action and field equations}\label{sec:f(R)}

From a phenomenological standpoint (and without the addition of more fields), one way to generalize the Einstein-Hilbert action \eqref{eq:actionGR} is to replace the Ricci scalar $R$ with a generic function $f(R)$
\begin{equation}\label{eq:actionfR}
	S = \int d^4x \sqrt{-g} \left[\kappa f(R)+\mathcal{L}_\text{m} \right]\,.
\end{equation}
where $\kappa = c^4/(16\pi G)$ and $\mathcal{L}_\text{m}$ is the Lagrangian of the matter fields. It is immediate that GR is recovered when $f(R)=(R-2\Lambda)$. We obtain the field equations of this action \eqref{eq:actionfR} using the same method as in GR, so that
\begin{equation}\label{eq:fieldfR}
	G_{\mu\nu}f'=\frac{1}{2}g_{\mu\nu}[f-Rf']+\Delta_{\mu\nu}f'+ \frac{1}{2\kappa}T_{\mu\nu}\,,
\end{equation}
where primes denotes differentiation with respect to the Ricci scalar, $G_{\mu\nu}$ is the Einstein tensor and $\Delta_{\mu\nu}\equiv\nabla_\mu \nabla_\nu-g_{\mu\nu}\square$, with $\square=\nabla_\mu \nabla^\mu$ the D'Alembertian operator.

Since the Ricci scalar involves first and second order derivatives of the metric, the presence of $\Delta_{\mu\nu}f'$ in the field equations \eqref{eq:fieldfR} makes them fourth order differential equations. If the action is linear in R, the fourth order terms vanish and the theory reduces to GR. There is also a differential relation between $R$ and the trace of the EMT $T\equiv g^{\mu\nu} T_{\mu\nu}$, given by the trace equation
\begin{equation}\label{eq:R-TfR}
	3\square f'-2f+Rf'=\frac{1}{2\kappa}T\,,
\end{equation}
rather than the algebraic relation found in GR when $\Lambda=0$, $R=-T/(2\kappa)$. This enables the admittance of a larger pool of solutions than GR, such as solutions that have scalar curvature, $R\neq0$, when $T=0$. The maximally symmetric solutions lead to a constant Ricci scalar, so for constant $R$ and $T_{\mu\nu}=0$, one obtains
\begin{equation}\label{eq:maxsymfR}
	Rf'-2f=0\,,
\end{equation}
which is an algebraic equation in $R$ for a given $f$. Here it becomes important to distinguish between singular ($R^{-n}~,~n>0$) and non-singular ($R^n~,~n>0$) $f(R)$ models \cite{Dick2004}.

For non-singular models, $R=0$ is always a possible solution, the field equations \eqref{eq:fieldfR} reduce to $R_{\mu\nu}=0$, and the maximally symmetric solution is Minkowski spacetime. When $R=C$ with $C$ a constant, this becomes equivalent to a cosmological constant, the field equations reduce to $R_{\mu\nu}=g_{\mu\nu}C/4$, and the maximally symmetric solution is a de Sitter or anti-de Sitter space, depending on the sign of $C$. For singular $f(R)$ theories, however, $R=0$ is no longer an admissible solution to Eq. \eqref{eq:maxsymfR}.

Similarly to GR, applying the Bianchi identities on the covariant derivative of the field equations yields the same conservation law for the energy-momentum tensor as in GR \eqref{eq:conservGR}, $\nabla^\mu T_{\mu\nu} = 0$.

It is also possible to write the field equations \eqref{eq:fieldfR} in the form of the Einstein equations with an effective stress-energy tensor
\begin{equation}\label{eq:fieldfReff}
	G_{\mu\nu}=\frac{1}{2\kappa f'}\left[T_{\mu\nu}+2\kappa \nabla_{\mu\nu} f'+\kappa g_{\mu\nu} \left(f-Rf'\right)\right]\equiv\frac{1}{2\kappa f'}\left[T_{\mu\nu}+T_{\mu\nu}^\text{(eff)}\right]\,,
\end{equation}
where $G_\text{eff}\equiv G/f'$ is an effective gravitational coupling strength, so that demanding that $G_\text{eff}$ be positive returns $f'>0$.

\section{$f(R)$ cosmology}\label{subsec:cosmofR}

As with any gravitational theory, for an $f(R)$ theory to be a suitable candidate for gravity, it must be compatible with the current cosmological evidence, explaining in particular the observed cosmological dynamics, and the evolution of cosmological perturbations must be compatible with the CMB, large scale structure formation and BBN.

To derive the modified Friedmann and Raychaudhuri equations we again assume a flat, homogeneous and isotropic universe described by the FLRW metric with line element 
\begin{equation}
	\label{eq:line_fR}
	ds^2=-dt^2+a^2(t)\left[dr^2 +r^2 d\theta^2 +r^2\sin^2\theta d\phi^2\right]\,.
\end{equation}
We shall assume that the universe is filled with a collection of perfect fluids with energy density $\rho$, pressure $p$, and energy-momentum tensor 
\begin{equation}
	\label{eq:energy-mom2}
	T_{\mu\nu} = - (\rho+p)U_\mu U_\nu + p g_{\mu\nu}\,.
\end{equation}

Inserting the metric and energy-momentum into the field equations \eqref{eq:fieldfR}, one obtains
\begin{equation}\label{eq:friedfR}
	H^2=\frac{1}{3f'}\left[\frac{1}{2\kappa}\sum_i \rho_i+\frac{Rf'-f}{2}-3H\dot{f'}\right]\,,
\end{equation}
\begin{equation}\label{eq:rayfR}
	2\dot{H}+3H^2=-\frac{1}{f'}\left[\frac{1}{2\kappa}\sum_i p_i+\ddot{f'}+2H\dot{f'}+\frac{f-Rf'}{2}\right]\,,
\end{equation}
where $\rho_i$ and $p_i$ are the rest energy density and pressure of each of the perfect fluids, respectively. Eq. \eqref{eq:fieldfReff} implies that one must have $f'>0$ so that $G_\text{eff}>0$. Adding to this condition, $f''>0$ is required in order to avoid ghosts \cite{Starobinsky2007} and the Dolgov-Kawasaki instability \cite{Dolgov2003}. As the stability of $f(R)$ theories (regardless of the coupling to matter) is not the focus of this thesis, we will not expand on this topic (see \cite{DeFelice2010,Sotiriou2010} for a more detailed analysis).

One can collect the extra geometric terms in the Friedmann equations by introducing an effective density and pressure, defined as
\begin{equation}\label{eq:rhofR}
	\rho_\text{eff}=\kappa\left(\frac{Rf'-f}{ f'}-\frac{6H\dot{f'}}{ f'}\right)\,,
\end{equation}
\begin{equation}\label{eq:pressfR}
	p_\text{eff}=\frac{\kappa}{ f'}\left(2\ddot{f'}+4H\dot{f'}+f-Rf'\right)\,,
\end{equation}
\noindent where $\rho_\text{eff}$ must be non-negative in a spatially flat FLRW spacetime for the Friedmann equation to have a real solution when $\rho\rightarrow0$. When $\rho_\text{eff}\gg\rho$, the Friedmann \eqref{eq:friedfR} and Raychaudhuri \eqref{eq:rayfR} equations take the form
\begin{equation}\label{eq:friedfReff}
	H^2=\frac{1}{6\kappa}\rho_\text{eff}\,,
\end{equation}
\begin{equation}\label{eq:rayfReff}
	\frac{\ddot{a}}{ a}=-\frac{q}{ 12\kappa}\left(\rho_\text{eff}+3p_\text{eff}\right)\,,
\end{equation}
where $q\equiv -a\ddot{a}/\dot{a}^2$ is the deceleration parameter.

If we now consider the case of $f(R)\propto R^n$ and consider a power law expansion characterized by $a(t)=a_0(t/t_0)^\alpha$, the effective EOS parameter $w_\text{eff}=p_\text{eff}/\rho_\text{eff}$ and the parameter $\alpha$ become (for $n\neq1$) \cite{Sotiriou2010}
\begin{equation}
	w_\text{eff}=-\frac{6n^2-7n-1 }{ 6n^2-9n+3}\,,
\end{equation}
\begin{equation}
	\alpha=\frac{-2n^2+3n-1 }{ n-2}\,.
\end{equation}
Now one can simply choose a value for $n$ such that $\alpha>1$ to obtain an accelerated expansion. Likewise, $n$ can be constrained using data from supernovae and CMB observations \cite{Capozziello2003}. Using the aforementioned model, one can constrain the values of $n$ that still have $\alpha>1$ and a negative deceleration parameter. From supernovae type Ia (SNeIa) we obtain the constraint  $-0.67\leq n\leq -0.37$ or $1.37\leq n \leq 1.43$, and from WMAP data we obtain $-0.450\leq n\leq -0.370$ or $1.366\leq n \leq 1.376$.

More recently, the Planck collaboration has also set limits on this type of $f(R)$ theories \cite{Ade2016a,Aghanim2020}. Solving the Friedmann equation in $f(R)$ gravity requires setting two initial conditions. One of them is usually set by requiring that 
\begin{equation}
	\lim_{R\rightarrow\infty}\frac{f(R)}{R}=0 \,,
\end{equation}
and the other, usually called $B_0$, is the present day value of 
\begin{equation}\label{eq:bound}
	B(z)=\frac{f'' }{ f'} \frac{H\dot{R} }{ \dot{H}-H^2}=\frac{2(n-1)(n-2) }{ -2n^2+4n-3}\,.
\end{equation}
Planck data implies that we must have $B_0 \lesssim 7.9\times 10^{-5}$, effectively restricting the value of $n$ to be very close to unity.

It is worthy of note that one can also use $f(R)$ theories to describe inflation, for example using the well-known Starobinsky model \cite{Starobinsky1980,DeFelice2010} given by
\begin{equation}\label{eq:starfR}
	f(R)=R+\frac{R^2}{ 6M^2}\,,
\end{equation}
where $M$ is a mass scale and where the presence of the linear term in $R$ is responsible for bringing inflation to an end. The field equations \eqref{eq:friedfR} and \eqref{eq:rayfR} return
\begin{equation}\label{eq:friedstar}
	\ddot{H}-\frac{\dot{H}^2}{ 2H}+\frac{1}{2} M^2 H=-3H\dot{H}\,,
\end{equation}
\begin{equation}\label{eq:raystar}
	\ddot{R}+3H\dot{R}+M^2 R=0\,.
\end{equation}
During inflation, the first two terms of Eq. \eqref{eq:friedstar} are much smaller than the others, and one obtains a linear differential equation for $H$ that can be integrated to give
\begin{align}\label{eq:starsolution}
	H&\simeq H_\text{i}-\frac{M^2 }{ 6}(t-t_\text{i})\,,\nonumber\\
	a&\simeq a_\text{i} \exp\left[H_i(t-t_\text{i})-\frac{M^2}{ 12}(t-t_\text{i})^2\right]\,,\nonumber\\
	R&\simeq 12H^2-M^2\,,
\end{align}
where $H_\text{i}$ and $a_\text{i}$ are the Hubble parameter and the scale factor at the onset of inflation ($t=t_\text{i}$), respectively.

\section{Equivalence with scalar field theories}\label{subsec:scalarfR}

An interesting property of $f(R)$ theories is that they are equivalently to Jordan-Brans-Dicke (JBD) theories \cite{Sotiriou2006,Faraoni2007b,DeFelice2010}, which can be written in both the Jordan and Einstein frames, \textit{i.e.} with the scalar field coupled directly with the Ricci scalar, or not, respectively. The first equivalence is drawn by rewriting the $f(R)$ action \eqref{eq:actionfR} as a function of an arbitrary scalar field $\chi$,
\begin{equation}\label{eq:actionfchi}
	S=\int d^4x \sqrt{-g}\left[\kappa\left(f(\chi)+f'(\chi)(R-\chi)\right)+\mathcal{L}_\text{m}\right]\,,
\end{equation}
where primes represent differentiation with respect to $\chi$. The null variation of this action with respect to $\chi$ returns
\begin{equation}\label{eq:r=chi}
	f''(\chi)(R-\chi)=0\,,
\end{equation}
which implies $\chi=R$ provided that $f''(\chi)\neq0$, and therefore the action \eqref{eq:actionfchi} takes the same form as Eq. \eqref{eq:actionfR}.

Defining a new field $\varphi\equiv f'(\chi)$, Eq. \eqref{eq:actionfchi}
can be expressed as
\begin{equation}\label{eq:actionfrJBD}
	S=\int d^4x \sqrt{-g}\left[\kappa\varphi R-V(\varphi)+\mathcal{L}_\text{m} \right]\,,
\end{equation}
where $V(\varphi)$ is a field potential given by
\begin{equation}\label{eq:potfrJBD}
	V(\varphi)=\kappa\left[\chi(\varphi)\varphi-f\left(\chi(\varphi)\right)\right]\,,
\end{equation}
which has the same form as the JBD action
\begin{equation}\label{eq:actionJBD}
	S=\int d^4x \sqrt{-g}\left[\frac{1}{2}\varphi R-\frac{\omega_\text{JBD}}{ 2\varphi}g^{\mu\nu}\partial_\mu \varphi \partial_\nu\varphi-V(\varphi)+\mathcal{L}_\text{m} \right]\,,
\end{equation}
when the JBD parameter $\omega_\text{JBD}$ is null.

It is also possible to write this action in the Einstein frame via a conformal transformation
\begin{equation}\label{eq:confmetric}
	\tilde{g}_{\mu\nu}=\Omega^2g_{\mu\nu}\,,
\end{equation}
where $\Omega^2$ is the conformal factor and the tilde represents quantities in the Einstein frame. The Ricci scalars in each of the frames $R$ and $\tilde{R}$ have the relation
\begin{equation}\label{eq:confRicci}
	R=\Omega^2\left(\tilde{R}+6\tilde{\square}\ln\Omega - 6\tilde{g}^{\mu\nu}\partial_\mu(\ln \Omega) \partial_\nu (\ln \Omega) \right)\,.
\end{equation}
Substituting in the action \eqref{eq:actionfrJBD} and using the relation $\sqrt{-g}=\Omega^{-4}\sqrt{-\tilde{g}}$ we can rewrite it as
\begin{align}\label{eq:actionfreins1}
	S = \int d^4x \sqrt{-\tilde{g}} \Big[\kappa \Omega^{-2} \varphi \Big(\tilde{R} &+6\tilde{\square}(\ln\Omega) - 6\tilde{g}^{\mu\nu}\partial_\mu(\ln \Omega) \partial_\nu (\ln \Omega) \Big) \nonumber \\ &-\Omega^{-4}\varphi^2 U+\Omega^{-4}\mathcal{L}_\text{m}(\Omega^{-2}\tilde{g}_{\mu\nu},\Psi_M) \Big]\,,
\end{align}
where now the matter Lagrangian is a function of the transformed metric $\tilde{g}_{\mu\nu}$ and the matter fields $\Psi_M$, and $U$ is a potential defined as
\begin{equation}\label{eq:fRpotential}
	U=\kappa\frac{\chi(\varphi)\varphi-f\left(\chi(\varphi)\right) }{ \varphi^2}\,.
\end{equation}

Careful observation of the previous equation makes it clear that one obtains the action in the Einstein frame for the choice of transformation
\begin{equation}\label{eq:conftransfR}
	\Omega^2=\varphi\,,
\end{equation}
where it is assumed that $\varphi>0$. We now rescale the scalar field as
\begin{equation}\label{eq:fRfield}
	\phi\equiv \sqrt{3 \kappa} \ln\left(\varphi\right)\,.
\end{equation}
Since the integral $\int d^4x \sqrt{-\tilde{g}}\tilde{\square}(\ln\Omega)$ vanishes due to Gauss's theorem, we can finally write the action \eqref{eq:actionfreins1} in the Einstein frame
\begin{equation}
	S = \int d^4x \sqrt{-\tilde{g}} \left[\kappa\tilde{R} - \frac{1}{2}\partial^\alpha\phi \partial_\alpha \phi  - U(\phi)+e^{-2\frac{\phi }{ \sqrt{3\kappa}}}\mathcal{L}_\text{m}\left(e^{-\frac{\phi }{ \sqrt{3\kappa}}}\tilde{g}_{\mu\nu},\Psi_M\right) \right]\,,
\end{equation}
where $e^{-\frac{\phi }{ \sqrt{3\kappa}}}\tilde{g}_{\mu\nu}$ is the physical metric.

These three representations are equivalent at the classical level, as long as one takes care to scale the fundamental and derived units when performing conformal transformations \cite{Faraoni2007c}, so one can choose to work in the most convenient representation. The scalar-field representation may be more familiar to particle physicists, whereas the geometric nature of $f(R)$ may appeal more to mathematicians and relativists.

\section{The weak-field limit and local constraints}

Modern measurements of solar system dynamics allow one to derive local constraints on modified gravity. This is usually done by performing a weak-field expansion of the field equations on a perturbed Minkowski or FLRW metric. In general, one must always ensure two conditions:

\textit{Condition 1:} $f(R)$ is analytical at background curvature $R=R_0$.

\textit{Condition 2:} The pressure of the local star-like object is approximately null, $p\simeq 0$. This implies that the trace of the energy-momentum tensor is simply $T\simeq -\rho$.

Additionally, a third condition is necessary if one wishes to avoid non-Newtonian terms in the weak-field expansion:

\textit{Condition 3:} $mr\ll1$, where $m$ is the effective mass of the scalar degree of freedom of the theory (defined in Eq. \eqref{eq:fR-mass-param}) and $r$ is the distance to the local star-like object.

If condition 3 is verified, then the extra scalar degree of freedom (the Ricci scalar) will have a range longer than the Solar System. In this case, one can perform a parametrized post-Newtonian (PPN) expansion \cite{Eddington:1922,Robertson:1962,Nordtvedt:1968b,Will:1972,Will:1973} around a background FLRW metric, and the resulting expansion will allow one to derive constraints from experimental measurements of the PPN parameters \cite{Chiba2007}.

On the other hand, if the range of the scalar degree of freedom is shorter than the typical distances in the Solar System, then the weak-field expansion will feature Yukawa-like terms, which can be constrained by experiment \cite{Naf2010}.

\subsection[Post-Newtonian Expansion]{Post-Newtonian expansion}
The PPN formalism provides a systematic approach in which one can analyse and compare different metric theories of gravity in the weak-field slow-motion limit. Essentially, this relies on performing a linear expansion of the metric and the field equations of the modified gravity theory and then comparing them against the post-Newtonian parameters, assigned to different gravitational potentials, whose values can be constrained by experiment.

For $f(R)$ gravity, one assumes that the scalar curvature can be expressed as the sum of two components
\begin{equation}
	R(r,t)\equiv R_0(t)+R_1(r) \,,
\end{equation}
where $R_0(t)$ is the background spatially homogeneous scalar curvature, and $R_1(r)$ is a time-independent perturbation to the background curvature. Since the timescales usually considered in Solar System dynamics are much shorter than cosmological ones, we can usually take the background curvature to be constant, \textit{i.e.} $R_0= \text{const.}$. We can therefore separate the source for the Ricci scalar into two different components, one cosmological and another local. So the trace Eq. \eqref{eq:R-TfR} reads
\begin{equation}\label{eq:R-Tf2R-sep}
	3\square f'-2f+Rf'=\frac{1}{2\kappa}(T^\text{cos}+T^\text{s})\,,
\end{equation}
where $T^\text{cos}$ and $T^\text{s}$ are the cosmological and local contributions, respectively. If one takes into account that $R_1\ll R_0$ (see \cite{Chiba2007} for more details) and that $R_0$ solves the terms of the expansion of Eq. \eqref{eq:R-Tf2R-sep} that are independent of $R_1$, one can write it as
\begin{equation}
	\label{eq:fR-trace-exp}
	3f''_0\square R_1(r)-\left(f'_0 R_0-f_0'\right)= \frac{1}{2\kappa}T^\text{s} \,,
\end{equation}
where $f_0\equiv f(R_0)$. For a static, spherically symmetric body, one can further write this as
\begin{equation}
	\label{eq:fR-trace-exp-2}
	\nabla^2R_1-m^2 R_1 = -\frac{1}{2\kappa} \frac{\rho^\text{s}}{f''_0} \,,
\end{equation}
where $\rho^\text{s}$ is the body's density and $m$ is the mass parameter, defined as
\begin{equation}
	\label{eq:fR-mass-param}
	m^2 \equiv \frac{1}{3} \left(\frac{f'_0}{f''_0}-R_0-3\frac{\square f''_0}{f''_0}\right) \,.
\end{equation}
If $mr\ll1$, one can solve Eq. \eqref{eq:fR-trace-exp-2} outside the star to obtain
\begin{equation}
	\label{eq:fR-R1-sol}
	R_1=\frac{1}{24\kappa\pi f''_0}\frac{M}{r}\,,
\end{equation}
where $M$ is the total mass of the source.

By considering a flat FLRW metric with a spherically symmetric perturbation
\begin{equation}
	\label{eq:pert-flrw}
	ds^2=-\left[1+2\Psi(r)\right] dt^2 + a^2(t)\left\{\left[1+2\Phi(r)\right]dr^2 +r^2 d\theta^2 +r^2\sin^2\theta d\phi^2\right\} \,,
\end{equation}
and solving the linearized field equations for $\Psi(r)$ and $\Phi(r)$ with the solution obtained for $R_1$ \eqref{eq:fR-R1-sol} and $a(t=t_0)=1$, one obtains \cite{Chiba2007}
\begin{equation}
	\label{eq:pert-flrw-2}
	ds^2=-\left(1-\frac{2GM}{r}\right) dt^2 + \left(1+\frac{GM}{r}\right)\left(dr^2 +r^2 d\theta^2 +r^2\sin^2\theta d\phi^2\right) \,,
\end{equation}
where $M$ is the total mass of the central source. Comparing Eq. \eqref{eq:pert-flrw-2} with the equivalent PPN metric
\begin{equation}
	\label{eq:metric-ppn-fr}
	ds^2=-\left(1-\frac{2GM}{r}\right) dt^2 + \left(1+\gamma\frac{2GM}{r}\right)\left(dr^2 +r^2 d\theta^2 +r^2\sin^2\theta d\phi^2\right) \,,
\end{equation}
where $\gamma$ is a PPN parameter, one can see that in $f(R)$ gravity $\gamma=1/2$. The tightest bound on this parameter comes from the tracking of the Cassini probe, where it was determined that $\gamma-1= (2.1\pm2.3)\times 10^{-5}$. Thus $f(R)$ gravity is generally incompatible with solar system tests, provided that the linearized limit is valid and that $m r\ll 1$ (see \cite{Chiba2007} for a more in-depth look at this constraint). Nevertheless, some $f(R)$ models can feature a so-called ``chameleon'' mechanism which hides the scalar degree of freedom in regions with high densities, and thus allows the theory to be constrained by local tests, rather than outright excluded \cite{Faulkner2007,Capozziello2008}.

\subsection{``Post-Yukawa'' expansion}

In \cite{Naf2010} the authors perform a more general weak-field expansion (\textit{i.e.} accounting for Yukawa terms) of the field equations on a perturbed asymptotically flat metric with components $g_{\mu\nu}$ given by
\begin{align}
	\label{eq:fR-asymp-flat-metric}
	g_{00} &= -1+ {}^{(2)}h_{00} + {}^{(4)} h_{00} + \mathcal{O}(c^{-6}) \,, \nonumber \\
	g_{0i} &= {}^{(3)}h_{0i}+\mathcal{O}(c^{-5}) \,, \nonumber \\
	g_{ij} &= \delta_{ij} + {}^{(2)} h_{ij} + \mathcal{O}(c^{-4}) \,,
\end{align}
where ${}^{(n)}h_{\mu\nu}$ denotes a quantity of order $\mathcal{O}(c^{-n})$. Since $R\sim \mathcal{O}(c^{-2})$ and assuming $f$ to be analytic at $R=0$ with $f'(0)=1$, it is sufficient to consider the expansion
\begin{equation}
	\label{eq:fR-Yuk}
	f(R) = R + aR^2 \,, \qquad a\neq 0 \,,
\end{equation}
where the presence of a cosmological constant is ignored due to considering an asymptotically flat background.

Introducing a scalar field $\varphi$ defined as
\begin{equation}
	\label{eq:phi-fr-yuk}
	f'(R) = 1+2a\varphi \,,
\end{equation}
one can rewrite the trace equation \eqref{eq:R-TfR} as
\begin{equation}
	\label{eq:fR-trace-phi}
	\square\varphi= \frac{1}{6a\kappa}T+\frac{1}{6a}\varphi \,.
\end{equation}
If one expands the Ricci tensor $R_{\mu\nu}$, scalar field $\varphi$ and Energy momentum tensor $T^{\mu\nu}$ to the necessary order, then Eq. \eqref{eq:fR-trace-phi} can be rewritten up to leading order as
\begin{equation}
	\label{eq:fR-trace-yuk}
	\nabla^2 {}^{(2)}\varphi -\alpha^2  {}^{(2)}\varphi= -\frac{\alpha^2}{2\kappa} {}^{(-2)}T^{00}\,
\end{equation}
where $\alpha^2\equiv 1/(6a)$, which has the solution
\begin{equation}
	{}^{(2)}\varphi(\vec{x},t)=\frac{1}{c^2}V(\vec{x},t) \,,
\end{equation}
where
\begin{equation}
	V(\vec{x},t)\equiv\frac{2G\alpha^2}{c^2}\int\frac{{}^{(-2)}T^{00}(\vec{x}',t)e^{-\alpha|\vec{x}-\vec{x}'|}}{|\vec{x}-\vec{x}'|}d^3x' \,.
\end{equation}
One can then use the field equations \eqref{eq:fieldfR} to obtain the solution for ${}^{(2)}h_{00}$
\begin{equation}
	{}^{(2)}h_{00}(\vec{x},t)=\frac{1}{c^2}\left[2U(\vec{x},t)-W(\vec{x},t)\right] \,,
\end{equation}
with the potentials
\begin{equation}
	\label{eq:fr-U}
	U(\vec{x},t)\equiv\frac{4G}{3c^2}\int\frac{{}^{(-2)}T^{00}(\vec{x}',t)}{|\vec{x}-\vec{x}'|}d^3x' \,,
\end{equation}
\begin{equation}
	\label{eq:fr-W}
	W(\vec{x},t)\equiv\frac{1}{12\pi}\int\frac{V(\vec{x}',t)}{|\vec{x}-\vec{x}'|}d^3x' \,.
\end{equation}
Whereas the potential $U$ corresponds to the standard Newtonian term, $W$ contains the Yukawa term in $V$. The remaining components of the perturbed metric can be calculated similarly (we direct the reader to \cite{Naf2010} for the complete calculation). Yukawa corrections to the Newtonian  potential are already tightly constrained in the literature, and in this case, the Eöt-Wash experiment \cite{Kapner2007} constrains the model parameter $a=1/(6\alpha^2)$ to
\begin{equation}
	a\lesssim 10^{-10}\text{ m}^2\,.
\end{equation}

\chapter{The Lagrangian of Cosmological Fluids} 
\label{chapter_lag} 

Many popular modified gravity theories feature a nonminimal coupling (NMC) between matter and gravity or another field. In addition to the rich dynamics these theories provide, the use of the variational principle in the derivation of the equations of motion (EOM) leads to the appearance of the on-shell Lagrangian of the matter fields in the EOM, in addition to the usual energy-momentum tensor (EMT). This is a significant departure from minimally coupled theories like general relativity (GR) and $f(R)$ theories, which are insensitive to the form of the matter Lagrangian leading to the appropriate EMT. The knowledge of the on-shell Lagrangian of the matter fields thus becomes paramount if one is to derive reliable predictions from the theory. In essence, this is the problem tackled in this chapter, and it comprises an important part of the original work published in Refs. \cite{Avelino2018,Ferreira2020,Avelino2020} (Sections \ref{sec.nmclag}, \ref{sec.whichlag} and \ref{sec.nmclagrole}).

\section{Perfect fluid actions and Lagrangians}\label{sec:fluid_lag}

The form of the on-shell Lagrangian of certain classes of minimally coupled perfect fluids has been extensively studied in the literature \cite{Schutz1977,Brown1993a}, but it is often misused in NMC theories. Before tackling this problem in NMC gravity, we will first briefly summarize some of these derivations when the perfect fluid is minimally coupled, like in GR. 

Consider a perfect fluid in GR, described locally by the following thermodynamic variables
\begin{align}\label{eq:thermo variables}
	n\,&:\quad \text{particle number density}\,,\\
	\rho\,&:\quad \text{energy density}\,,\\
	p\,&:\quad \text{pressure}\,,\\
	\mathcal{T}\,&:\quad \text{temperature}\,,\\
	s\,&:\quad \text{entropy per particle}\,,	
\end{align}
defined in the rest frame of the fluid, and with four-velocity $U^\mu$. Assuming that the total number of particles $N$ is conserved, the first law of thermodynamics reads
\begin{equation}
	\label{eq:first_law_1}
	d\mathcal{U}=\mathcal{T}dS-pdV\,,
\end{equation}
where $V$, $\mathcal{U}=\rho V$, $S = sN$ and $N=nV$ are the physical volume, energy, entropy and particle number, respectively. In terms of the variables defined above, it can be easily rewritten as
\begin{equation}
	\label{eq:1st law thermo}
	d\rho=\mu dn+n\mathcal{T}ds\,,
\end{equation}
where
\begin{equation}
	\label{eq:chemical potential}
	\mu=\frac{\rho+p}{n}
\end{equation}
is the chemical potential. Eq. \eqref{eq:chemical potential} implies that the energy density may be written as a function of $n$ and $s$ alone, that is $\rho=\rho(n,s)$. Building a perfect-fluid action from these variables alone, and ignoring possible microscopic considerations as to the composition of the fluid, requires the addition of a few constraints, as shown by Schutz \cite{Schutz1977}. These constraints are particle number conservation and the absence of entropy exchange between neighbouring field lines,
\begin{equation}\label{eq:fluid con1}
	(n U^\mu)_{;\mu}=0\,,
\end{equation}
\begin{equation}
	\label{eq:fluid con2}
	(nsU^\mu)_{;\mu}=0\,,
\end{equation}
respectively, and the fixing of the fluid flow lines on the boundaries of spacetime. 

These constraints can be easily satisfied if one uses a set of scalar fields $\alpha^A\,,\,A=1,\,2,\,3$ as Lagrangian coordinates for the fluid \cite{Lin1963,Serrin1959}. Consider a spacelike hypersurface with a coordinate system $\alpha^A$. By fixing the coordinates on the boundaries of spacetime and labelling each fluid flow line by the coordinate at which it intersects the hypersurface, one automatically satisfies the last constraint. The remaining two constraints can be satisfied through the use of Lagrange multipliers in the action.

Let's introduce a contravariant vector density $J^\mu$, interpreted as the particle number density flux vector and defined implicitly via the four-velocity as
\begin{equation}
	\label{eq:j vector}
	U^\mu = \frac{J^\mu}{|J|}\,,
\end{equation}
where $|J|=\sqrt{-J^\mu J_\mu}$ is its magnitude, and the particle number density is given by $n=|J|/\sqrt{-g}$.

One can now build the fluid action as a functional of the vector $J^\mu$, the metric $g_{\mu\nu}$, the entropy per particle $s$, the Lagrangian coordinates $\alpha^A$, and spacetime scalars $\varphi$, $\theta$ and $\beta_A$, which will act as Lagrange multipliers for the particle number and entropy flow constraints. The off-shell action reads \cite{Brown1993a}
\begin{equation}
	\label{eq:action Brown}
	S_\text{off-shell} = \int d^4 x \left[-\sqrt{-g}\rho(n,s)+J^\mu(\varphi_{,\mu}+s\theta_{,\mu}+\beta_A{\alpha^A}_{,\mu})\right]\,.
\end{equation}
The EMT derived from the action has the form
\begin{equation}
	\label{eq:Brown EMT}
	T^{\mu\nu}=\rho U^\mu U^\nu + \left(n\frac{\partial\rho}{\partial n}-\rho\right)(g^{\mu\nu}+U^\mu U^\nu)\,,
\end{equation}
which represents a perfect fluid with EOS
\begin{equation}
	\label{eq:Brown EOS}
	p = n\frac{\partial\rho}{\partial n}-\rho \,.
\end{equation}

The EOM of the fluid, derived via a null variation of the action \eqref{eq:action Brown} with respect to each of the fields, are
\begin{align}	
	J^\nu\,&:\quad \mu U_\nu + \varphi_{,\nu}+s\theta_{,\nu}+\beta_A {\alpha^A}_{,\nu}=0\,,\label{eq:Brown EOM J}	\\
	\varphi\,&:\quad {J^\mu}_{,\mu} = 0\,,\label{eq:Brown EOM phi}	\\
	\theta\,&:\quad (sJ^\mu)_{,\mu}=0 \,,\label{eq:Brown EOM theta}	\\
	s\,&:\quad \sqrt{-g}\frac{\partial\rho}{\partial s}-\theta_{,\mu}J^\mu=0 \,,\label{eq:Brown EOM s}\\
	\alpha^A\,&:\quad (\beta_A J^\mu)_{,\mu}=0 \,,\label{eq:Brown EOM alpha}\\
	\beta_A\,&:\quad {\alpha^A}_{,\mu}J^\mu = 0 \,.\label{eq:Brown EOM beta}
\end{align}

It is useful at this stage to derive some relations between the fields and thermodynamic quantities. Comparing the first law of thermodynamics \eqref{eq:1st law thermo} with Eq. \eqref{eq:Brown EOM s} one obtains
\begin{align}
	\label{eq:theta T relation}
	&\sqrt{-g}\frac{\partial \rho}{\partial s} - \theta_{,\nu} U^\nu=0  \nonumber\\
	\Rightarrow&\sqrt{-g}n\mathcal{T}-\theta_{,\nu}  \sqrt{-g} n U^\nu = 0 \nonumber\\
	\Rightarrow&\frac{1}{n}\frac{\partial \rho}{\partial s}=\theta_{,\nu} U^\nu=\mathcal{T} \,.
\end{align}
Similarly, one can contract Eq. \eqref{eq:Brown EOM J} with $U^\nu$ and  
use the EOM and Eq. \eqref{eq:theta T relation} to obtain
\begin{align}
	\label{eq:phi mu sT  relation}
	&- \mu + \varphi_{,\nu}U^\nu+s\theta_{,\nu}U^\nu+\beta_A {\alpha^A}_{,\nu}U^\nu=0 \nonumber\\
	\Rightarrow&-\mu+\varphi_{,\nu}U^\nu+s\mathcal{T}=0 \nonumber\\
	\Rightarrow&\varphi_{,\nu}U^\nu = \mu-s\mathcal{T} = \mathpzc{f} \,,
\end{align}
where $\mathpzc{f}$ is the chemical free-energy density.

One can now apply the EOM on the off-shell action \eqref{eq:action Brown} in order to obtain its expression on-shell. Using Eq. \eqref{eq:Brown EOM J}, and the definitions of $J^\mu$ and its magnitude, one can write
\begin{align}
	\label{eq:Brown on shell 1}
	S_\text{on-shell} &= \int d^4 x\left[-\sqrt{-g}\rho-\mu J^\nu U_\nu\right] \nonumber\\
	&= \int d^4 x \left[-\sqrt{-g}\rho+\mu |J|\right]\nonumber\\
	&=\int d^4 x\sqrt{-g}\left[-\rho+\mu n\right] \,,
\end{align}
and substituting in the chemical potential \eqref{eq:chemical potential}, one finally obtains
\begin{equation}
	\label{eq:action brown p}
	S_\text{on-shell} = \int d^4 x \sqrt{-g}\,p \,.
\end{equation}

Note, however, that one always has the liberty of adding boundary terms to the off-shell action, without affecting the EOM. Two particular interesting choices are the  integrals
\begin{equation}
	\label{eq:surface int 1}
	-\int d^4x (\varphi J^\mu)_{,\mu} \,,
\end{equation}
and
\begin{equation}
	\label{eq:surface int 2}
	-\int d^4x(\theta s J^\mu)_{,\mu} \,,
\end{equation}
Adding both of these integrals to the off-shell action \eqref{eq:action Brown} returns
\begin{equation}
	\label{eq:action Brown rho off shell}
	S_\text{off-shell}=\int\left[-\sqrt{-g}\rho+J^\nu \beta_A \alpha^A_{,\nu}-\theta(sJ^\nu)_{,\nu}-\varphi J^\nu_{\,,\nu}\right] \,,
\end{equation}
and applying the EOM and relations \eqref{eq:theta T relation} and \eqref{eq:phi mu sT  relation} we obtain the on-shell action
\begin{align}
	\label{eq:fluid_lag_rho}
	S_\text{on-shell}&=\int d^4 x\left[-\sqrt{-g}\rho- J^\nu (\mu U_\nu+ \varphi_{,\nu} + s\theta_{,\nu})\right] \nonumber \\
	&=\int d^4 x \sqrt{-g}(-\rho) \,.
\end{align}
Thus, the addition of surface integrals to the off-shell action changes the on-shell perfect fluid Lagrangian from $p$ to $-\rho$, even though it does not change the EOM or the form of the EMT. Likewise, further additions of these or other integrals would allow one to write the on-shell Lagrangian in a multitude of ways. For example, we can  add the integrals \eqref{eq:surface int 1} and \eqref{eq:surface int 2} $c_1$ times. In this case the off-shell action reads
\begin{align}
	\label{eq:action_Brown_T_off}
	S_\text{off-shell}=\int d^4x\big[-\sqrt{-g}\rho + J^\nu \beta_A \alpha^A_{,\nu} &+(1-c_1)J^\nu\left(\varphi_{,\nu} + s\theta_{,\nu}\right) \nonumber\\
	&-c_1\theta(sJ^\nu)_{,\nu} -c_1\varphi J^\nu_{\,,\nu}\big] \,,
\end{align}
and applying the EOM we obtain
\begin{align}
	\label{eq:action_Brown_T_on}
	S_\text{on-shell} =\int d^4 x \sqrt{-g}\left[\left(1-c_1\right)p-c_1\rho\right] \,.
\end{align}

Nevertheless, since the source term for the Einstein field equations is the EMT of the perfect fluid and not its on-shell Lagrangian, the dynamics of gravity in GR and minimally coupled theories are insensitive to this choice.

\section{The barotropic fluid Lagrangian}
\label{sec.barlag}

The degeneracy of the on-shell Lagrangian presented in Section \ref{sec:fluid_lag} can be broken if one makes further assumptions. In particular, this is the case  if one derives the Lagrangian for a barotropic perfect fluid whose pressure $p$ depends only on the fluid's particle number density $n$ \footnote{In some works in the literature the barotropic fluid is defined as a fluid whose pressure is only a function of the rest mass density $\rho_\text{m}$. However, assuming that the particles are identical and have conserved rest mass, then the rest mass density of the fluid is proportional to the particle number density $\rho_\text{m}\propto n$, and the two descriptions are equivalent.}, \textit{i.e.} $p =p(n)$, and assumes that the off-shell Lagrangian is only a function of the particle number density ${\mathcal L}_{\rm m}^{\rm off}=\mathcal{L}_{\rm m}^{\rm off}(n)$ \cite{Harko2010a,Minazzoli2012,Minazzoli2013,Arruga2021}.

\subsection{Derivation with $\mathcal{L}_\text{m}^\text{off}=\mathcal{L}_\text{m}^\text{off}(n)$}
\label{subsec.harko}
To show this, one can simply start from the definition of the EMT,
\begin{equation}
	\label{eq:bar-emt}
	T_{\mu\nu}=-\frac{2}{\sqrt{-g}}\frac{\partial(\sqrt{-g}\mathcal{L}_\text{m}^\text{off})}{\partial g^{\mu\nu}} =\mathcal{L}_\text{m}^\text{off}g_{\mu\nu}-2\frac{\partial \mathcal{L}_\text{m}^\text{off}}{\partial g^{\mu\nu}} \,.
\end{equation}
If one considers that the mass current is conserved, \textit{i.e.} $\nabla_\mu(n U^\mu)=0$, where $U^\mu$ are the components of the four-velocity of a fluid element, then one can show that \cite{Harko2010a}
\begin{equation}
	\label{eq:var-sig}
	\delta n = \frac{1}{2} n(g_{\mu\nu} - U_\mu U_\nu)\delta g^{\mu\nu} \,.
\end{equation}
Using Eq. \eqref{eq:var-sig} and assuming that the off-shell Lagrangian is a function of the number density (or equivalently, the rest mass density), one can rewrite Eq. \eqref{eq:bar-emt} as \cite{Harko2010a}
\begin{equation}
	\label{eq:bar-emt2}
	T^{\mu\nu}=n \frac{d \mathcal{L}_\text{m}^\text{off}}{dn} U^\mu U^\nu+\left(\mathcal{L}_\text{m}^\text{off} - n\frac{d \mathcal{L}_\text{m}^\text{off}}{dn}\right) g^{\mu\nu} \,.
\end{equation}
Comparing Eq. \eqref{eq:bar-emt2} with the EMT of a perfect fluid
\begin{equation}
	\label{eq:pf-emt}
	T^{\mu\nu}=-(\rho+p)U^\mu U^\nu + p g^{\mu\nu} \,,
\end{equation}
we identify
\begin{align}
	\mathcal{L}_\text{m}^\text{off}(n) &= -\rho(n)\,, \label{eq:bar-lag}\\
	\frac{d\mathcal{L}_\text{m}^\text{off}(n)}{dn}&=-\frac{\rho(n)+p(n)}{n} \,.
\end{align}
Hence,
\begin{equation}
	\label{eq:mina-eos}
	\frac{d\rho(n)}{dn}=\frac{\rho(n)+p(n)}{n} \,.
\end{equation}
Eq. \eqref{eq:mina-eos} is exactly the equation of state obtained in Section \ref{sec:fluid_lag}, in Eq. \eqref{eq:Brown EOS}. Eq. \eqref{eq:mina-eos} has the general solution \cite{Minazzoli2012}
\begin{equation}
	\label{eq:mina-rho}
	\rho(n) = C n + n\int\frac{p(n)}{n^2}dn \,,
\end{equation}
where $C$ is an integration constant. Therefore the on-shell and off-shell Lagrangians are the same and unique, and determined by the rest mass density of the fluid
\begin{equation}
	\label{eq:Harko-Min-lag}
	\mathcal{L}_\text{m}^\text{on} = \mathcal{L}_\text{m}^\text{off} = -\rho(n) = -Cn-n\int\frac{p(n)}{n^2}dn \,.
\end{equation}


\subsection{Derivation with $\mathcal{L}_\text{m}^\text{off}=\mathcal{L}_\text{m}^\text{off}(j^\mu,\phi)$}
\label{subsec.fer}

We can perform a derivation of the on-shell Lagrangian for a barotropic fluid in a similar way to the treatment applied in Section \ref{sec:fluid_lag} with a simplified action, as performed in Ref. \cite{Ferreira2020}. Consider a fluid characterized by the following intensive variables
\begin{align}\label{eq:bar-variables}
	n\,&:\quad \text{particle number density}\,,\\
	\rho\,&:\quad \text{energy density}\,,\\
	p\,&:\quad \text{pressure}\,,\\
	s\,&:\quad \text{entropy per particle}\,,	
\end{align}
defined in the rest frame of the fluid, and with four-velocity $U^\mu$. Much like in Section \ref{sec:fluid_lag}, assuming that the total number of particles is conserved, we can rewrite the first law of thermodynamics \eqref{eq:first_law_1} as
\begin{equation}
	\label{eq:1st-law-bar}
	d\left(\frac{\rho}{n}\right)=-p d\left(\frac{1}{n}\right)+\mathcal{T}ds\,.
\end{equation}
If the flow is isentropic, $ds=0$, and therefore Eq. \eqref{eq:1st-law-bar} simplifies to 
\begin{equation}
	\label{eq:1st-law-bar2}
	d\left(\frac{\rho}{n}\right)=-p d\left(\frac{1}{n}\right)\,.
\end{equation}
If we consider that the fluid's energy density and pressure depend only on its number density, $\rho=\rho(n)$ and $p=p(n)$, then we can solve Eq. \eqref{eq:1st-law-bar2} to obtain
\begin{equation}
	\label{eq:fer-rho}
	\rho(n) = C n + n\int\frac{p(n)}{n^2}dn \,,
\end{equation}
where $C$ is an integration constant.

Consider now a minimally coupled fluid described by the action
\begin{equation}
	\label{eq:fer-act}
	S=\int d^4x\sqrt{-g}\mathcal{L}_\text{m}^\text{off}(j^\mu,\phi) \,,
\end{equation}
where
\begin{equation}
	\label{eq:fer-lag}
	\mathcal{L}_\text{m}^\text{off}(j^\mu,\phi) = F(|j|)-\phi\nabla_\mu j^\mu \,,
\end{equation}
$j^\mu$ are the components of a timelike vector field, $\phi$ is scalar field, and $F$ is a function of $|j|=\sqrt{-j^\mu j_\mu}$. Using the variational principle on the action \eqref{eq:fer-act} with respect to $j^\mu$ and $\phi$ one obtains the EOM
\begin{align}
	\frac{\delta S}{\delta j^\mu} &= 0 = -\frac{1}{|j|}\frac{dF}{d|j|}j_\mu + \nabla_\mu \phi \,, \label{eq:j-eq-mot} \\
	\frac{\delta S}{\delta \phi} &= 0 = \nabla_\mu j^\mu \,. \label{eq:phi-eq-mot}
\end{align}

Varying the matter action with respect to the metric components one obtains
\begin{equation}
	\label{eq: EMT0}
	\delta S=\int d^{4}x \frac{\delta\left(\sqrt{-g}\mathcal{L}_\text{m}^\text{off}\right)}{\delta g_{\mu\nu}} \delta g_{\mu \nu}= \frac12 \int d^{4}x {\sqrt{-g}} \, T^{\mu \nu} \delta g_{\mu \nu}\,,
\end{equation}
where
\begin{eqnarray}\label{eq: EMT1}
	\delta\left(\sqrt{-g}\mathcal{L}_\text{m}^\text{off}\right)&=& \sqrt{-g} \delta \mathcal{L}_\text{m}^\text{off} + \mathcal{L} \delta \sqrt{-g} \nonumber \\
	&=& \sqrt{-g} \delta \mathcal{L}_\text{m}^\text{off} +  \frac{\mathcal{L}_\text{m}^\text{off}}{2}  \sqrt{-g} g^{\mu\nu} \delta g_{\mu\nu}\,,
\end{eqnarray}
with
\begin{equation}\label{eq: EMT2}
	\delta \mathcal{L}_\text{m}^\text{off} =- \frac{1}{2}\frac{dF}{d|j|}\frac{j^{\mu}j^{\nu}}{|j|} \delta g_{\mu\nu} - \phi \delta (\nabla_\sigma j^\sigma) \,,
\end{equation}
and
\begin{eqnarray}\label{eq: EMT3}
	\phi \delta \left(\nabla_\sigma j^\sigma\right)&=&\phi \delta \left(\frac{\partial_{\sigma}\left(\sqrt{-g} j^{\sigma}\right)}{\sqrt{-g}} \right) \nonumber \\
	&=& -\frac12 g^{\mu\nu} \delta g_{\mu\nu} \nabla_\sigma \left(\phi j^\sigma\right)+ \frac12\nabla_\sigma\left( \phi j^\sigma g^{\mu\nu} \delta g_{\mu\nu}\right)\,.
\end{eqnarray}
Discarding the last term in Eq. \eqref{eq: EMT3} --- this term gives rise to a vanishing surface term in Eq. \eqref{eq: EMT0} ($\delta g_{\mu \nu}=0$ on the boundary) --- and using Eqs. \eqref{eq:j-eq-mot} and \eqref{eq:phi-eq-mot} it is simple to show that the EMT associated with the  Lagrangian defined in \eqref{eq:fer-lag} given by \eqref{eq:bar-emt} is
\begin{equation}
	\label{eq:fer-emt}
	T^{\mu\nu}=-\frac{dF}{d|j|}\frac{j^\mu j^\nu}{|j|}+\left(F-|j|\frac{dF}{d|j|}\right)g^{\mu\nu} \,.
\end{equation}
Comparing Eq. \eqref{eq:fer-emt} with the EMT of a perfect fluid \eqref{eq:pf-emt} we immediately obtain the identifications
\begin{align}
	n &= |j| \,,\\
	\rho(n) &= -F \,,\\
	p(n) &= F-n\frac{dF}{dn}\,,\\
	U^\mu &= \frac{j^\mu}{n} \,.
\end{align}
With these identifications, then it is immediate to see the condition implied by Eq. \eqref{eq:phi-eq-mot} is 
\begin{equation}
	\nabla_\mu j^\mu = \nabla_\mu(n U^\mu) = 0 \,,
\end{equation}
\textit{i.e.} that the particle number density is conserved. In this case the on-shell Lagrangian is equal to
\begin{equation}
	\mathcal{L}_\text{m}^\text{on}=F=-\rho(n)\,.
\end{equation}
Using this result, in combination with Eq. \eqref{eq:fer-rho}, it is possible to write the on-shell Lagrangian as
\begin{equation}
	\label{eq: Harko Lagrangian}
	\mathcal{L}_\text{m}^\text{on}= -\rho(n)= -Cn - n\int^{n}\frac{p\left(n'\right)}{n^{2}}dn \,,
\end{equation}
just like in Eq. \eqref{eq:Harko-Min-lag}.

Let us now assume another off-shell Lagrangian, obtained from Eq. \eqref{eq:fer-lag} via the transformation
\begin{equation}
	\label{eq:fer-lag-p}
	\mathcal{L}_\text{m}^\text{off}\rightarrow \mathcal{L}_\text{m}^\text{off}+\nabla_\mu(\phi j^\mu)= F(|j|)+j^\mu \nabla_\mu \phi \, .
\end{equation}
Note that the Lagrangians in Eqs. \eqref{eq:fer-lag} and \eqref{eq:fer-lag-p} differ only in a total derivative, so the EOM \eqref{eq:j-eq-mot} and \eqref{eq:phi-eq-mot} remain valid. Substituting the Lagrangian \eqref{eq:fer-lag-p} in Eq. \eqref{eq:bar-emt}, one once again obtains the EMT in Eq. \eqref{eq:fer-emt}, and therefore the identifications between the fluid variables and the fields remain the same. However, when one now use the EOM on the off-shell Lagrangian, one obtains a different on-shell value
\begin{equation}
	\label{eq:fer-lag-p-on}
	\mathcal{L}_\text{m}^\text{on} = F + j^\mu j_\mu \frac{1}{|j|}\frac{dF}{d|j|} = p(n) \,.
\end{equation}

The on-shell Lagrangians in Eqs. \eqref{eq: Harko Lagrangian} and \eqref{eq:fer-lag-p-on} both represent a barotropic perfect fluid with the usual energy-momentum tensor, and in the case of a minimal coupling can even be used to describe the same physics. The reason one can only obtain the result of $\mathcal{L}_\text{m}^\text{on}= -\rho$ in Section \ref{subsec.harko} is because the form of the assumed off-shell Lagrangian is less general than the one assumed in Eq. \eqref{eq:fer-lag}. Likewise, if we had assumed \textit{a priori} that the off-shell Lagrangian in the action in Eq. \eqref{eq:fer-act} only depended on $n\equiv |j|$, we would have obtained the same result: that the only possible on-shell Lagrangian would be $\mathcal{L}_\text{m}^\text{on}= -\rho$.

\section{The solitonic-particle fluid Lagrangian}
\label{sec.nmclag}

It was shown in Section \ref{sec:fluid_lag} that the form of the on-shell Lagrangians can take many different forms and has no impact on the EOM, as long as the fluid is minimally coupled with gravity. However, this degeneracy of the Lagrangian is lost in theories that feature an NMC between matter and gravity, or in fact between matter and any other field \cite{Avelino2018,Avelino2020,Ferreira2020}.

In this section, we show that the on-shell Lagrangian of a fluid composed of particles with fixed rest mass and structure must be given by the trace of the EMT of the fluid. We will consider two different approaches: one where the fluid can be described as a collection of single point particles with fixed rest mass, and another where fluid particles can be described by localized concentrations of energy (solitons).

\subsection{Point-particle fluid}
\label{subsec:single_part}
In many situations of interest, a fluid (not necessarily a perfect one) may be simply described as a collection of many identical point particles undergoing quasi-instantaneous scattering from time to time \cite{Avelino2018,Avelino2018a}. Hence, before discussing the Lagrangian of the fluid as a whole, let us start by considering the action of a single point particle with mass $m$
\begin{equation}
	S=-\int d \tau \, m \,,
\end{equation}
and EMT 
\begin{equation} \label{eq: particle EM tensor}
	T_*^{\mu\nu} = \frac{1}{\sqrt {-g}}\int d \tau \, m \, u^\mu u^\nu \delta^4\left(x^\sigma-\xi^\sigma(\tau)\right)\,,
\end{equation}
where the $*$ indicates that the quantity refers to a single particle, $\xi^\mu(\tau)$ represents the particle worldline and $u^\mu$ are the components of the particle 4-velocity. If one considers its trace $T_*=T_*^{\mu \nu} g_{\alpha \nu}$ and integrates over the whole of spacetime, we obtain
\begin{eqnarray}
	\int d^{4}x \sqrt{-g} \, T_* &=&- \int d^4x \,d\tau\, m\, \delta^4\left(x^\sigma-\xi^\sigma(\tau)\right) \nonumber\\
	&=&- \int d\tau \,m \, ,
\end{eqnarray}
which can be immediately identified as the action for a single massive particle, and therefore implies that
\begin{equation}
		\int d^{4}x \sqrt{-g} \,\mathcal{L}_*^{\rm on}=	\int d^{4}x \sqrt{-g} \,T_*\,.
\end{equation}

If a fluid can be modelled as a collection of point particles, then its on-shell Lagrangian at each point will be the average value of the single-particle Lagrangian over a small macroscopic volume around that point
\begin{eqnarray}
	\langle \mathcal{L}_*^{\rm on} \rangle &=& \frac{\int d^4 x \sqrt{-g} \, \mathcal{L}_*^{\rm on}}{\int  d^4 x \sqrt{-g}}\\
	&=& \frac{\int d^4 x \sqrt{-g} \, T_*}{\int  d^4 x \sqrt{-g}} = \langle T_* \rangle\,,
\end{eqnarray}
where $\langle T_* \rangle = T$ is the trace of the EMT of the perfect fluid. This provides a further possibility for the on-shell Lagrangian of a perfect fluid:
\begin{equation}
	\langle \mathcal{L}_*^{\rm on} \rangle\equiv\mathcal{L}^{\rm on}= T =-\rho+3p\,,
\end{equation}
where $p=\rho \langle v^2 \rangle/3=\rho \mathcal T$, $\sqrt{\langle v^2 \rangle}$ is the root-mean-square velocity of the particles and $\mathcal T$ is the temperature. Notice that in the case of dust ($p=0$) we recover the result obtained in Eq. \eqref{eq:fluid_lag_rho} ($\mathcal{L}^{\rm on}=-\rho$).

\subsection{Solitonic-particle fluid and Derrick's argument}

In addition to the previous interpretation, particles can also be seen as localized energy concentration, \textit{i.e.} solitons, with fixed rest mass and structure and that are not significantly impacted by their self-induced gravitational field (see also \cite{Polyakov2018,Avelino2019}). We shall assume that the spacetime is locally Minkowski on the particle's characteristic length scale. For this interpretation to hold, one must ensure that these concentrations of energy are stable.

Consider a real scalar-field multiplet $\{\phi^1,\dots, \phi^D\}$ in $D+1$ dimensional Minkowski spacetime, with the Lagrangian 
\begin{equation}
	\mathcal{L}_\phi(\phi^a,X^{bc}) = f(X^{bc})-V(\phi^a) \,,
\end{equation}
where $f(X^{bc})$ and $V(\phi^a)$ are the kinetic and potential terms, respectively, and
\begin{equation}
	\label{eq:scalar_kin}
	X^{bc}= -\frac{1}{2}\partial_\mu\phi^b \partial^\mu\phi^c \,.
\end{equation}
The EMT is given by
\begin{align}
	\label{eq:scalar_emt_gen}
	T_{\mu\nu}&=-2\frac{\delta\mathcal{L}_\phi}{\delta g^{\mu\nu}} + g_{\mu\nu}\mathcal{L}_\phi \nonumber \\
	&=\frac{\partial\mathcal{L}_\phi}{\partial X^{ab}}\partial_\mu\phi^a \partial_\nu\phi^b + g_{\mu\nu}\mathcal{L}_\phi\,,
\end{align}
and the total energy $E$ can be calculated via the integral
\begin{equation}
	\label{eq:field_energ}
	E= \int T_{00}\, d^D x \,.
\end{equation}

It has been shown by Hobart and Derrick \cite{Hobart1963,Derrick1964} that a maximally symmetric concentration of energy is not stable when $D>1$ if the kinetic term is given by the usual form
\begin{equation}
	\label{eq:scalar_kin_stand}
	f(X^{bc})= \delta_{bc}X^{bc}=X \,.
\end{equation}
However, one can show that these solutions do exist if the kinetic part of the Lagrangian is not given by Eq. \eqref{eq:scalar_kin_stand}, as long as certain conditions are met \cite{Avelino2011,Avelino2018a}. 

Consider a static solution $\phi^a=\phi^a(x^i)$ and a rescalling of the spatial coordinates $x^i\rightarrow\tilde{x}^i=\lambda x^i$. Under this transformation, a necessary condition for the existence of this solution is that the transformed static concentration of energy $E(\lambda)$ must satisfy
\begin{equation}
	\label{eq:Derrick_1}
	\left[\frac{\delta E(\lambda)}{\delta\lambda}\right]_{\lambda=1}=0 \,.
\end{equation}
In addition, stability of this solution demands that
\begin{equation}
	\label{eq:Derrick_2}
	\left[\frac{\delta^2E(\lambda)}{\delta\lambda^2}\right]_{\lambda=1}\geq0 \,.
\end{equation}
Eqs. \eqref{eq:Derrick_1} and \eqref{eq:Derrick_2} succinctly summarize Derrick's argument.

Under the spherical transformation $x^i\rightarrow\tilde{x}^i=\lambda x^i$ the line element of the metric may be rewritten as
\begin{equation}
	ds^2=-dt^2+\delta_{ij}dx^i dx^j = -dt^2+\tilde{g}_{ij}d\tilde{x}^i d\tilde{x}^j \,.
\end{equation}
where $\delta_{ij}$ is the Kronecker delta and  $\tilde{g}_{ij}=\lambda^{-2}\delta_{ij}$ is the spatial part of the transformed metric, with determinant $\tilde{g}=-\lambda^{-2D}$.

Since we assume a static solution, in the particle's proper frame
\begin{equation}
	\frac{\delta\mathcal{L}_\phi}{\delta g^{00}}=0 \,,
\end{equation}
so that the $00$ component of Eq. \eqref{eq:scalar_emt_gen} reads
\begin{equation}
	\label{eq:EMT_00}
	T_{00}=-\mathcal{L}_\phi \,.
\end{equation}
The transformed static concentration of energy can therefore be written as
\begin{equation}
	E(\lambda)=-\int \sqrt{-\tilde{g}} \mathcal{L}_\phi(\tilde{g}^{ij},x^k)d^D\tilde{x} \,,
\end{equation}
and so a variation with respect to $\lambda$ returns
\begin{align}
	\frac{\delta E}{\delta\lambda}&=-\int \frac{\delta\left(\sqrt{-\tilde{g}} \mathcal{L}_\phi\right)}{\delta\lambda} d^D\tilde{x} \nonumber \\
	&=-\int \left[2\frac{\delta\mathcal{L}_\phi}{\delta\tilde{g}^{ij}}\tilde{g}^{ij}-D\mathcal{L}_\phi\right]\lambda^{-1-D} d^D\tilde{x} \,,
\end{align}
Setting $\lambda=1$ and applying the first condition \eqref{eq:Derrick_1}, we obtain
\begin{align}
	\label{eq:Derrick_cond_1}
	\left[\frac{\delta E}{\delta\lambda}\right]_{\lambda=1}&=-\int \left[2\frac{\delta\mathcal{L}_\phi}{\delta g^{ij}}g^{ij}-D\mathcal{L}_\phi\right] d^Dx \nonumber \\
	&= \int T^i_{\;\;i} d^Dx= 0 \,.
\end{align}
Likewise,
\begin{equation}
	\frac{\delta^2E}{\delta\lambda^2}=-\int \left[4\frac{\delta^2\mathcal{L}_\phi}{\delta(\tilde{g}^{ij})^2}(\tilde{g}^{ij})^2 +(2-4D)\frac{\delta\mathcal{L}_\phi}{\delta\tilde{g}^{ij}}\tilde{g}^{ij} +(D+D^2)\mathcal{L}_\phi\right]\lambda^{-2-D}d^D\tilde{x} \,,
\end{equation}
and applying Eq. \eqref{eq:Derrick_2} we obtain
\begin{equation}
	\int \left[4\frac{\delta^2\mathcal{L}_\phi}{\delta(g^{ij})^2}(g^{ij})^2 +(2-4D)\frac{\delta\mathcal{L}_\phi}{\delta g^{ij}}g^{ij} +D(D+1)\mathcal{L}_\phi\right]d^Dx \leq 0	 \,,
\end{equation}

The on-shell Lagrangian of a fluid composed of solitonic particles $\mathcal{L}_\text{fluid}$ can be written as volume average of the Lagrangian of a collection of particles, \textit{i.e.}
\begin{equation}
	\label{eq:Lag_vol_average}
	\mathcal{L}_\text{fluid}\equiv\langle \mathcal{L}_\phi \rangle = \frac{\int \mathcal{L}_\phi d^D x}{\int d^D x} \,.
\end{equation}
Taking into account Eqs. \eqref{eq:EMT_00} and \eqref{eq:Derrick_cond_1}, and that one can write the trace of the EMT as $T=T^0_{\;\;0}+T^i_{\;\;i}$, we obtain
\begin{equation}
	\label{eq:Lag_EMT_sol}
	\mathcal{L}_\text{fluid}= \frac{\int T^0_{\;\;0} d^D x}{\int d^D x} =\frac{\int T d^D x - \int T^i_{\;\;i}d^D x}{\int d^D x} = \langle T\rangle \equiv T_\text{fluid} \,,
\end{equation}
where $T_\text{fluid}$ is the trace of the EMT of the fluid. Note that 
\begin{equation}
	\label{eq:Lag_EMT_sol2}
	\mathcal{L}_\text{fluid}= T_\text{fluid} \,,
\end{equation}
is a scalar equation, and therefore invariant under any Lorentz boost, despite being derived in a frame where the particles are at rest. This also implies that Eq. \eqref{eq:Lag_EMT_sol2} is valid regardless of individual soliton motion, as long as they preserve their structure and mass.

\section{Which Lagrangian?}
\label{sec.whichlag}
We have shown that the models represented in Sections \ref{sec:fluid_lag}, \ref{sec.barlag} and \ref{sec.nmclag}, characterized by different Lagrangians, may be used to describe the dynamics of a perfect fluid. If the matter fields couple only minimally to gravity, then these models may even be used to describe the same physics. However, this degeneracy is generally broken in the presence of an NMC either to gravity \cite{Bertolami2007,Bertolami2008a,Sotiriou2008,Faraoni2009,Harko2010,Bertolami2010,Ribeiro2014,Azizi2014,Bertolami2014} or to other fields \cite{Bekenstein1982,Sandvik2002,Anchordoqui2003,Copeland2004,Lee2004,Koivisto2005,Avelino2008,Ayaita2012,Pourtsidou2013,Bohmer2015,Barros2019,Kase2020a}, in which case the identification of the appropriate on-shell Lagrangian may become essential in order to characterize the overall dynamics (note that this is not an issue if the form of the off-shell Lagrangian is assumed \textit{a priori}, as in \cite{Bohmer2015,Bettoni2011,Bettoni2015,Dutta2017,Koivisto2015,Bohmer2015a,Brax2016,Tamanini2016}). The models in Sections \ref{sec:fluid_lag} and \ref{sec.barlag} imply both the conservation of particle number and entropy. However, both the entropy and the particle number are in general not conserved in a fluid described as a collection of point particles or solitons. Hence, the models in Section \ref{sec.nmclag} can have degrees of freedom that are not accounted for by the models in Section \ref{sec:fluid_lag} and \ref{sec.barlag}, since they take into account microscopic properties of the fluid. The models in Section \ref{sec.nmclag} have an equation of state parameter $w=p/\rho$ in the interval $[0,1/3]$, which while appropriate to describe a significant fraction of the energy content of the Universe, such as  dark matter, baryons and photons, cannot be used to describe dark energy.

\section{The role of the Lagrangian in nonminimally-coupled models}
\label{sec.nmclagrole}

As previously discussed, the EMT does not provide a complete characterization of NMC matter fields, since the Lagrangian will also in general explicitly appear in the EOM. To further clarify this point, we present here a few examples of models in which there is an NMC between matter or radiation with dark energy (DE) \cite{Ferreira2020}, and we will explore an NMC with gravity in Chapter \ref{chapter_nmc}.

Consider the model described by the action
\begin{equation}
	\label{eq: NMC DE}
	S = \int d^4 x \sqrt{-g}\left[R+\mathcal{L}+\mathcal{L}_{\text{F}\phi}\right]\,,
\end{equation}
where $R$ is the Ricci scalar, $\phi$ is the DE scalar field described by the Lagrangian
\begin{equation}
	\label{eq: L DE}
	\mathcal{L} = X - V(\phi)\,,
\end{equation}
and $\mathcal{L}_{\text{F}\phi}$ is the Lagrangian of the matter term featuring an NMC with DE \cite{Wetterich1995,Amendola2000,Zimdahl2001,Farrar2003}
\begin{equation}
	\label{eq: matter DE nmc}
	\mathcal{L}_{\text{F}\phi}=f(\phi)\mathcal{L}_\text{F}\,.
\end{equation}
Here, $f(\phi)>0$ is a regular function of $\phi$ and $\mathcal{L}_\text{F}$ is the Lagrangian that would describe the matter component in the absence of an NMC to gravity (in which case $f$ would be equal to unity). Using the variational principle, it is straightforward to derive the EOM for the gravitational and scalar fields. They are given by
\begin{equation}
	\label{eq: eom metric NMC DE}
	G^{\mu\nu} = f\, T_\text{F}^{\,\mu\nu} +\nabla^\mu\phi\nabla^\nu\phi -\frac{1}{2}g^{\mu\nu}\nabla_\alpha\phi\nabla^\alpha\phi -g^{\mu\nu}\, V\,,
\end{equation}
\begin{equation}
	\label{eq: eom DE NMC}
	\square\phi -\frac{d V}{d \phi} + \frac{d f}{d \phi}\mathcal{L}_\text{F}=0\,,
\end{equation}
respectively, where $G^{\mu\nu}$ is the Einstein tensor, $\square \equiv \nabla_\mu\nabla^\mu$ is the Laplace-Beltrami operator, and 
\begin{equation}
	\label{eq: Em tensor NMC}
	T_\text{F}^{\mu\nu}=\frac{2}{\sqrt{-g}}\frac{\delta\left(\sqrt{-g}\mathcal{L}_\text{F}\right)}{\delta g_{\mu\nu}}
\end{equation}
are the components of the EMT associated with the Lagrangian $\mathcal{L}_\text{F}$. Note that the Lagrangian of the matter fields is featured explicitly in the EOM for $\phi$. Thus, knowledge of the EMT alone is not enough to fully describe the dynamics of any of the fields.

Consider the coupled matter EMT defined by $T_{\text{F}\phi}^{\mu\nu}=f(\phi)T_\text{F}^{\mu\nu}$. By taking the covariant derivative of Eq. \eqref{eq: eom metric NMC DE} and using the Bianchi identities one obtains
\begin{equation}
	\label{eq: EM tensor cons 1}
	\nabla_\nu T_{\text{F}\phi}^{\mu\nu} = -\nabla_\nu(\partial^\nu\phi \partial^\mu\phi) + \frac{1}{2}\nabla^\nu(\partial_\alpha\phi \partial^\alpha\phi) + \frac{d V}{d\phi}\partial^\mu\phi \,,
\end{equation}
thus showing that the coupled matter EMT is in general not conserved. Using  Eq. \eqref{eq: eom DE NMC} it is possible to rewrite this equation in such a way as to highlight the explicit dependence on the Lagrangian
\begin{equation}
	\nabla_\nu T_{\text{F}\phi}^{\mu\nu} = \frac{d f}{d\phi}\mathcal{L}_\text{F}\partial^\nu\phi \,. \label{EMCF1}
\end{equation}
If $\mathcal{L}_\text{F}$ describes a fluid of particles with fixed rest mass $m_\text{F}$, then one must have $\mathcal{L}_\text{F}=T_\text{F}$, as per Sec. \ref{subsec:single_part}. Also, $\mathcal{L}_{\text{F}\phi} =f(\phi)\mathcal{L}_\text{F}$ will describe a fluid with particles of variable rest mass $m(\phi) = f(\phi)m_\text{F}$. In this case, Eq.~\eqref{EMCF1} may also be written as
\begin{equation}
	\nabla_\nu T_{\text{F}\phi}^{\mu\nu} = -\beta T_\text{F}\partial^\mu\phi \,, \label{EMCF2}
\end{equation}
where
\begin{equation}
	\beta(\phi) = -\frac{d \ln m(\phi)}{d \phi}\,. \label{betadef}
\end{equation}
In the present example, we shall focus on the macroscopic fluid dynamics, but the NMC between matter and DE also affects the dynamics of the individual particles (see, for example, \cite{Ayaita2012} for more details).

\subsection{Coupling between dark energy and neutrinos}

A related model featuring an NMC between neutrinos and DE, so-called growing neutrino quintessence, where the neutrinos are described by the Lagrangian
\begin{equation}
	\label{eq: neutrinos L}
	\mathcal{L}_{\mathcal V} = i\bar{\psi}\left(\gamma^\alpha\nabla_\alpha +m(\phi)\right)\psi  \,,
\end{equation}
has been investigated in \cite{Ayaita2012}. Here, $\bar{\psi}$ is the Dirac conjugate, $m (\phi)$ is a DE-field dependent neutrino rest mass, the quantities $\gamma^\alpha(x)$ are related to the usual Dirac matrices $\gamma^a$ via $\gamma^\alpha =\gamma^a e^\alpha_a$ where $e^\alpha_a$ are the vierbein, with $g^{\alpha\beta}=e_a^\alpha e_b^\beta \eta^{ab}$ and $\eta^{ab} =\text{diag}(-1,1,1,1)$, and $\nabla_\alpha$ is the covariant derivative that now takes into account the spin connection (see \cite{Brill1957} for more details on the vierbein formalism). The classical EOM for the neutrinos, derived from the action
\begin{equation}
	\label{eq: neutrino action}
	S = \int d^4 x \sqrt{-g}\left[R+\mathcal{L}+\mathcal{L}_{\mathcal V}\right]\,,
\end{equation}
may be written as
\begin{align}
	\label{eq: neutrino eom}
	\gamma^\alpha\nabla_\alpha \psi+m(\phi)\psi &=0\,, \\
	\nabla_\alpha\bar{\psi}\gamma^\alpha-m(\phi)\bar{\psi} &= 0\,.
\end{align}
The components of the corresponding EMT are
\begin{equation}
	\label{eq: neutrino em tensor}
	T_{\mathcal V}^{\alpha\beta} = -\frac{i}{2}\bar{\psi}\gamma^{(\beta}\nabla^{\alpha)}\psi +\frac{i}{2}\nabla^{(\alpha}\bar{\psi}\gamma^{\beta)}\psi \,,
\end{equation}
where the parentheses represent a symmetrization over the indices $\alpha$ and $\beta$. The trace of the EMT is given by
\begin{equation}
	\label{eq: EM neutrino trace}
	T_{\mathcal V} = i\bar\psi\psi m(\phi) = - m(\phi)  \widehat n \,,
\end{equation}
where $\widehat n= -i\bar\psi\psi$ is a scalar that in the nonrelativistic limit corresponds to the neutrino number density.

Taking the covariant derivative of Eq. \eqref{eq: neutrino em tensor} one obtains
\begin{equation}
	\label{eq: EM neut tensor cons1}
	\nabla_\mu T_{\mathcal V}^{\alpha\mu} = -\beta (\phi) T_{\mathcal V}\partial^\alpha\phi \,,
\end{equation}
where $\beta(\phi)$ is defined in Eq. \eqref{betadef}. A comparison between Eqs. \eqref{EMCF2} and \eqref{eq: EM neut tensor cons1} implies that $\mathcal{L}_{F\phi}$ and $\mathcal{L}_{\mathcal V}$ provide equivalent on-shell descriptions of a fluid of neutrinos in the presence of an NMC to the DE field. The same result could be achieved by analyzing the dynamics of individual neutrino particles \cite{Ayaita2012}. 

\subsection{Coupling between dark energy and the electromagnetic field}

Consider now a model described by Eqs. \eqref{eq: NMC DE} and \eqref{eq: matter DE nmc} with
\begin{equation}
	\label{eq: electro lagrangian}
	\mathcal{L}_\text{F} = \mathcal{L}_\text{EM} = -\frac{1}{4}F_{\mu\nu}F^{\mu\nu} \,,
\end{equation}
where $F_{\alpha\beta}$ is the electromagnetic field tensor \cite{Bekenstein1982,Sandvik2002,Avelino2008}. This model will naturally lead to a varying fine-structure ``constant"
\begin{equation}
	\label{eq: fine struct}
	\alpha(\phi) = \frac{\alpha_0}{f(\phi)} \,,
\end{equation}
whose evolution is driven by the dynamics of the DE scalar field $\phi$. Equation \eqref{eq: eom DE NMC} implies that the corresponding EOM is given by 
\begin{equation}
	\label{eq: eom electro scalar}
	\square\phi -\frac{d V}{d \phi} + \frac{\alpha_0}{4
		\alpha^2}\frac{d \alpha}{d \phi}F_{\mu\nu}F^{\mu\nu}= 0 
\end{equation}
or, equivalently,
\begin{equation}
	\label{eq: eom electro scalar 2}
	\square\phi -\frac{d V}{d \phi} - \frac{\alpha_0}{    \alpha^2}\frac{d \alpha}{d \phi}\mathcal{L}_\text{EM}= 0 \,.
\end{equation}
Electromagnetic contributions to baryon and lepton mass mean that in general $\mathcal{L}_\text{EM} \neq 0$. However, $\mathcal{L}_\text{photons}=(E^2-B^2)_\text{photons} = 0$ (here, $E$ and $B$ represent the magnitude of the electric and magnetic fields, respectively) and, therefore, electromagnetic radiation does not contribute to $\mathcal{L}_\text{EM}$. Note that the last term on the left-hand side of Eq. \eqref{eq: eom electro scalar} is constrained, via the equivalence principle, to be small \cite{Olive2002}. Therefore, the contribution of this term to the dynamics of the DE field is often disregarded (see, e.g., \cite{Anchordoqui2003,Copeland2004,Lee2004}).

It is common, in particular in cosmology, to describe a background of electromagnetic radiation as a fluid of point particles whose rest mass is equal to zero (photons). In this case, one should use the appropriate on-shell Lagrangian of this fluid in Eq. \eqref{eq: eom electro scalar 2}. In Sec. \ref{sec.nmclag} we have shown that if the fluid is made of particles of fixed mass, then the appropriate on-shell Lagrangian is $\mathcal{L}_\text{EM}= T = 3p-\rho$. For photons (with $p=\rho/3$) this again implies that the on-shell Lagrangian $\mathcal{L}_\text{EM}$ vanishes, thus confirming that photons do not source the evolution of the DE scalar field $\phi$. 
 

\chapter[Nonminimally Coupled $f(R)$ Theories]{Nonminimally Coupled $f(R)$ Theories} 
\label{chapter_nmc} 

A popular group of modified gravity theories are those that feature a nonminimal coupling (NMC) between the matter fields and gravity or another field. As mentioned in Chapter \ref{chapter_lag}, this leads to the appearance of the on-shell Lagrangian of the matter fields in the equations of motion (EOM), in addition to the usual energy-momentum tensor (EMT). In this chapter, we will analyse the impact of this modification in the background dynamics of the Universe, its thermodynamics and the behaviour of particles and fluids. While these theories have been studied before in the literature, the use of the appropriate Lagrangian for the matter fields and the study of its consequences are presented in Sections \ref{sec.cosmo}, \ref{sec:emNMC}, \ref{sec.scalar},  \ref{sec.firstlaw}, \ref{sec.seclaw} and \ref{sec.htheo}, and published in Refs. \cite{Avelino2018,Azevedo2019b,Avelino2020,Azevedo2020,Avelino2022}.

\section{Action and equations of motion}

The family of theories investigated in this thesis is $f(R)$-inspired, featuring an NMC between curvature and matter. It can be defined through the action
\begin{equation}\label{eq:actionf1f2}
	S = \int d^4x \sqrt{-g} \left[\kappa f_1(R)+f_2(R)\mathcal{L}_\text{m} \right] \,,
\end{equation}
where $f_1(R)$ and $f_2(R)$ are arbitrary functions of the Ricci scalar $R$, $\mathcal{L}_\text{m}$ is the Lagrangian of the matter fields and $\kappa = c^4/(16\pi G)$ \cite{Allemandi2005,Nojiri2004,Bertolami2007,Sotiriou2008,Harko2010,Harko2011,Nesseris2009,Thakur2013,Bertolami2008a,Harko2013,Harko2015}. One recovers GR by taking $f_1(R)=R-2\Lambda$ and $f_2(R)=1$. The field equations are obtained as usual by imposing a null variation of the action with respect to the metric,
\begin{equation}\label{eq:fieldNMC}
	FG_{\mu\nu}=\frac{1}{2}f_2 T_{\mu\nu} + \Delta_{\mu\nu}F+\frac{1}{2}g_{\mu\nu}\kappa f_1 - \frac{1 }{2}g_{\mu\nu} RF \,,
\end{equation}
where
\begin{equation}
	\label{eq:F}
	F=\kappa f'_1+f'_2\mathcal{L}_\text{m}\,,
\end{equation}
the primes denote a differentiation with respect to the Ricci scalar, $G_{\mu\nu}$ is the Einstein tensor and $\Delta_{\mu\nu}\equiv\nabla_\mu \nabla_\nu-g_{\mu\nu}\square$, with $\square=\nabla_\mu \nabla^\mu$ being the D'Alembertian operator, and $T_{\mu\nu}$ are the components of the EMT given by
\begin{equation}
	\label{eq:energy-mom3}
	T_{\mu\nu} = - \frac{2}{\sqrt{-g}}\frac{\delta\left(\sqrt{-g}\mathcal{L}_\text{m}\right)}{\delta g^{\mu\nu}}\,.
\end{equation}

An alternative way to write the field equations is
\begin{equation}
	\label{eq:eff_eom}
	G_{\mu\nu}=\lambda\left(T_{\mu\nu}+\hat{T}_{\mu\nu}\right)\,,
\end{equation}
where $\lambda$ is an effective coupling and $\hat{T}_{\mu\nu}$ is an effective EMT. Comparing Eqs. \eqref{eq:fieldNMC} and \eqref{eq:eff_eom} immediatly sets
\begin{equation}
	\lambda = \frac{f_2}{2\kappa f'_1+2\mathcal{L}_\text{m} f'_2}\,,
\end{equation}
and
\begin{equation}
	\hat{T}_{\mu\nu} = \left(\frac{\kappa f_1}{f_2}-\frac{RF}{f_2}\right)g_{\mu\nu}+\frac{2\Delta_{\mu\nu} F}{f_2}\,.
\end{equation}
Demanding that gravity remain attractive requires a positive effective coupling, that is \cite{Bertolami2009}
\begin{equation}
	\label{eq:pos_grav}
	\lambda=\frac{f_2}{2\kappa f'_1+2\mathcal{L}_\text{m} f'_2} > 0\,.
\end{equation}

A common requirement for the NMC functions $f_1$ and $f_2$ is that the theory remain free of Dolgov-Kawasaki instabilities. This criterion is similar to the one found for minimally-coupled $f(R)$ theories and was initially determined in \cite{Faraoni2007} and later extended in \cite{Bertolami2009}. It is expressed by
\begin{equation}
	\label{eq:dolgov}
	\kappa f''_1+f''_2\mathcal{L}_\text{m}\geq0\,.
\end{equation}

A crucial feature of these theories is that the EMT is no longer covariantly conserved: in fact, applying the Bianchi identities to the EOM \eqref{eq:fieldNMC} leads to
\begin{equation}\label{eq:conservNMC}
	\nabla^\mu T_{\mu\nu} = \frac{f'_2}{ f_2}\left(g_{\mu\nu}\mathcal{L}_\text{m}-T_{\mu\nu}\right)\nabla^\mu R \,.
\end{equation}
This energy-momentum nonconservation associated with the NMC to gravity suggests that a more general definition of the EMT, including a yet to be defined gravitational contribution, may be worthy of consideration. This has not yet been sufficiently explored in the context of NMC theories, but has proven to be quite problematic in the context of GR \cite{Bonilla1997,Clifton2013,Sussman2014}.

Likewise, the fluids will suffer an additional acceleration due to the NMC to gravity. Consider a perfect fluid with EMT
\begin{equation}
	\label{eq:EMT_fluid}
	T^{\mu\nu}=(\rho+p)U^\mu U^\nu + p g^{\mu\nu} \,,
\end{equation}
where $\rho$ and $p$ are the proper energy density and pressure of the fluid, respectively, and $U^\mu$ is the 4-velocity of a fluid element, satisfying $U_\mu U^\mu = -1$. The 4-acceleration of a perfect fluid element may be written as \cite{Bertolami2007,Bertolami2008a}
\begin{align}
	\label{eq:nmc_fluid_acc}
	\mathfrak{a}^\mu_\text{[fluid]}&=\frac{dU^\mu}{d\tau}+\Gamma^\mu_{\alpha\beta}U^\alpha U^\beta \nonumber\\
	&=\frac{1}{\rho+p}\left[\frac{f'_2}{ f_2}\left(\mathcal{L}_\text{m}-p\right)\nabla_\nu R + \nabla_\nu p\right]h^{\mu\nu}_\text{[fluid]}\,,
\end{align}
where $h^{\mu\nu}_\text{[fluid]}\equiv g^{\mu\nu}+U^\mu U^\nu$ is the projection operator.

It is then clear that the knowledge of the Lagrangian of the matter fields, regardless of their macroscopic or microscopic description, is fundamental to the study of these theories.

\section{Equivalence with scalar field theories}\label{subsec:scalarNMC}

As shown before for $f(R)$ theories, one can rewrite NMC theories similarly with an action with two scalar fields, and do a conformal transformation to the Einstein frame \cite{Bertolami2008b}. For completeness, we make a brief detour to show explicitly how this equivalence manifests. For the former, it is enough to write the action
\begin{equation}\label{eq:actionscalarNMC1}
	S=\int d^4x \sqrt{-g}\left[\kappa f_1(\chi)+\varphi(R-\chi)+f_2(\chi)\mathcal{L}_\text{m} \right]\,,
\end{equation}
where variation with respect to $\varphi$ and $\chi$ give, respectively,
\begin{equation}\label{eq:NMCphi}
	\chi=R,
\end{equation}
\begin{equation}\label{eq:NMCpsi}
	\varphi=\kappa f'_1(\chi)+f'_2(\chi)\mathcal{L}_\text{m}\,.
\end{equation}
We can rewrite the action \eqref{eq:actionscalarNMC1} in the form of a JBD type theory with $\omega_{\rm JBD}=0$
\begin{equation}\label{eq:actionscalarNMC2}
	S=\int d^4x \sqrt{-g}\left[\varphi R-V(\chi,\varphi)+f_2(\chi)\mathcal{L}_\text{m} \right]\,,
\end{equation}
with a potential
\begin{equation}\label{eq:potentialscalarNMC}
	V(\chi,\varphi)=\varphi\chi-\kappa f_1(\chi)\,.
\end{equation}

Alternatively, one can make a conformal transformation $g_{\mu\nu}\rightarrow \tilde{g}_{\mu\nu}=\Omega^2 g_{\mu\nu}$ so that the action \eqref{eq:actionscalarNMC2} reads
\begin{align}\label{eq:actionscalarNMC3}
	S = \int d^4x \sqrt{-\tilde{g}} \Big[\kappa \Omega^{-2}\varphi \Big(&\tilde{R} +6\tilde{\square}(\ln\Omega) - 6\tilde{g}^{\mu\nu}\partial_\mu(\ln \Omega) \partial_\nu (\ln \Omega) \Big) \nonumber \\ &-\Omega^{-4}\varphi^2 U+\Omega^{-4}f_2(\chi)\mathcal{L}_\text{m}(\Omega^{-2}\tilde{g}_{\mu\nu},\Psi_M) \Big]\,,
\end{align}
where the potential $U$ is given by  
\begin{equation}\label{eq:potentialscalarNMC2}
	U=\kappa\frac{\varphi\chi-\kappa f_1(\chi) }{ \varphi^2}\,.
\end{equation}
To express the action in the Einstein frame, the conformal factor must now obey
\begin{equation}\label{eq:conftransNMC}
	\Omega^2=\varphi,
\end{equation}
where it is assumed that $f'_1(R)>0$. We now rescale the two scalar fields as
\begin{equation}\label{eq:NMCfield1}
	\phi\equiv \sqrt{3 \kappa} \ln\left(\varphi\right)\,,
\end{equation}
\begin{equation}\label{eq:NMCfield2}
	\psi\equiv f_2(\chi)\,.
\end{equation}
Once again Gauss's theorem allows us to finally write the action \eqref{eq:actionscalarNMC3} as
\begin{align}
	S = \int d^4x \sqrt{-\tilde{g}} \bigg[\kappa\tilde{R} &- \frac{1}{2}\partial^\alpha\phi \partial_\alpha \phi  - U(\phi,\psi) \nonumber \\
	&+\psi e^{-2\frac{\phi }{ \sqrt{3\kappa}}}\mathcal{L}_\text{m}\left(e^{-\frac{\phi }{ \sqrt{3\kappa}}}\tilde{g}_{\mu\nu},\Psi_M\right) \bigg]\,,
\end{align}
where $e^{-\frac{\phi }{ \sqrt{3\kappa}}}\tilde{g}_{\mu\nu}$ is the physical metric.

\section{NMC cosmology}
\label{sec.cosmo}
Consider a homogeneous and isotropic universe described by a Friedmann-Lemaître-Robertson-Walker (FLRW) metric, represented by the line element
\begin{equation}
	\label{eq:line-nmc}
	ds^2=-dt^2+a^2(t)\left[\frac{dr^2}{1-kr^2} +r^2 d\theta^2 +r^2\sin^2\theta d\phi^2\right]\,,
\end{equation}
where $a(t)$ is the scale factor, $k$ is the spatial curvature of the universe (which is observationally constrained to be very close to zero), $t$ is the cosmic time, and $r$, $\theta$ and $\phi$ are polar comoving coordinates, filled by a collection of perfect fluids, with EMT given by Eq. \eqref{eq:EMT_fluid}.

The modified Friedmann equation (MFE) becomes
\begin{equation}\label{eq:fried-f1f2-1}
	H^2+\frac{k}{a^2}=\frac{1}{6F}\left[FR- \kappa f_1+f_2\sum_i\rho_i-6H\dot{F}\right]\,,
\end{equation}
and the modified Raychaudhuri equation (MRE) becomes
\begin{equation}\label{eq:ray-f1f2-1}
	2\dot{H}+3H^2+\frac{k}{a^2}=\frac{1}{2F}\left[FR-\kappa f_1-f_2 \sum_i p_i-4H\dot{F}-2\ddot{F}\right] \,,
\end{equation}
where $\rho_i$ and $p_i$ are the energy density and pressure of each of the fluids, respectively, a dot represents a derivative with respect to the cosmic time, $H\equiv\dot{a}/a$ is the Hubble factor, $F'\equiv \kappa f''_1+f''_2 \mathcal{L}_\text{m}$, and $\mathcal{L}_\text{m}$ is now the on-shell matter Lagrangian. It should be noted that the presence of both standard and NMC $f(R)$ terms can produce very interesting dynamics, both at late and early times.

It was shown in Chapter \ref{chapter_lag} that the on-shell Lagrangian of a perfect fluid composed of particles with fixed rest mass and structure is given by
\begin{equation}
	\label{eq:lag-nmc}
	\mathcal{L}_\text{m}=T=T^{\mu\nu}g_{\mu\nu}=3p-\rho \,.
\end{equation}
Therefore, the $t$ component of Eq. \eqref{eq:conservNMC} becomes
\begin{equation}
	\label{eq:dens-cons_nmc}
	\dot{\rho}+3H(\rho+p)=-(\mathcal{L}_\text{m}+\rho)\frac{\dot{f}_2}{ f_2}=-3p\frac{\dot{f}_2}{ f_2}\,,
\end{equation}
where $w\equiv p/\rho$ is the equation of state (EOS) parameter. For a single fluid $i$, Eq. \eqref{eq:dens-cons_nmc} can be directly integrated to give
\begin{equation}
	\label{eq:densityevo}
	\rho_i=\rho_{i,0} a^{-3(1+w_i)} f_2^{-3w_i}\,,
\end{equation}
where $\rho_{i,0}$ is the energy density at the present time, when $a(t)=a_0=1$. It is then immediate to show that in the case of dust ($w=0$) the usual conservation law $\rho \propto a^{-3}$ holds, while in the case of photons ($w=1/3$) the NMC generally leads to a significant change to the evolution of the photon energy density ($\rho \propto a^{-4}f_2^{-1}$ instead of $\rho \propto a^{-4}$). The relative change to the conservation laws of photons and baryons may be used to derive strong constraints on the form of $f_2$, as we shall see in Chapter \ref{chapter_constr}.

Taking into account that the proper pressure of the fluid is given by $p=\rho  v^2/3$ (assuming, for simplicity, that $v$ is the same for all particles) and requiring that the number of particles per comoving volume be conserved, or equivalently that $\rho \propto \gamma a^{-3}$, (where $\gamma \equiv 1/\sqrt{1-v^2}$), and substituting into Eq. \eqref{eq:dens-cons_nmc}, we find that the velocity of fluid particles in FLRW spacetime is given by
\begin{equation}
	\label{eq:dotvel}
	{\dot v} +\left( H + \frac{{\dot f}_2}{f_2} \right) (1-v^2) v =0 \,.
\end{equation}
Hence, the momentum of such a particle evolves as \cite{Avelino2018}
\begin{equation}
	\label{eq:momev}
	m \gamma v \propto (a f_2)^{-1}\,.
\end{equation}

\section{Energy-momentum constraints in general relativity}
\label{sec.em-cons}

It is worth to take a deeper dive on how the nonconservation of the EMT influences the dynamics of particles and fluids, and how such dynamics constrain the allowed on-shell Lagrangians. In preparation for a subsequent analysis in the broader context of NMC gravity, in this section we shall present five different derivations of the equation for the evolution of the speed of free localized particles of fixed mass and structure in a homogeneous and isotropic FLRW universe, relying solely on the conservation of linear momentum and energy in the context of GR. 

Let us start by considering the Einstein-Hilbert action
\begin{equation}
S=\int {\sqrt {-g}}(\kappa R+{\mathcal L}_{\rm m}) d^4 x \,,
\end{equation}
In general relativity the EMT of the matter fields, whose components are given by Eq. \eqref{eq:energy-mom3}, is covariantly conserved, so that
\begin{equation}
\nabla_{\nu} {T{^\nu}}_\mu= 0\,. \label{emconservation}
\end{equation}
Throughout this section we shall consider either the EMT of the individual particles with components ${T_*}^{\mu \nu}$ or the EMT of a perfect fluid composed of many such particles. The components of the latter are given by Eq. \eqref{eq:EMT_fluid} --- notice the use of $*$ to identify the EMT of a single particle. Energy-momentum conservation implies that
\begin{equation}
	h^{\mu \beta} \nabla_{\alpha} {T{^\alpha}}_\beta = (\rho+p)U^\nu  \nabla_\nu U^\mu  + h^{\mu \beta}  \nabla_\beta p= 0 \,,
\end{equation}
where $h^{\mu \nu}=g^{\mu \nu}+ U^\mu U^\nu$ is the projection operator. In the case of dust $p=0$ and, therefore, 
\begin{equation}
	U^\nu  \nabla_\nu U^\mu  = 0 \,.
\end{equation}

\subsection{Particles in a Minkowski spacetime}

Consider a single particle and a rest frame where its EMT is static. Assuming that the gravitational interaction plays a negligible role on the particle structure,  the spacetime in and around the particle may be described by a Minkowski metric element
\begin{equation}
	ds^2=-dt^2 +d {\vb r} \cdot d {\vb r} =-dt^2  +dx^2+dy^2+dz^2 \,,
\end{equation}
where $t$ is the physical time and ${\vb r} = (x,y,z)$ are Cartesian coordinates.

The particle's proper frame is defined by
\begin{equation}
	\int {{T_*}^i}_{0[\rm prop]} \, d^3 r_{[\rm prop]}= -\int {{T_*}^0}_{i[\rm prop]} \, d^3 r_{[\rm prop]} =0 \,, \label{Ti0}
\end{equation}
where Latin indices take the values $1,\dots,3$, with 
\begin{equation}
	E_{[\rm prop]}  = - \int {{T_*}^0}_{0[\rm prop]} \, d^3 r  \label{Eprop}
\end{equation}
being the proper energy of the particle (the subscript $[\rm prop]$ is used to designate quantities evaluated in the proper frame). On the other hand, the generalized von Laue conditions \cite{doi:10.1002/andp.19113400808,Avelino2018a},
\begin{equation}
	\int {{T_*}^1}_{1[\rm prop]}  \, d^3 r_{[\rm prop]}  = \int {{T_*}^2}_{2[\rm prop]}  \, d^3 r_{[\rm prop]}  \int {{T_*}^3}_{3[\rm prop]}  \, d^3 r _{[\rm prop]} = 0 \,, \label{vonlaue}
\end{equation}
are required for particle stability.

Consider a Lorentz boost in the $x$ direction defined by
\begin{eqnarray}
	t&=&\gamma(t_{[\rm prop]}+vx_{[\rm prop]})\,,\\
	x&=&\gamma(x_{[\rm prop]}+vt_{[\rm prop]})\,,\\
	y&=&y_{[\rm prop]}\,,\\
	z&=&z_{[\rm prop]}\,,
\end{eqnarray}
where $\gamma=\left(1-v^{2}\right)^{-1/2}$ is the Lorentz factor and $v$ is the particle velocity. Under this boost, the components of the EMT ${T_*}_{\mu\nu}$ transform
as
\begin{equation}
	{{T_*}^{\mu}}_{\nu}={\Lambda^{\mu}}_{\alpha}\,{\Lambda_{\nu}}^{\beta}\,{{T_*}^{\alpha}}_{\beta[\rm prop]}\,,\label{emtransf}
\end{equation}
where the non-zero components of ${\Lambda^\mu}_\alpha$ and ${\Lambda_\nu}^\beta$ are
\begin{eqnarray}
	{\Lambda^0}_0&=&{\Lambda^1}_1={\Lambda_0}^0={\Lambda_1}^1=\gamma\,,\label{lorentz1}\\
	{\Lambda^0}_1&=&{\Lambda^1}_0=-{\Lambda_0}^1=-{\Lambda_1}^0=\gamma v\,,\label{lorentz2}\\
	{\Lambda^2}_2&=&{\Lambda^3}_3={\Lambda_2}^2={\Lambda_3}^3=1\label{lorentz3}\,,
\end{eqnarray}
with all other components vanishing. In the moving frame the energy and linear momentum of the particle are given, respectively, by
\begin{eqnarray}
	E &=& - \int {{T_*}^0}_0 \, d^3 r =E_{[\rm prop]}  \gamma  \,, \label{Eeq}\\
	\mathfrak{p} &=& \int {{T_*}^1}_0 \, d^3 r=E_{[\rm prop]} \gamma v =E v \label{peq}\,,
\end{eqnarray}
where Eqs. (\ref{Ti0}),  (\ref{Eprop}), (\ref{emtransf}),  (\ref{lorentz1}), (\ref{lorentz2}), as well as Lorentz contraction, have been taken into account in the derivation of Eqs. (\ref{Eeq}) and  (\ref{peq}). These two  equations imply that $E^2-\mathfrak{p}^2=E_{\rm [prop]}^2$ and
\begin{equation}
	\dot{\mathfrak{p}}  = \dot E  \frac{E}{\mathfrak{p}}=  \frac{\dot E}{v}=E_{[\rm prop]} \dot v \gamma^3\label{dotp}\,.
\end{equation}

On the other hand, using Eqs.  (\ref{emtransf}), (\ref{lorentz2}) and  (\ref{lorentz3}) one finds
\begin{eqnarray}
	\int {{T_*}^1}_1 \, d^3 r&=& E_{[\rm prop]}  \gamma v^2 = E v^2\,,\label{T11}\\
	\int {{T_*}^2}_2 \, d^3 r &=& \int {{T_*}^3}_3 \, d^3 r = 0  \,,
\end{eqnarray}
so that
\begin{equation}
	\int {{T_*}^i}_i\, d^3 r=E_{[\rm prop]}\gamma v^2 = E v^2\,. \label{trace}
\end{equation}
Also notice that
\begin{equation}
	\int {T_*} \, d^3 r= \int {{T_*}^\mu}_\mu\, d^3 r=-\frac{E_{[\rm prop]}}{\gamma} = -\frac{E}{\gamma^2}\,. \label{traceT}
\end{equation}

\subsection{Free particles in an FLRW spacetime}

In a flat, homogeneous and isotropic universe, described by the FLRW metric, the line element may be written as
\begin{equation}
	ds^2=a^2(\zeta)(-d\zeta^2+d \vb{q} \cdot d \vb{q}\,) \,,
\end{equation}
where $a(\zeta)$ is the scale factor, $\zeta = \int a^{-1} dt$ is the conformal time and $\vb q$ are comoving Cartesian coordinates. In an FLRW spacetime the nonvanishing components of the connection are given by 
\begin{equation}
	\Gamma_{0 0}^0=\mathscr{H}\,, \quad \Gamma_{i j}^0=\mathscr{H} \,\delta_{ij}\,, \quad \Gamma_{0 j}^i=\mathscr{H} \,{\delta^i}_j \,,
\end{equation}
where $\mathscr{H} = \grave{a}/a$ and a grave accent denotes a derivative with respect to the conformal time $\zeta$.

\subsubsection*{Linear momentum conservation}

Consider again a single free particle moving along the $x$-direction. The $x$-component of Eq. (\ref{emconservation}) describing momentum conservation in an FLRW spacetime then implies that
\begin{equation}
	0=\nabla_\nu {{T_*}^\nu}_1   =  \partial_0 {{T_*}^0}_1 + \partial_i {{T_*}^i}_1  + 4  \mathscr{H} {{T_*}^0}_1\,.
\end{equation}
Integrating over the spatial volume one finds that
\begin{equation}
	\grave{\mathfrak{p}}+ \mathscr{H} \mathfrak{p}=0\,, \label{pev}
\end{equation}
where
\begin{equation}
	\mathfrak{p}=\int {{T_*}^0}_1 \, d^3 r=a^3 \int {{T_*}^0}_1 \, d^3 q \,.
\end{equation}
In this derivation we have assumed that the particle is isolated so that the EMT vanishes outside it. Hence,
\begin{equation}
	\int \partial_i {{T_*}^\mu}_\nu \, d^3 q =0 \label{isolated}
\end{equation}
for any possible value of $\mu$, $\nu$ and $i$. Notice that Eq. (\ref{pev}) implies that $\mathfrak{p} = E_{[\rm prop]} \gamma v \propto a^{-1}$. Dividing Eq. (\ref{pev})  by $E_{[\rm prop]} $, taking into account Eq. (\ref{dotp}),  one obtains the  equation for the evolution of the free particle velocity in a homogeneous and isotropic FLRW universe:
\begin{equation}
	{\grave v}+ \mathscr{H}(1-v^2) v=0\,. \label{vev}
\end{equation}

\subsubsection*{Energy conservation}

Energy conservation, on the other hand, implies that
\begin{equation}
	0=\nabla_\nu {{T_*}^\nu}_0   =  \partial_0 {{T_*}^0}_0 + \partial_i {{T_*}^i}_0  +  3  \mathscr{H} {{T_*}^0}_0 -  \mathscr{H} {{T_*}^i}_i \,.
\end{equation}
Integrating over the spatial volume, and using Eqs. (\ref{trace}) and (\ref{isolated}), one finds that
\begin{equation}
	{\grave E}+ \mathscr{H} v^2 E=0\,, \label{Eev}
\end{equation}
where
\begin{equation}
	E=-\int {{T_*}^0}_0 \,d^3 r=-a^3 \int {{T_*}^0}_0 \,d^3 q \,.
\end{equation}
Dividing Eq. (\ref{Eev}) by $v$, taking into account Eq. (\ref{dotp}), once again one obtains Eq. (\ref{pev}) for the evolution of linear momentum in a homogeneous and isotropic FLRW universe.

\subsection{Perfect fluids in an FLRW spacetime}

We shall now derive the dynamics of free particles assuming that they are part of a homogeneous perfect fluid (see Eq. \eqref{eq:EMT_fluid}) with the proper energy density $\rho$ and the proper pressure $p$ depending only on time.

\subsubsection*{Linear momentum conservation: dust}

In the case of a perfect fluid with vanishing proper pressure $p$, the components of the EMT are
\begin{equation}
	\label{eq:pfemt}
	T^{\mu\nu}=\rho \, U^\mu U^\nu \,,
\end{equation}
If the fluid moves in the positive $x$-direction, then
\begin{eqnarray}
	U^0&=&\frac{d\zeta}{d\tau} = \frac{\gamma}{a}\,,\\
	U^1&=&\frac{dq^1}{d\tau} = {\grave q}^1\frac{d\zeta}{d\tau}=v \, U^0 =v\frac{\gamma}{a} \,,\\
	U^2&=&U^3=0\,.
\end{eqnarray}
The $x$-component of Eq. (\ref{emconservation}), describing momentum conservation, implies that
\begin{equation}
\grave U^1 U^0+ 2 \Gamma^1_{1 0} U^0 U^1=0 \,.
\end{equation}
Multiplying this equation by $E_{[\rm prop]} a/U^0$, taking into account that $U^1=\gamma v/a$ and that $\Gamma^0_{1 1}=\mathscr{H}$, one obtains once again Eq. (\ref{pev}) for the evolution of linear momentum of a free particle in a homogeneous and isotropic FLRW universe. 

\subsubsection*{Energy conservation: dust}

The time component of Eq. (\ref{emconservation}), describing energy conservation, is given by
\begin{equation}
\grave U^0 U^0+  \Gamma^0_{0 0} U^0 U^0 + \Gamma^0_{1 1} U^1 U^1=0 \,.
\end{equation}
Multiplying this equation by $E_{[\rm prop]} a/U^0$, taking into account that $U^0=\gamma/a$,  $U^1=\gamma v/a$, and that $\Gamma^0_{0 0}=\Gamma^0_{1 1}=\mathscr{H}$, one obtains once again Eq. (\ref{Eev}) for the evolution of the energy of a free particle in a homogeneous and isotropic FLRW universe, which has been shown to be equivalent to Eq. (\ref{pev}) for the evolution of the linear momentum. 

\subsubsection*{Energy conservation: homogeneous and isotropic fluid}
\label{sec:emGRF}

We shall now consider a homogeneous and isotropic perfect fluid (at rest in the comoving frame, so that $U^i=0$)  made up of free particles all with the same speed $v$. This fluid can be pictured as the combination of six equal density dust fluid components moving in the positive/negative $x$, $y$, and $z$ directions. The time component of Eq. (\ref{emconservation}), describing energy conservation, implies that
\begin{equation}
	{\grave \rho} +  3 \mathscr{H} (\rho +p) =0\,. \label{rhoEQ}
\end{equation}
If the number $N$ of particles in a volume $V=a^3$ is conserved then
\begin{equation}
\rho=\frac{NE}{V} = \frac{N E}{a^3}=N E_{[\rm prop]}  \frac{\gamma}{a^3}\propto \frac{\gamma}{a^3}\,. \label{rhoEV}
\end{equation}
On the other hand, if the perfect fluid is an ideal gas then its proper pressure is given by 
\begin{equation}
p=\rho\frac{v^2}{3}\,. \label{Pideal}
\end{equation}
Substituting the conditions given in Eqs. (\ref{rhoEV}) and (\ref{Pideal}) into Eq. (\ref{rhoEQ}) multiplied by $a/N$, one again arrives at Eq. (\ref{Eev}), the same as the one derived considering energy conservation for individual free particles.

\section{Energy-momentum constraints in NMC gravity}
\label{sec:emNMC}

In this section we shall again present five different derivations of the equation for the evolution of the speed of individual localized particles of fixed mass and structure in a homogeneous and isotropic FLRW universe, but now considering the possibility of a coupling between gravity and the matter fields. We shall demonstrate that consistency between the results obtained uniquely defines the correct form of the corresponding on-shell Lagrangians \cite{Avelino2022}. 

Let us once again consider the action \eqref{eq:actionf1f2}, allowing for a NMC between gravity and the matter fields. In this and other NMC theories the energy momentum tensor of the matter fields, whose components are given in Eq. (\ref{eq:energy-mom3}), is not in general covariantly conserved. Instead one has
\begin{equation}
\nabla_{\nu} {T_{\mu}}^\nu=  S_{\mu}\,, \label{Tncons}
\end{equation}
where
\begin{equation}
S_{\mu} = ({\mathcal L}_{\rm m} \delta_{\mu}^\nu -{T_{\mu}}^\nu) \frac{\nabla_\nu f_2}{f_2} \,.\label{Tncons2}
\end{equation}

Here, we shall again consider either the EMT of the individual particles with components ${T_*}^{\mu \nu}$ or the EMT of a perfect fluid composed of many such particles whose components are given in Eq. \eqref{eq:EMT_fluid}.

In the case of a perfect fluid Eq. (\ref{Tncons}), with $S_\mu$ given by Eq. (\ref{Tncons2}), implies that \cite{Bertolami2007}
\begin{equation}
	U^\nu  \nabla_\nu U^\mu  = \frac{1}{\rho+p} \left[\left(\mathcal{L}_{\rm f} -p\right) \frac{\nabla_\nu f_2}{f_2} -\nabla_\nu p\right]h^{\mu \nu}\,, \label{UevNMC}
\end{equation}
where ${\mathcal L}_{\rm f}$ and $h^{\mu \nu}=g^{\mu \nu}+ U^\mu U^\nu$ are the  on-shell Lagrangian of the perfect fluid and the projection operator, respectively. In the following we shall also consider the particular case of dust with $\mathcal{L}_\text{f}=\mathcal{L}_\text{dust}$ (characterized by $p_{\rm dust}=0$ and $\rho_{\rm dust}=-T_{\rm dust}$), for which
\begin{equation}
	U^\nu  \nabla_\nu U^\mu  = \frac{\mathcal{L}_{\rm dust} }{\mathcal \rho_{\rm dust}} \frac{\nabla_\nu f_2}{f_2} h^{\mu \nu}\,, \label{UevNMCdust}
\end{equation}

\subsection{Free particles in FLRW spacetimes}

Consider once again the motion of localized particles of fixed mass and structure in an FLRW background, but this time in the context of NMC gravity. Given that the EMT is no longer covariantly conserved, the presence of additional dynamical terms, dependent on the matter Lagrangian, will need to be taken into account.

\subsubsection*{Linear momentum evolution}

For a single isolated particle moving along the $x$-direction in an FLRW background, the $x$-component of Eq. (\ref{Tncons2}) implies that
\begin{equation}
\int S_1 \, d^3 r =  -\mathfrak{p}   \frac{\grave f_2}{f_2} \,, \label{S1int}
\end{equation}
Hence, considering the $x$-component of Eq. (\ref{Tncons}), and following the same steps of the previous section, the equation for the evolution of the linear momentum of the particle can now be generalized to 
\begin{equation}
	\grave{\mathfrak{p}}+ \Theta  \, \mathfrak{p}=0\,, \label{pev1}
\end{equation}
where $\Theta$ is defined by
\begin{equation}
	\Theta = \frac{\grave b}{b}= \frac{\grave a}{a}+ \frac{\grave  f_2}{f_2} = \mathscr{H} + \frac{\grave  f_2}{f_2} \,, \label{pev1theta}
\end{equation}
and $b=a f_2$. Notice that Eq. (\ref{pev1})  was obtained without making any assumptions about the specific form of the on-shell Lagrangian.

\subsubsection*{Energy evolution}

Of course, one must be able to arrive at the same result using the time component of Eq. (\ref{Tncons}) --- otherwise there would be an inconsistency. The time component of Eq. (\ref{Tncons2}) requires that
\begin{equation}
\int S_0 \, d^3 r =  \left(\int {\mathcal L}_{\rm m} \,  d^3 r +E\right) \frac{\grave  f_2}{f_2} \,, \label{S0int}
\end{equation}
Following the same steps of the previous section but now using the time component of Eq. (\ref{Tncons}) and Eq. (\ref{S0int}) one obtains
\begin{equation}
	{\grave E}+ \mathscr{H}  v^2 E=-\left(\int {\mathcal L}_{\rm m} \,  d^3 r +E\right) \frac{\grave  f_2}{f_2} \,. \label{Eev1}
\end{equation}
Dividing Eq. (\ref{Eev1}) by $v$, taking into account Eqs. (\ref{peq}) and (\ref{dotp}), one finds that 
\begin{equation}
	\grave{\mathfrak{p}}+ \mathscr{H}  \mathfrak{p}=-\frac{\int {\mathcal L}_{\rm m} \,  d^3 r +E}{v} \frac{\grave  f_2}{f_2} \,.
\end{equation}
Consistency with Eqs. (\ref{pev1})  and  (\ref{pev1theta}) then requires that
\begin{equation}
	\mathfrak{p}=\frac{\int {\mathcal L}_{\rm m} \,  d^3 r +E}{v}\,,
\end{equation}
Taking into account that $\mathfrak{p}=v E=E_{[\rm prop]} \gamma v$ and Eq. (\ref{traceT}), this in turn implies that 
\begin{equation}
	\int {\mathcal L}_{\rm m} d^3 r = -\frac{E}{\gamma^2} =  -\frac{E_{[\rm prop]}}{\gamma} = \int {T_*} \, d^3 r\,.
\end{equation}
Hence, the volume average of the on-shell Lagrangian of a particle of fixed mass and structure is equal to the volume average of the trace of its EMT, independently of the particle structure and composition.

\subsection{Perfect fluids in FLRW spacetimes}

Here, we shall derive the dynamics of moving localized particles with fixed proper mass and structure in an FLRW assuming that they are part of a homogeneous perfect fluid, but now in the context of NMC gravity.

\subsubsection*{Linear momentum constraints: dust}

In the case of dust, a perfect fluid with $p_{\rm dust}=0$, the $x$-component of Eq. (\ref{UevNMC}) may be written as
\begin{equation}
\grave U^1 U^0+ 2 \Gamma^1_{1 0} U^0 U^1= \frac{\mathcal{L}_{\rm dust} }{\rho_{\rm dust}} \frac{\grave f_2}{f_2} U^0 U^1 \,.
\end{equation}
Multiplying this equation by $E_{[\rm prop]} a/U^0$, taking into account that $U^1=\gamma v/a$ and that $\Gamma^0_{1 1}=\mathscr{H}$, one obtains 
\begin{equation}
\grave{\mathfrak{p}} + \mathscr{H} \mathfrak{p} = \frac{\mathcal{L}_{\rm dust} }{\rho_{\rm dust}} \frac{\grave f_2}{f_2} \mathfrak{p} \,. \label{pevNMC}
\end{equation}
Consistency with Eqs. (\ref{pev1})  and  (\ref{pev1theta}) requires that
\begin{equation}
	\mathcal{L}_{\rm dust}=\mathcal -\rho_{\rm dust}= T_{\rm dust} \,. \label{dustlag}
\end{equation}

\subsubsection*{Energy constraints: dust}

The time component of Eq. (\ref{UevNMCdust}) is given by
\begin{equation}
\grave U^0 U^0+  \Gamma^0_{0 0} U^0 U^0 + \Gamma^0_{1 1} U^1 U^1=\frac{\mathcal{L}_{\rm dust} }{\mathcal \rho_{\rm dust}} \frac{\grave f_2}{f_2} (g^{00}+U^0 U^0)  \,,
\end{equation}
Multiplying this equation by $E_{[\rm prop]} a/U^0$, taking into account that $g^{00}=-1/a^2$, $U^0=\gamma/a$,  $U^1=\gamma v/a$, $v^2\gamma^2=\gamma^2-1$, and that $\Gamma^0_{0 0}=\Gamma^0_{1 1}=\mathscr{H}$, one obtains
\begin{equation}
	{\grave E}+ \mathscr{H} v^2 E=\frac{\mathcal{L}_{\rm dust} }{\mathcal \rho_{\rm dust}} \frac{\grave f_2}{f_2} E v^2\,. \label{EevNMC}
\end{equation}
Dividing Eq. (\ref{Eev1}) by $v$, taking into account Eqs. (\ref{peq}) and (\ref{dotp}), one again arrives at Eq. (\ref{pevNMC}) for the evolution of linear momentum. 

\subsubsection*{Energy constraints: homogeneous and isotropic fluid}

Consider a homogeneous and isotropic perfect fluid (at rest in the comoving frame, so that $U^i=0$) made up of localized particles of fixed mass and structure all with the same speed $v$. The time component of Eq. (\ref{Tncons}), is given by
\begin{equation}
{\grave \rho_{\rm f}} +  3 \mathscr{H} (\rho_{\rm f} +p_{\rm f}) =  -({\mathcal L}_{\rm f}  + \rho_{\rm f}) \frac{\grave f_2}{f_2}\,,\label{Tcons2}
\end{equation}
where ${\mathcal L}_{\rm f}$, $\rho_f$ and $p_{\rm f}$ are the on-shell Lagrangian, proper energy density and proper pressure of the fluid, respectively. If the number of particles is conserved then Eq. (\ref{rhoEV}) is satisfied. On the other hand, if the perfect fluid is an ideal gas then its proper pressure is given by Eq. (\ref{Pideal}): $p_{\rm f}=\rho_{\rm f} v^2/3$. Substituting the conditions given in Eqs. (\ref{rhoEV}) and (\ref{Pideal}) into Eq. (\ref{Tcons2}) and multiplying it by $a^3/N$, one obtains
\begin{equation}
	{\grave E}+ \mathscr{H}  v^2 E=  -\left(\frac{{\mathcal L}_{\rm f}}{\rho_{\rm f}}  + 1\right) \frac{\grave f_2}{f_2}E\,. \label{Eevf}
\end{equation}
As in Sec. \ref{sec:emGRF}, this homogeneous and isotropic perfect fluid can be pictured as the combination of six equal density dust fluid components moving in the positive/negative $x$, $y$, and $z$ directions. Therefore, in the proper frame of the resulting perfect fluid, the evolution of particle energy and linear momentum of each of its dust components and of the total combined fluid must be the same, \textit{i.e.} Eqs. (\ref{EevNMC}) and \eqref{Eevf} must result in the same equation of motion. This implies that
\begin{equation}
	-\frac{\mathcal{L}_{\rm dust} }{\rho_\text{dust}}v^2=\frac{\mathcal{L}_{\rm f} }{\rho_\text{f}}+1 \,. \label{finald}
\end{equation}
We can therefore write the on-shell Lagrangian of the perfect fluid as
\begin{align}
	\mathcal{L}_\text{f} &= -\rho_\text{f}\left(\frac{\mathcal{L}_\text{dust}}{\rho_{\rm dust}}v^2+1\right)=\rho_\text{f}\left(v^2-1\right)  \nonumber \\
	\Rightarrow \mathcal{L}_\text{f} &= 3p_{\rm f}-\rho_\text{f} = T_{\rm f} \,,\label{lagperf}
\end{align}
where we have taken into account Eqs. \eqref{Pideal} and \eqref{dustlag}. Naturally, in the case of dust ($v=0$) Eq. \eqref{lagperf} again implies that $\mathcal{L}_\text{dust}=T_{\rm dust}=-\rho_\text{dust}$.

In the derivation of this result the crucial assumption is that the fluid can be described by the ideal-gas equation of state --- no assumptions have been made regarding the role of gravity on the structure of the particles in this case. Notice that this result is not in contradiction with the findings of Refs. \cite{Harko2010a,Minazzoli2012}, according to which the on-shell Lagrangian of a fluid with 1) a conserved number of particles and 2) an off-shell Lagrangian dependent solely on the particle number number density is ${\mathcal L}_{\rm m}^{\rm on}=-\rho$, since the second condition does not apply to an ideal gas.

\section{Scalar matter fields in NMC gravity}
\label{sec.scalar}
It is interesting to analyse the case where the matter fields are given by a real scalar field $\phi$ governed by a generic Lagrangian of the form
\begin{equation}
	\mathcal{L}_\text{m} =\mathcal{L}_\text{m}(\phi,X) \,,
\end{equation}
where
\begin{equation}
	X= -\frac{1}{2}\nabla^\mu  \phi \partial_\mu \phi\,,
\end{equation}
is the kinetic term. Therefore Eq. \eqref{eq:energy-mom3} implies that the components of the EMT are given by
\begin{equation}
	T_{\mu\nu}=\mathcal{L}_{\text{m},X}\partial_\mu \phi \partial^\nu \phi + \mathcal{L}_\text{m} g_{\mu\nu}\, .
\end{equation}

We can now look at a few matter Lagrangians and examine their significance in NMC gravity \cite{Avelino2018}. 

\subsection{Perfect fluid with $\mathcal{L}_\text{m}=p$}

For timelike $\partial_\mu \phi$, it is possible to write the EMT in a perfect fluid form
\begin{equation}\label{eq:fluid2}
	T^{\mu\nu} = (\rho + p) U^\mu U^\nu + p g^{\mu\nu} \,,
\end{equation}
by means of the following identifications
\begin{equation}\label{eq:new_identifications}
	u_\mu = \frac{\partial_\mu \phi}{\sqrt{2X}} \,,  \quad \rho = 2 X p_{,X} - p \, ,\quad p =  \mathcal{L}_\text{m}(\phi,X)\, .
\end{equation}
In Eq.~(\ref {eq:fluid2}), $U^\mu$ are the components of the 4-velocity field describing the motion of the fluid, while $\rho$ and $p$ are its proper energy density and pressure, respectively. Observe that in this case $\mathcal{L}_\text{m}=p$, which in the context of GR is one of the possible choices considered in the literature for the on-shell Lagrangian of a perfect fluid. Note that, since the 4-velocity is a timelike vector, the correspondence between scalar field models of the form  $\mathcal{L}_\text{m}=\mathcal{L}_\text{m} (\phi,X)$ and perfect fluids breaks down whenever $\partial_\mu \phi$ is spacelike, as is the case of non-trivial static solutions. In the case of a homogeneous and isotropic universe filled with a perfect fluid with arbitrary density $\rho$ and $p=0$, Eqs. (\ref{eq:fluid2}) and  (\ref{eq:new_identifications}) imply that the dynamics of this fluid may be described by a matter Lagrangian whose on-shell value is equal to zero everywhere (a simple realization of this situation would be to take $\mathcal{L}_\text{m}=X-V= {\dot \phi}^2/2-V(\phi)$ with the appropriate potential $V$ and initial conditions, so that $V(\phi)$ is always equal to ${\dot \phi}^2/2$).

\subsection{Solitons in 1+1 dimensions: $\mathcal{L}_\text{m}=T$}

In this section, our particle shall be modelled once again as a topological soliton of the field $\phi$, but in 1+1 dimensions. For concreteness, assume that the matter fields may be described by a real scalar field with Lagrangian 
\begin{equation}
	\mathcal{L}_\text{m}= -\frac12 \partial_\mu \phi \partial^\mu \phi  - V(\phi)\,,
\end{equation}
where $V(\phi) \ge 0$ is a real scalar field potential
\begin{equation}
	V(\phi)=\frac{\lambda}{4} \left(\phi^2-\varepsilon^2\right)^2\,,
\end{equation}
which has two degenerate minima at $\phi=\pm \varepsilon$.

In this case, ${\mathcal L}_{\text{m},X}=1$ and the EMT of the matter fields is given by
\begin{equation}
	\label{eq:fluid1}
	T^{\mu\nu} = \partial^\mu \phi \partial^\nu \phi + \mathcal{L}_\text{m} g^{\mu \nu}\,.
\end{equation}
On the other hand, the equation of motion for the scalar field $\phi$ is 
\begin{equation}
	\Box \phi=- \frac{f'_2}{f_2} \partial_\mu R \partial^\mu \phi +V_{,\phi}\,. \label{dalphi}
\end{equation}
Multiplying Eq. \eqref{dalphi} by $\partial^\nu \phi$, and taking into account that \eqref{eq:fluid1}, one recovers Eq. \eqref{eq:conservNMC}.

\subsubsection*{Minkowski spacetime}

In a $1+1$ dimensional Minkowski space-time the line element can be written as $ds^2=-dt^2+dz^2$.  Hence, neglecting the self-induced gravitational field, the Lagrangian and the equation of motion of the scalar field $\phi$ are given respectively by
\begin{eqnarray}
	\mathcal{L}_\text{m}&=& \frac{{\left(\phi_{,t}\right)^2}}{2} -\frac{\left(\phi_{,z}\right)^2}{2}-V(\phi)\,, \label{Lagphi}\\ 
	\phi_{,tt} - \phi_{,zz}&=& -\frac{dV}{d \phi}\,, \label{phieqmM}
\end{eqnarray}
where the $\phi_{,t}$, $\phi_{,z}$ and $\phi_{,tt}$, $\phi_{,zz}$ represent the first and second derivatives with respect to the physical time $t$ and the spatial coordinate $z$.

The components of the EMT of the particle can now be written as
\begin{eqnarray}
	\rho_\phi=-{T^{0}}_0&=&\frac{\left(\phi_{,t}\right)^2}{2}+\frac{\left(\phi_{,z}\right)^2}{2}+V(\phi)\,,\\
	T^{0z}&=&-\phi_{,t}\phi_{,z}\,,\\
	p_\phi={T^{z}}_z&=&\frac{\left(\phi_{,t}\right)^2}{2}+\frac{\left(\phi_{,z}\right)^2}{2}-V(\phi)\,,
\end{eqnarray}
so that the trace $T$ of the EMT is given by
\begin{equation}
	T={T^\mu}_\mu= {T^{0}}_0+{T^{z}}_z =-\rho_\phi+p_\phi=-2 V(\phi)\,. \label{Ttrace}
\end{equation}

Consider a static soliton with $\phi=\phi(z)$. In this case Eq. (\ref{phieqmM}) becomes
\begin{equation}
	\phi_{,zz}= \frac{dV}{d\phi} \label{phieqmM1}\,,
\end{equation}
and it can be integrated to give
\begin{equation}
	\frac{\left(\phi_{,z}\right)^2}{2} = V\,,\label{KeqU}
\end{equation}
assuming that $|\phi| \to \varepsilon$ for $z \to \pm \infty$. If the particle is located at $z=0$,  Eq. (\ref{phieqmM1}) has the following solution
\begin{equation}
	\phi = \pm \varepsilon \tanh\left(\frac{z}{{\sqrt 2}R}\right)\,,
\end{equation}
with
\begin{equation}
	R=\lambda^{-1/2} \varepsilon^{-1}\,.
\end{equation}
The rest mass of the particle is given by
\begin{eqnarray}
	m&=&\int_{- \infty}^{\infty} \rho dz = 2 \int_{- \infty}^{\infty} V dz = \frac{8 {\sqrt 2}}{3} V_\text{max} R  = \nonumber \\
	&=& \frac{2{\sqrt 2}}{3}\lambda^{1/2}  \varepsilon^3\,,
\end{eqnarray}
where $V_\text{max} \equiv V(\phi=0) = \lambda \varepsilon^4/4$. Here we have taken into account that Eq. (\ref{KeqU}) implies that in the static case the total energy density is equal to $2V$. On the other hand, from Eqs. (\ref{Lagphi}) and (\ref{Ttrace}), one also has that
\begin{equation}
	\mathcal{L}_\text{m}=T\,, \label{LT}
\end{equation}
where this equality is independent of the reference frame, and, consequently, it does not depend on whether the particle is moving or at rest. Also note that this result also applies to collections of particles and, in particular, to one which can be described as perfect fluid. However, unlike the result obtained for a homogeneous scalar field described by a matter Lagrangian of the form $\mathcal{L}_\text{m}(\phi,X)$, according to which the on-shell Lagrangian of a perfect fluid with proper pressure $p=0$ is $\mathcal{L}_\text{m}^{\rm on} = 0$ (independently of its proper density $\rho$), one finds that a perfect fluid with $p=0$ made of static solitonic particles would have an on-shell Lagrangian given by $\mathcal{L}_\text{m}^{\rm on} = T =  -\rho$. This is an explicit demonstration that the Lagrangian of a perfect fluid depends on microscopic properties of the fluid not specified by its EMT.

\subsubsection*{FLRW spacetime}

Consider a $1+1$ dimensional FLRW space-time with line element $ds^2=-dt^2+a^2(t) dq_z^2$, where $q_z$ is the comoving spatial coordinate and $a(t)$ is the scale factor. Taking into account that 
\begin{equation}
	{ \phi^{,\mu}}_{,\mu} = \left(- \Gamma^\mu_{\mu \nu} +  \frac{f'_2}{f_2} \partial_\nu R  \right)   \phi^{,\nu}\,,
\end{equation}
one obtains
\begin{equation}
	{\ddot \phi}  +   \left(H + \frac{{\dot f}_2}{f_2}\right){\dot \phi} - \nabla^2 \phi= -\frac{dV}{d\phi}\,, \label{dynphi}
\end{equation}
where $H \equiv {\dot a} / a$ is the Hubble parameter and $\nabla^2 \equiv  d^2 /dz^2 = a^{-2} d^2 / d q_z^2$ is the physical Laplacian.

The dynamics of $p$-branes in $N+1$-dimensional FRW universes has been studied in detail in \cite{Sousa2011a,Sousa2011b} (see also \cite{Avelino2016}). There, it has been shown that the equation for the velocity $v$ of a $0$-brane in a $1+1$-dimensional FRW spacetime implied by Eq. (\ref{dynphi}) is given by
\begin{equation}
	{\dot v} +\left( H + \frac{{\dot f}_2}{f_2} \right) (1-v^2) v =0 \,,
\end{equation}
which is exactly the same result we obtained in Eq.~\eqref{eq:dotvel} for the velocity of the fluid particles with the on-shell Lagrangian $\mathcal{L}_\text{m}^{\rm on}=T$.

\section{The first law of thermodynamics and particle creation/decay}\label{sec.firstlaw}

Naturally, the absence of energy-momentum conservation has significant implications to the laws of thermodynamics, namely the possibility of a violation of the second law of thermodynamics. To study these effects, we shall consider the thermodynamics of a universe filled with a perfect fluid, in the presence of an NMC between geometry and matter described by the action given in Eq.~\eqref{eq:actionf1f2}. 

Particle creation or decay via an NMC to gravity would require significant perturbations to the FLRW geometry on the relevant microscopic scales since the FLRW metric is essentially Minkowskian on such scales. The constraints on gravity on microscopic scales are extremely weak and it might be possible to construct viable modified theories of gravity in which the gravitational interaction on such scales is significantly enhanced with respect to GR (see, for example, \cite{Avelino2012a,Avelino2012}). However, these small scale perturbations have not been considered in the derivation of Eq.~\eqref{eq:dens-cons_nmc} and have not been explicitly taken into account in previous works when considering particle creation or decay via an NMC to gravity. Consequently, the only consistent interpretation for the change to the evolution of the energy density of a fluid made of soliton-like particles associated with the term on the right-hand-side of Eq.~\eqref{eq:dens-cons_nmc} is the modification to the evolution of the linear momentum of such particles described by Eq.~\eqref{eq:momev}. Here, we shall start by considering the thermodynamics of a homogeneous and isotropic universe in the absence of significant small-scale perturbations, and then describe phenomenologically the case in which microscopic perturbations to the FLRW geometry result in particle creation or decay.

\subsection{Perfect fluid with $\mathcal{L}_\text{m}=T$}

We start by treating the Universe as a system where the average number of particles per comoving volume is conserved, for which the first law of thermodynamics takes the form
\begin{equation}
	\label{eq:conservenergy}
	d (\rho a^3)= dQ_{\rm NMC}-pd(a^3) \, ,
\end{equation} 
where $dQ_{\rm NMC}$ is the ``heat'' received by the system over the interval of time $dt$ due to the NMC between the gravitational and the matter fields \cite{Azevedo2019b}. As previously mentioned, in the literature \cite{Prigogine1988, Prigogine1989, Lima2014, Harko2015} an adiabatic expansion ($dQ/dt=0$) is usually considered, and therefore an extra term associated with particle creation due to the NMC between the gravitation and matter fields is added to Eq.~\eqref{eq:conservenergy}. However, as in previous work, an FLRW metric is assumed. Hence, no perturbations to the background geometry that could be responsible for spontaneous particle creation are considered. In this scenario,  we are left with associating the NMC with the non-adiabaticity of the expansion.

Eq.~\eqref{eq:conservenergy} may be rewritten as
\begin{equation}
	\label{eq:heat1}
	\dot{\rho}+3H(\rho+p)=\frac{{\dot{Q}}_{\rm NMC}}{ a^3} \, .
\end{equation}
Using Eq.~\eqref{eq:dens-cons_nmc}, one obtains the ``heat'' transfer rate with $\mathcal{L}_\text{m}=3p-\rho$
\begin{eqnarray}
	\label{eq:heat_transf}
	{\dot{Q}}_{\rm NMC}&=&-({\mathcal L}_\text{m}  + \rho) a^3 \frac{\dot f_2}{f_2}=-3p a^3\frac{\dot f_2}{f_2} \nonumber \\
	&=&-\rho v^2 a^3\frac{\dot f_2}{f_2} \, .
\end{eqnarray}
This implies that for non-relativistic matter ($v\ll1$), such as baryons and cold dark matter, ${\dot{Q}}_{\rm NMC}\sim 0$ so that the usual energy-momentum conservation approximately holds. On the other hand, relativistic matter is strongly impacted by this energy-momentum transfer.

\subsection{Particle creation/decay and effective Lagrangians}

Here, we consider the possibility that the perturbations to the FLRW geometry on microscopic scales may be responsible for particle creation or decay. Discussing particle creation/decay with the matter Lagrangian $\mathcal{L}_\text{m}=T$ in great detail would of course require a microscopic description of the particle structure, which we leave purposefully generic, and its interaction with gravity on microscopic scales. While such analysis is beyond the scope of this thesis, we can treat particle creation/decay phenomenologically, by introducing a modification to the energy-momentum conservation equation. If particle number is not conserved due to the NMC, an additional term, associated to particle creation/decay, should therefore be added to the right-hand side of Eq.~\eqref{eq:dens-cons_nmc} 
\begin{equation}
	\label{eq:densitycreation}
	{\dot \rho} +  3 H (\rho +p) =  -({\mathcal L}_\text{m}  + \rho) \frac{\dot f_2}{f_2} -\mathcal{L}_\Gamma \frac{\dot f_2}{f_2} \, .
\end{equation}
Note that $\mathcal{L}_\Gamma$ is not a true Lagrangian, but rather a phenomenological term associated to the effect of the NMC between matter and gravity on microscopic scales. If the mass and structure of the particles does not change due to the NMC to gravity, except for (almost) instantaneous particle creation or decay, the on-shell Lagrangian of the perfect fluid is still well described by $\mathcal{L}_\text{m}^{\rm on}=T$ (we also allow for almost instantaneous scattering events which do not have an impact in the form of the perfect-fluid Lagrangian). Hence, Eq.~\eqref{eq:momev} still describes the cosmological contribution to the evolution of the linear-momentum of the particles. Equation~\eqref{eq:densitycreation} may then be rewritten as
\begin{equation}
	\label{eq:densitycreation2}
	{\dot \rho} +  3 H (\rho +p) =  -({\mathcal L}_\text{eff}  + \rho) \frac{\dot f_2}{f_2} \, ,
\end{equation} 	
where
\begin{equation}
	\label{eq:efflagrangian}
	\mathcal{L}_\text{eff} = \mathcal{L}_\text{m} + \mathcal{L}_\Gamma \, .
\end{equation}

In this case Eq.~\eqref{eq:conservenergy} is changed to \cite{Prigogine1989}
\begin{equation}
	\label{eq:conservenergycreation}
	d (\rho a^3)= dQ_{\rm NMC}-pd(a^3) + \frac{h}{n}d (n a^3) \, ,
\end{equation} 
where $n$ is the particle number density and $h=\rho+p$ is the enthalpy per unit volume. For simplicity, we have also implicitly assumed that all particles are identical and that the corresponding perfect fluid is always in thermodynamic equilibrium. This is a natural assumption if the rate of particle creation/decay is much smaller than the particle scattering rate, a case in which thermalization following particle creation/decay occurs (almost) instantaneously.

Equation~\eqref{eq:conservenergycreation} may be rewritten as as
\begin{equation}
	\label{eq:creation}
	\dot{\rho}+3H(\rho+p)=\frac{{\dot{Q}}_{\rm NMC}}{ a^3} + \frac{h}{n}(\dot{n}+3Hn) \, ,
\end{equation}
and using Eq.~\eqref{eq:densitycreation2} one finds that
\begin{equation}
	\label{eq:creation2}
	\frac{{\dot{Q}}_{\rm NMC}}{ a^3} + \frac{h}{n}(\dot{n}+3Hn)=-(\mathcal{L}_\text{eff}+\rho) \frac{\dot f_2}{f_2} \, .
\end{equation}
Equations~\eqref{eq:heat_transf}, \eqref{eq:efflagrangian} and \eqref{eq:creation2} also imply that
\begin{equation}
	\label{eq:creation3}
	\frac{\rho+p}{n}(\dot{n}+3Hn)=-\mathcal{L}_{\Gamma} \frac{\dot f_2}{f_2}\, .
\end{equation}
Introducing the particle creation/decay rate 
\begin{equation}
	\label{eq:gamma}
	\Gamma  = \frac{\dot{n}}{n}+3H \, ,
\end{equation}
and using Eq.~\eqref{eq:creation3} one obtains
\begin{equation}
	\label{eq:creation4}
	\Gamma= -\frac{\mathcal{L}_\Gamma}{ \rho+p}\frac{\dot f_2}{f_2}\, .
\end{equation}

Alternatively, particle creation/decay may be described as an extra effective creation/decay pressure $p_\Gamma$ of the perfect fluid that must be included in the continuity equation \cite{Prigogine1988}
\begin{equation}
	\label{eq:conteqpres}
	\dot{\rho} +3H(\rho+p+p_\Gamma)= -({\mathcal L}_\text{m}  + \rho) \frac{\dot f_2}{f_2} \, ,
\end{equation}
where
\begin{equation}
	\label{eq:creationpressure}
	p_\Gamma = \frac{\mathcal{L}_{\Gamma}}{3H} \frac{\dot{f}_2}{ f_2}\,,
\end{equation}
may be obtained from Eq. \eqref{eq:densitycreation2}.

We have argued that the correct on-shell form of the Lagrangian of a perfect fluid composed of solitonic particles is $\mathcal{L}_\text{m}=T$, even in the presence of (almost) instantaneous particle scattering and/or particle creation/decay, and when $\mathcal{L}_\text{eff}=\mathcal{L}_\text{m}$, one trivially recovers the results of the previous subsection. Nevertheless, one may ask whether or not the Lagrangians suggested in previous work to describe such a perfect fluid could play the role of effective Lagrangians. Let us then consider the particular cases with $\mathcal{L}_\text{eff}=-\rho$ and $\mathcal{L}_\text{eff}=p$. 

If $\mathcal{L}_\text{eff}=-\rho$ then 
\begin{equation}
	\label{eq:Lgammarho}
	\mathcal{L}_\Gamma = \mathcal{L}_\text{eff} - \mathcal{L}_\text{m} = -3p \, ,
\end{equation}
where we have used Eq. \eqref{eq:efflagrangian} and taken into account that $\mathcal{L}_\text{m}=T=3p-\rho$. Hence, in this case 	
\begin{equation}
	\label{eq:creationpressurerho}
	p_\Gamma = -\frac{p}{ H} \frac{\dot{f}_2}{f_2}\, ,
\end{equation}
and there is a particle creation/decay rate given by
\begin{equation}
	\label{eq:gammarho}
	\Gamma= \frac{3p}{\rho+p}\frac{\dot f_2}{f_2}\, .
\end{equation}
Notably, if $\mathcal{L}_\text{eff}=-\rho$ the standard conservation equation for the energy density is recovered.

If $\mathcal{L}_\text{eff}=p$ then
\begin{equation}
	\label{eq:Lgammap}
	\mathcal{L}_\Gamma = \rho-2p \, .
\end{equation}
In this case, the effective pressure is equal to
\begin{equation}
	\label{eq:creationpressurep}
	p_\Gamma = \frac{\rho-2p}{ 3H} \frac{\dot{f}_2}{ f_2} \, ,
\end{equation}
and the particle creation/decay rate is
\begin{equation}
	\label{eq:gammap}
	\Gamma= -\frac{\rho-2p}{ \rho+p}\frac{\dot f_2}{f_2}\, .
\end{equation}
Note that if ${\mathcal L}_\text{eff}=p$ the standard evolution equation for the density is not recovered, unless $p=-\rho$.

In both cases, $\mathcal{L}_\text{eff}=-\rho$ and $\mathcal{L}_\text{eff}=p$, the particle creation/decay rate $\Gamma$ would not in general be a constant. Rather than depending on the particle properties and on the way these are affected by the NMC to gravity on microscopic scales, for a given choice of the function $f_2$ the evolution of $\Gamma$ given in Eqs. \eqref{eq:gammarho} and \eqref{eq:gammap} would depend essentially on the cosmology and the macroscopic properties of the fluid. As discussed before, the FLRW metric is essentially Minkowski on the microscopic scales relevant to particle creation/decay. Consequently, one should not expect such a cosmological dependence of the particle creation/decay rate $\Gamma$, which questions the relevance of the effective Lagrangians $\mathcal{L}_\text{eff}=-\rho$ and $\mathcal{L}_\text{eff}=p$ in this context.

\section{The second law of thermodynamics}\label{sec.seclaw}

Consider the fundamental thermodynamic relation
\begin{equation}
	\label{eq:ftr}
	\mathcal{T}dS = d(\rho a^3) + pda^3\, ,
\end{equation}
where $S$ is the entropy of the matter content and $\mathcal{T}$ is the temperature. Equations \eqref{eq:dens-cons_nmc}, \eqref{eq:heat1} and \eqref{eq:ftr} imply that \cite{Azevedo2020}
\begin{equation}
	\label{eq:entropy}
	\mathcal{T}dS = dQ_{\rm NMC}= -3p a^3 \frac{d f_2}{f_2} \, ,
\end{equation}
as opposed to GR, where $dS = 0$.

Consider a universe filled with dust and radiation, both represented by perfect fluids with proper energy density $\rho_{{\rm dust}}$ and $\rho_{{\rm r}}$, and pressure $p_{{\rm dust}}=0$ and $p_{{\rm r}}=\rho_{{\rm r}}/3$. Eq. {\eqref{eq:entropy} therefore implies that the comoving entropy of the dust component is conserved ($dS_{\rm dust}=0$). In this case the total comoving entropy $S$ is equal to the comoving entropy of the radiation component $S_{\rm r}$ ($S=S_{\rm r}$). Its proper pressure and density --- be it bosonic, fermionic or both --- satisfy $p_{\rm r}=\rho_{\rm r}/3$ and $\rho_{\rm r} \propto \mathcal{T}^4$, respectively. Here, the scattering timescale is implicitly assumed to be much smaller than the characteristic timescale of the change of the NMC coupling function $f_2$, so that the radiation component can always be taken to be in thermodynamic equilibrium at a temperature $\mathcal{T}$ (we are also assuming that the chemical potential is zero). Hence, in the case of radiation, Eq.~\eqref{eq:dens-cons_nmc} may be written as
	\begin{equation}
		\label{eq:conservenergypart}
		\frac{d\rho_{\rm r}}{d\mathcal{T}}\, \dot{\mathcal{T}}+3H(\rho_{\rm r}+p_{\rm r})=-3p_{\rm r}\frac{\dot{f}_2}{f_2} \,,
	\end{equation}
	or equivalently,
	\begin{equation}
		\label{eq:Tdot1}
		\dot{\mathcal{T}}=- \frac{3H\left(\rho_{\rm r}+p_{\rm r}\right)+3p_{\rm r}\frac{\dot{f}_2}{f_2}}{{\frac{d\rho_{\rm r}}{d\mathcal{T}}}} \,,
	\end{equation}
	with $p_{\rm r}=\rho_{\rm r}/3$ and $\rho_{\rm r} \propto \mathcal{T}^4$.
	
	Equation~\eqref{eq:Tdot1} is easily integrated and returns
	\begin{equation}
		\label{eq:Tradeq}
		\mathcal{T} \propto a^{-1} f_2^{-1/4} \,,
	\end{equation}
	so that 
	\begin{equation}
		\label{eq:densrad}
		\rho_{\rm r} \propto a^{-4}f_2^{-1} \,.
	\end{equation}
	Taking into account Eq. \eqref{eq:Tradeq} and the fact that $p_{\rm r}=\rho_{\rm r}/3 \propto \mathcal{T}^4$, Eq.~\eqref{eq:entropy} can be easily integrated to give
	\begin{equation}
		\label{eq:entropyevoeq}
		S\propto f_2^{-3/4} \,.
	\end{equation}

	Imposing the second law of thermodynamics
	\begin{equation}
		\label{eq:entropycreation}
		\mathcal{T}\dot{S} = \dot{Q}_{\rm NMC} = -3p a^3\frac{f'_2}{f_2}\dot{R} \geq 0\, ,
	\end{equation}
	would prove quite restrictive, in particular in the case of a universe in which the time derivative of the Ricci scalar changes sign, as we will demonstrate in the next subsection. In fact, the only function $f_2$ that would verify Eq. \eqref{eq:entropycreation} with all generality would be $f_2=\text{const.}$, which corresponds to the minimally coupled $f(R)$ limit. Conversely, in minimally coupled gravity $f_2=1$, and the right hand term of Eq. \eqref{eq:entropycreation} vanishes, leaving the second law of thermodynamics unchanged.
	
	The preservation of the second law of thermodynamics would require its generalization to take into account a gravitational entropy contribution. Even though some work has been done in this context for GR \cite{Bonilla1997,Clifton2013,Sussman2014,Acquaviva2018}, it remains a subject of much discussion and debate. However, the need for such a generalized description appears to be much greater in modified gravity models that inherently violate the second law of thermodynamics in its standard form.
	
	\subsection{Entropy in a universe with positive curvature}\label{sec:background}
	
	Here, we shall consider a homogeneous and isotropic universe with positive curvature ($k=1$) filled with dust and radiation. This provides a fair representation of the energy content of the Universe from early post-inflationary times until the onset of dark energy domination. The addition of a positive curvature will allow us to consider expanding and contracting phases of the evolution of the universe and to contrast the behaviour of the comoving entropy of the matter fields in these periods.
	
	The total proper energy density and pressure are given, respectively, by
	\begin{equation}
		\label{eq:densitypressure}
		\rho_{\rm total}=\rho_{\rm r}+\rho_{\rm dust}\,, \qquad \qquad p_{\rm total} =\frac{1}{3}\rho_{\rm r} \,,
	\end{equation}
	with
	\begin{equation}
		\label{eq:realdensities}
		\rho_{\rm r} =\rho_{{\rm r},0} \, a^{-4}f_2^{-1}\,, \qquad \qquad \rho_{\rm dust} =\rho_{\rm dust} \, a^{-3} \,.
	\end{equation}
	On the other hand, Eqs. \eqref{eq:lag-nmc} , \eqref{eq:densitypressure}, and \eqref{eq:realdensities},  imply that the on-shellr Lagrangian of the matter fields is equal to
	\begin{equation}
		\label{eq:lagrangianrpcdm}
		\mathcal{L}_\text{m}=3p_{\rm total}-\rho_{\rm total}=-\rho_{{\rm dust},0} \, a^{-3}\,.
	\end{equation}
	Here, the subscripts `$\rm r$' and `$\rm dust$' again denote the radiation and cold dark matter components, respectively, and the subscript `$0$' refers to an arbitrary initial time $t=0$. 
	
	Using Eqs. \eqref{eq:F}, \eqref{eq:densitypressure}, \eqref{eq:realdensities}, \eqref{eq:lagrangianrpcdm}, along with
	\begin{align}
		\label{eq:deltaF}
		& \Delta_{tt} F = -3HF' \dot{R} - 9H^2f'_2 \rho_{\rm dust} \nonumber\\
		&=-18HF' \left(\ddot{H}+4H\dot{H} -  2Ha^{-2} \right) - 9H^2f'_2 \rho_{\rm dust}\,,\\
		&\Delta_{ii}F=g_{ii}\left(2H\dot{F}+\ddot{F}\right)\,,
	\end{align}
	it is straightforward to rewrite the MFE \eqref{eq:fried-f1f2-1} and the MRE \eqref{eq:fried-f1f2-1} as
	\begin{align}
		\label{eq:friedmann}
		\left(H^2 +a^{-2}\right)F =& \frac{1}{6}\left(f'_1 R-f_1\right)+ \frac{1}{6} \rho_{{\rm r},0}a^{-4}- HF'\dot{R}\nonumber\\
		&+\frac{1}{6}\left[f_2-f'_2 \left(R+18H^2\right)\right]  \rho_{{\rm dust},0}a^{-3} \,,
	\end{align}
	\begin{align}
		\label{eq:ray}
		\left(\dot{H}+H^2\right)F=&-2\left(H^2+a^{-2}\right)F+\frac{1}{2} f_1 \nonumber \\
		&+\frac{1}{6}\rho_{{\rm r},0}a^{-4}f_2^{-1} +2H\dot{F} +\ddot{F} \,,
	\end{align}
	where we have chosen units such that $\kappa= 1$. Since the time derivatives of $F$ are
	\begin{align}
		\label{eq:Fdot}
		\dot{F}&=F'\dot{R} + 3H f'_2 \rho_{\rm dust}\,, \\
		\ddot{F} &= F'\ddot{R}+F''\dot{R}^2 +3\left(2H\dot{R}f''_2+\dot{H}f'_2-3H^2f'_2\right)\rho_{\rm dust}\,,
	\end{align}
	where
	\begin{align}
		\label{eq:R}
		R&=6\left(\dot{H}+2H^2+a^{-2}\right)\,, \\
		\dot{R}&=6\left(\ddot{H}+4H\dot{H}-2Ha^{-2}\right)\,, \\
		\ddot{R}&=6\left[\dddot{H}+4H\ddot{H}+4\dot{H}^2+2\left(2H^2-\dot{H}\right)a^{-2}\right] \,,
	\end{align}
	the MFE and MRE are third and fourth order nonlinear differential equation for the scale factor $a$ with respect to time, respectively.
	
	Here, we consider the functions $f_1 = R$ and $f_2 = \alpha R^\beta$, with constant $\alpha$ and $\beta$ (where $\alpha$ has units of $R^{-\beta}$), so that Eq. (\ref{eq:friedmann}) becomes
	\begin{align}
		\label{eq:friedmann1}
		&\left(H^2 +a^{-2}\right)F = \frac{1}{6} \rho_{{\rm r},0}a^{-4}\nonumber\\
		&+ \frac{\alpha}{6} R^{\beta} \left(1-\beta- \frac{18H^2}{R}\right)  \rho_{{\rm dust},0}a^{-3} \nonumber\\
		& + 6 \alpha \beta (\beta-1) R^{\beta-2}  \rho_{{\rm dust},0}a^{-3} H^2 \left(\frac{\ddot{H}}{H}+4\dot{H}-2a^{-2}\right)\,.
	\end{align}
	Starting from an arbitrary initial time ($t=0$), we integrate Eq. \eqref{eq:ray} using a 5th-order backwards differentiation formula, first backwards up to the Big Bang and then forward up to the Big Crunch. Since it is a fourth order differential equation it requires setting three further initial conditions ($H_0$, $\dot{H}_0$ and $\ddot{H}_0$) in addition to $a_0=1$ (as well as $\rho_{{\rm dust},0}$ and $\rho_{{\rm r},0}$).
	
	Notice that the comoving entropy may change only if $\rho_{\rm dust} \neq 0$ and $\rho_{\rm r} \neq 0$. If the universe was assumed to be filled entirely with cold dark matter, then the proper pressure would vanish and, therefore, so would the right-hand side of Eq.~\eqref{eq:entropy}. Hence, there would be no change to the comoving entropy content of the universe. Conversely, if the universe was composed only of radiation, then  Eq.~\eqref{eq:friedmann1} would reduce to the standard Friedmann equation found in GR. Hence, the Ricci scalar $R$ would vanish and, Eq.~\eqref{eq:entropy} would again imply the conservation of the comoving entropy. 
	
	In the remainder of this section we shall consider cosmologies with $\rho_{{\rm dust},0}=5.94$, $\rho_{{\rm r},0}=0.06$ and $H_0=0$ in the context either of GR or of NMC gravity models with  $\alpha=0.95$ and $\beta=0.01$. In the case of GR ($\alpha=1$, $\beta=0$), these conditions are sufficient to determine the full evolution of the universe. In the context of NMC gravity, Eq.~\eqref{eq:friedmann1} acts as an additional constraint, and with $a_0=1$ and $H_0=0$ becomes
	\begin{equation}
		\label{eq:MFEcond}
		\rho_{{\rm r},0}+\alpha\left[R_0^\beta+\beta(6-R_0)R_0^{\beta-1} \right]\rho_{{\rm dust},0}
		=6 \,,
	\end{equation}
	and therefore sets $\dot{H}_0$ at the initial time, leaving only one additional initial condition, $\ddot{H}_0$ .
	
	Fig.~\ref{fig:asyma} displays the evolution of the scale factor $a$ as a function of the physical time $t$ in the context of two distinct cosmological models computed assuming either GR (dashed blue line) or  NMC gravity (solid orange line). In the context of GR one may observe the exact symmetry between the expanding and contracting phases of the universe, which is verified independently of the initial conditions. The orange solid line shows the evolution of $a$ with $t$ in the context of an NMC gravity model with  $\ddot{H}_0=0.5$. Fig.~\ref{fig:asyma} shows that, in this case, the symmetry between the expanding and contracting phases of the universe is no longer preserved. It also reveals the presence of oscillations on the evolution of the scale factor of variable amplitude and frequency, as well as multiple local maxima of the scale factor (two, for this particular parameter choice). These features are common in the context of NMC gravity and are associated with the increased complexity of the higher-order nonlinear equations which rule the evolution of the universe in that context. Moreover, many NMC models (such as the present one for the chosen parameters) are subject to the Dolgov-Kawasaki instability \cite{Dolgov2003,Faraoni2007,Bertolami2009}. However, this oscillatory behaviour in the cosmological evolution of the universe has also been previously discussed for $f(R)$ models \cite{Appleby2010,Motohashi2010,Motohashi2011,Motohashi2012}, even when they satisfy the former and other stability criteria. A detailed analysis of such oscillations was not a focus of this thesis, as they do not affect this critical result --- that the second law of thermodynamics does not generally hold in the context of NMC gravity. Fig.~\ref{fig:asymH} displays the evolution of the Hubble parameter $H$ as a function of the physical time $t$ for the same models shown in Fig.~\ref{fig:asyma}. Notice the three zeros of $H$, as well as its sharp variations at specific values of the physical time $t$ in the case of NMC gravity. 
	
	\begin{figure}
		\centering
		\includegraphics[width=0.85\textwidth]{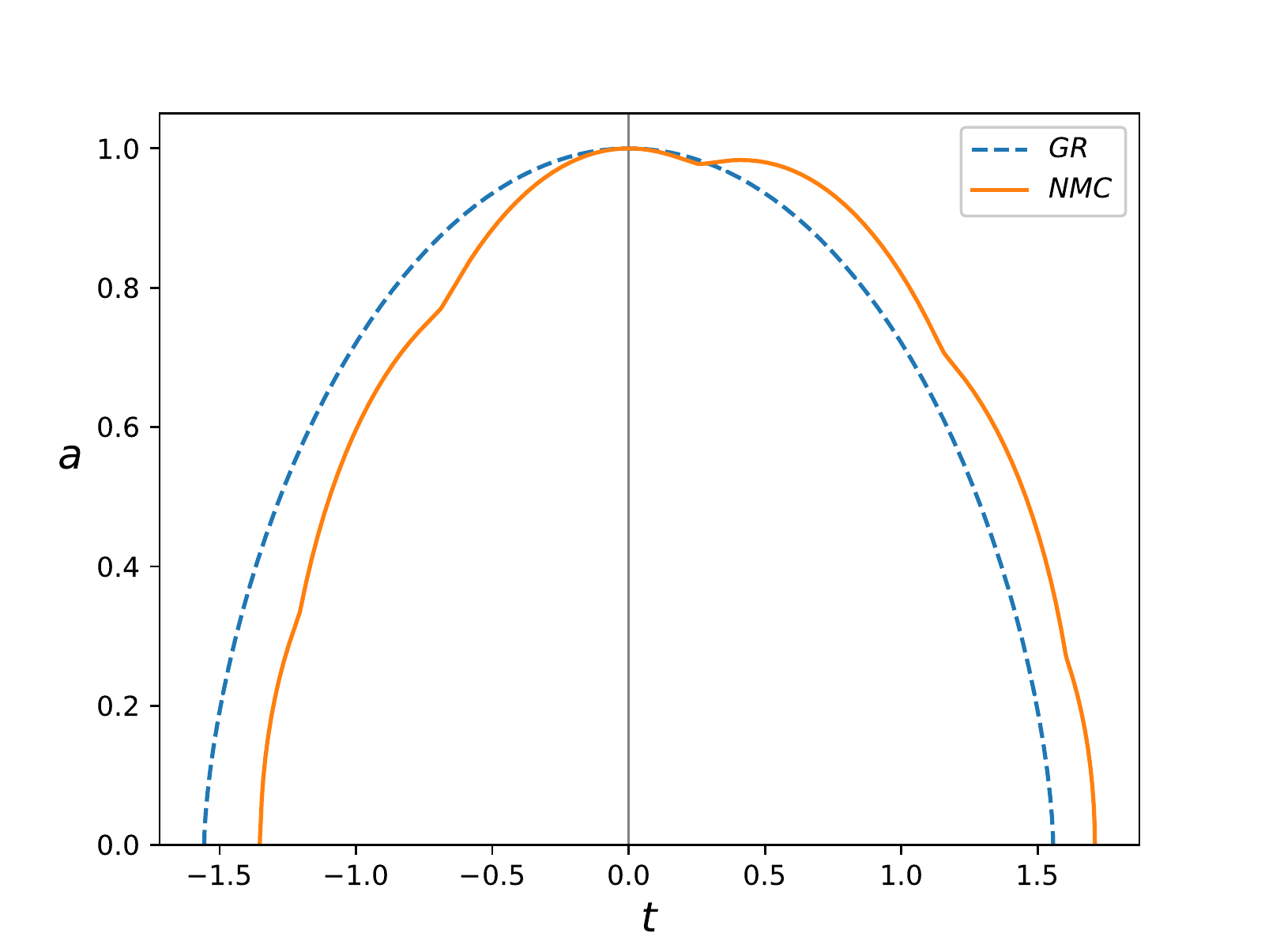}
		\caption[Evolution of the scale factor $a$ as a function of the physical time $t$ in the context of GR and of a NMC model]{Evolution of the scale factor $a$ as a function of the physical time $t$ in the context of GR (dashed blue line) and of a NMC gravity model with $\alpha=0.95$ and $\beta=0.01$ (solid orange line), having $\rho_{{\rm dust},0}=5.94$ and $\rho_{{\rm r},0}=0.06$ and $H_0=0$ as initial conditions. Notice the asymmetric evolution of the universe in the context of NMC gravity (in contrast with GR), and the presence of oscillations of variable amplitude and frequency, as well as two local maxima of $a$.
			\label{fig:asyma}}
	\end{figure}
	\begin{figure}
		\centering
		\includegraphics[width=0.85\textwidth]{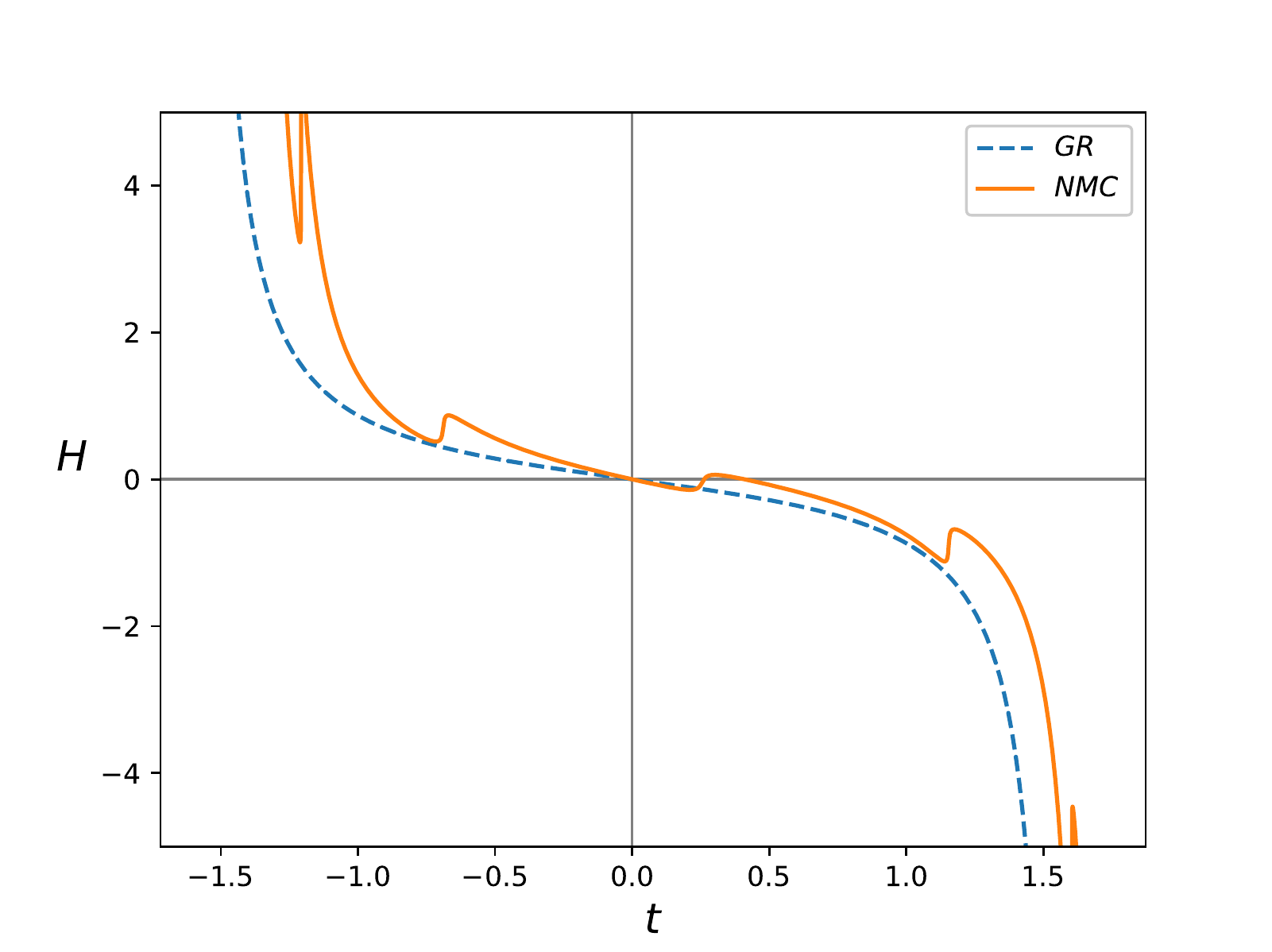}
		\caption[Evolution of the Hubble parameter $H$ as a function of the physical time $t$ in the context of GR and of a NMC model]{Same as in Fig.~\ref{fig:asyma} but for the evolution of the Hubble parameter $H$. Notice the three zeros of $H$, as well as its sharp variation at specific values of the physical time $t$ in the context of NMC gravity. \label{fig:asymH}}
	\end{figure}
	
	Although an asymmetry between the expanding and contracting phases is generic in the context of NMC gravity, one can use the freedom in the choice of initials conditions to impose a symmetric expansion and contraction by choosing $\ddot{H}_0 = 0$.
	
	The results for the symmetric case can be found in Figs.~\ref{fig:syma} through \ref{fig:symS}, which show, respectively, the evolution of the scale factor $a$, the Hubble parameter $H$, the Ricci scalar $R$ and the entropy $S$ as a function of the physical time $t$. The results presented in Figs.~\ref{fig:syma} and \ref{fig:symH} for the evolution of $a$ and $H$ with the physical time, display an exact symmetry between the expanding and contracting phases of the universe, both in the case of GR and NMC gravity. Also, note that in this case there is a single maximum of $a$ (zero of $H$). Otherwise, the results are similar to those shown in Figs.~\ref{fig:asyma} and \ref{fig:asymH} for an asymmetric evolution of the universe. 
	
	\begin{figure}
		\centering
		\includegraphics[width=0.85\textwidth]{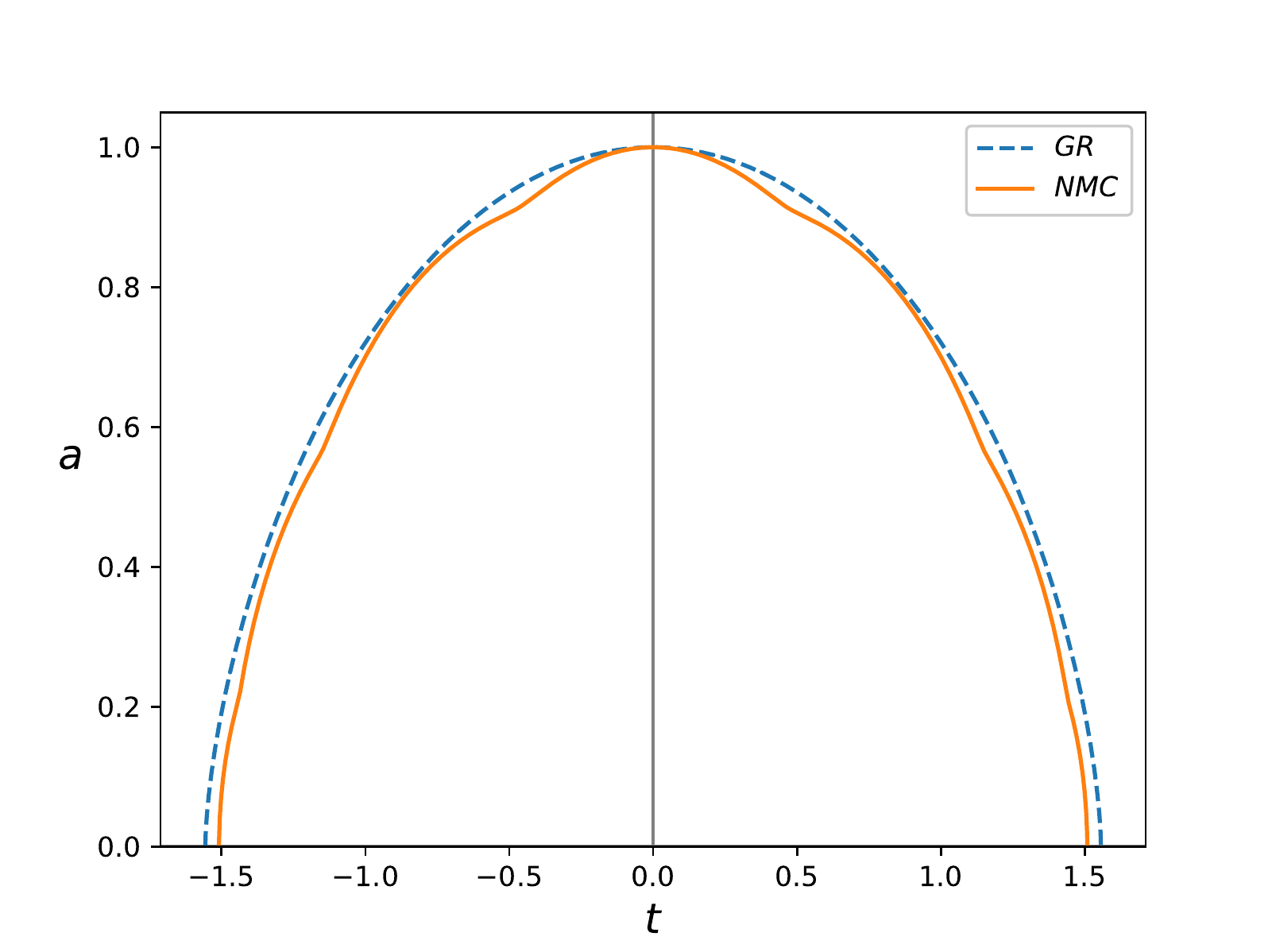}
		\caption[Evolution of the scale factor $a$ as a function of the physical time $t$ in the context of GR and of a NMC model]{Evolution of the scale factor $a$ as a function of the physical time $t$ in the context of GR (dashed blue line) and of a NMC gravity model with $\alpha=0.95$ and $\beta=0.01$ (solid orange line), having $\rho_{{\rm dust},0}=5.94$, $\rho_{{\rm r},0}=0.06$ and $H_0=0$ as initial conditions. An extra initial condition is required in the context of NMC gravity which we take to be ${\ddot H}_0=0$ in order to guarantee a symmetric evolution of the universe. 
			\label{fig:syma}}
	\end{figure}
	\begin{figure}
		\centering
		\includegraphics[width=0.85\textwidth]{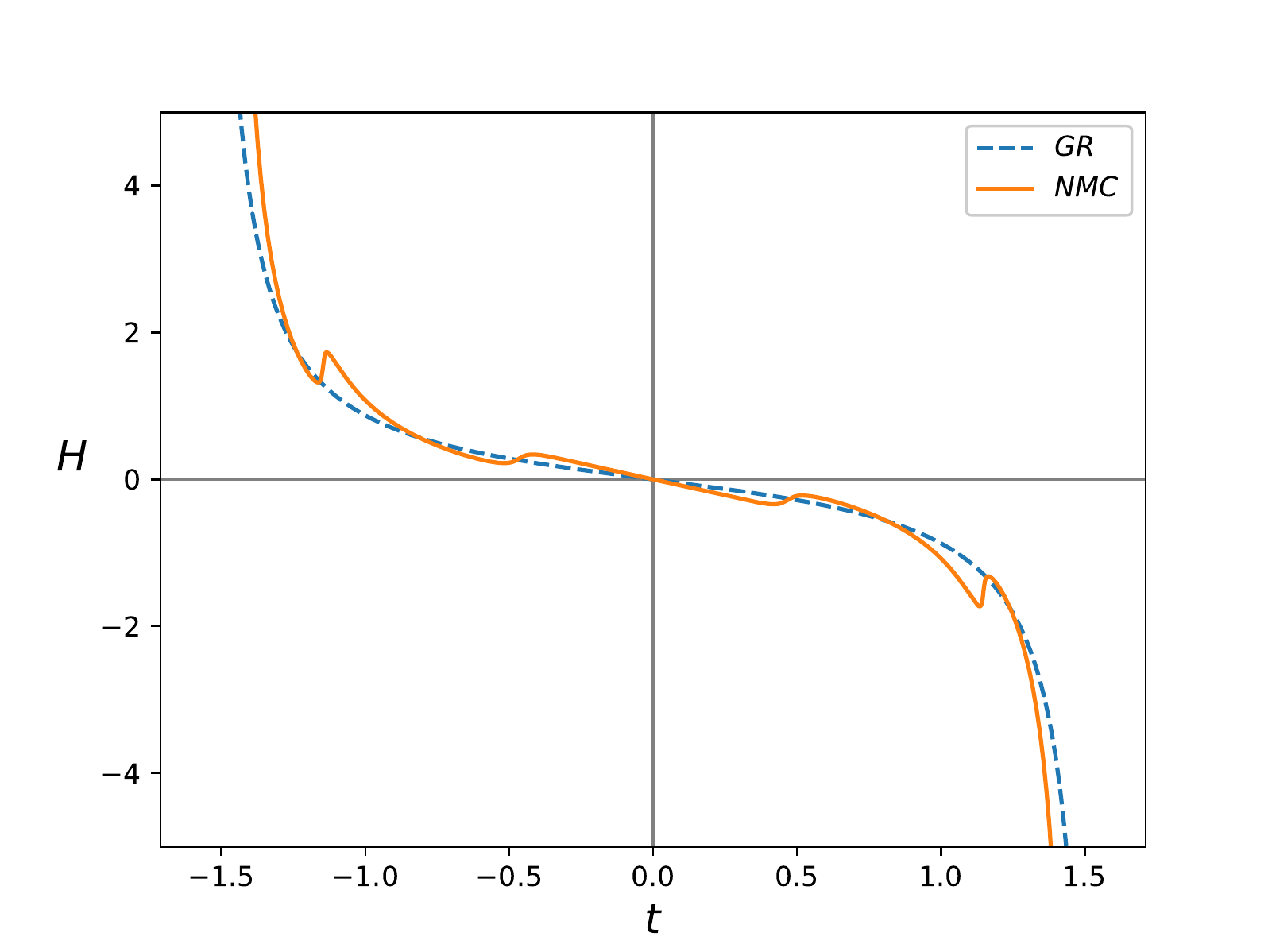}
		\caption[Evolution of the Hubble parameter $H$ as a function of the physical time $t$ in the context of GR and of a NMC model]{Same as in Fig.~\ref{fig:syma} but for the evolution of the Hubble parameter $H$. \label{fig:symH}}
	\end{figure}
	
	Figs.~\ref{fig:symR} and \ref{fig:symS} display the evolution of the Ricci scalar  $R$ and of the comoving entropy $S$ again for the symmetric case. Apart from the oscillations of variable amplitude and frequency, Fig.~\ref{fig:symR} shows that, on average, $R$ decreases during the expanding phase and increases during the contracting phase, while Fig. \ref{fig:symS} show that the comoving entropy has the opposite behaviour. This illustrates the coupling between the evolution of the comoving entropy and the dynamics of the universe, which generally exists in cosmological models with an NMC between gravity and the matter fields, linking the thermodynamic arrow of time to the cosmological evolution of the universe.
	
	\begin{figure}
		\centering
		\includegraphics[width=0.85\textwidth]{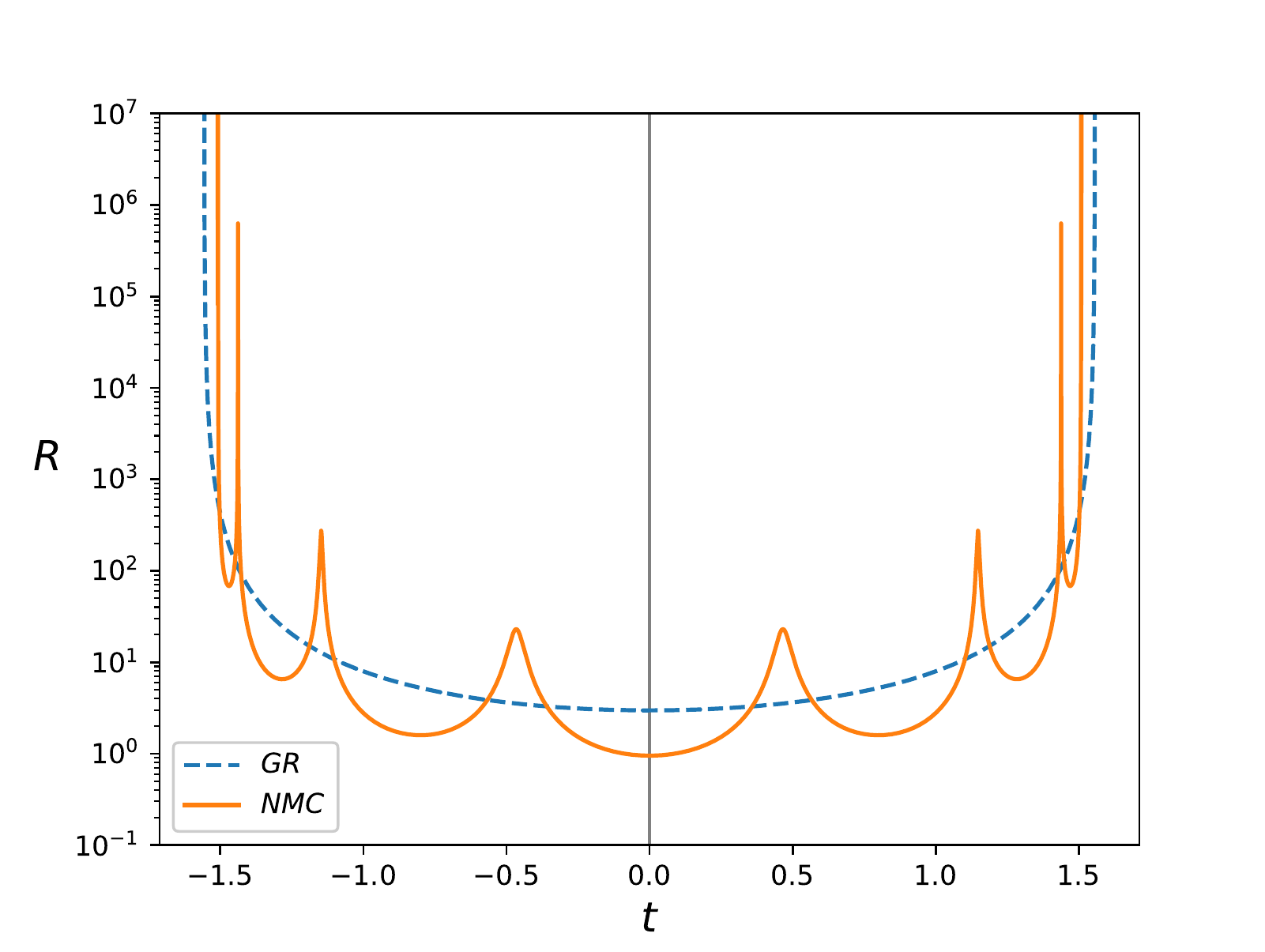}
		\caption[Evolution of the Ricci scalar $R$ as a function of the physical time $t$ in the context of GR and of a NMC model]{Same as in Fig.~\ref{fig:syma} but for the evolution of the Ricci scalar $R$. \label{fig:symR}}
	\end{figure}
	\begin{figure}
		\centering
		\includegraphics[width=0.85\textwidth]{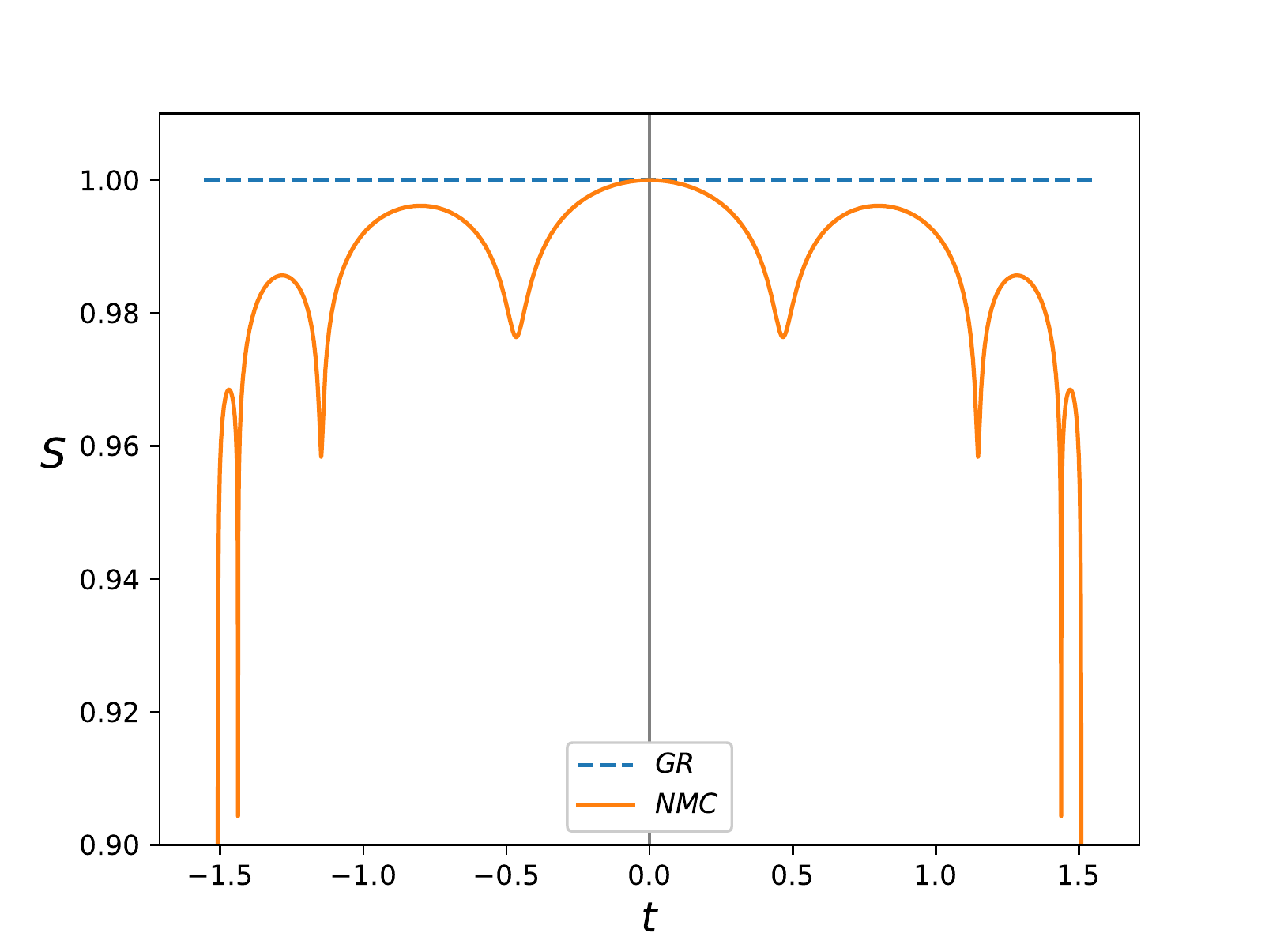}
		\caption[Evolution of the comoving entropy $S$ as a function of the physical time $t$ in the context of GR and of a NMC model]{Same as in Fig.~\ref{fig:syma} but for the evolution of the comoving entropy $S$, normalized so that $S=1$ at $t=0$. \label{fig:symS}}
	\end{figure}
	
	\section{Boltzmann's $\mathcal{H}$-theorem, entropy and the strength of gravity}
	\label{sec.htheo}
	
	In the late nineteenth century, Boltzmann almost single-handedly developed the foundations of modern statistical mechanics. One of his major contributions, Boltzmann's $\mathcal{H}$-theorem, implies that, under generic conditions, the entropy of a closed system is a non-decreasing function of time \cite{Jaynes1965}. However, Boltzmann's $\mathcal{H}$-theorem in its standard form does not necessarily hold in theories with an NMC to gravity, as we will show in this section \cite{Avelino2020}.
	
	\subsection{4-force on point particles}
	Consider the action of a single point particle
	\begin{equation}
		\label{eq:actionpp}
		S=-\int d \tau \, m \,,
	\end{equation}
	with energy momentum tensor
	\begin{equation}
		T^{\mu \nu} = \frac{m}{\sqrt {-g}}\int d \tau \, u^\mu u^\nu \delta^4(x^\sigma-\xi^\sigma(\tau)) \,,
	\end{equation}
	where $\delta^4(x^\sigma-\xi^\sigma(\tau))$ denotes the four–dimensional Dirac delta function, $\xi^\sigma(\tau)$ represents the particle worldline, $\tau$ is the proper time, $u^\mu$ are the components of the particle 4-velocity ($u^\mu u_\mu=-1$) and $m$ is the proper particle mass. If one considers its trace $T$ and integrates over the whole of space-time, we obtain
	\begin{eqnarray}
		\int d^{4}x \sqrt{-g} \, T &=&- \int d^4x \,d\tau\, m\, \delta^4\left(x^\sigma-\xi^\sigma(\tau)\right) \nonumber\\
		&=&- \int d\tau \,m \, ,
	\end{eqnarray}
	which can be immediately identified as the action for a single massive particle, and therefore implies that
	\begin{equation}
		\label{eq:lag}
		{\mathcal L}_\text{m} = T= -  \frac{m}{\sqrt {-g}}\int d \tau \, \delta^4(x^\sigma-\xi^\sigma) \,,
	\end{equation}
	is the particle Lagrangian, as we showed in Chapter \ref{chapter_lag}. The covariant derivative of the EMT may be written as
	\begin{eqnarray}
		\label{eq:en_mom_part_cons}
		\nabla_\nu T^{\mu \nu} &=& \frac{1}{\sqrt {-g}} \partial_\nu \left(\sqrt {-g}  T^{\mu \nu}\right)\nonumber\\
		&=& \frac{m}{\sqrt {-g}}\int d \tau \left(\nabla_\nu u^\mu\right) u^\nu \delta^4(x^\sigma-\xi^\sigma(\tau))\,.
	\end{eqnarray}
	By using \eqref{eq:en_mom_part_cons} and \eqref{eq:lag} in Eq.~\eqref{eq:conservNMC} we obtain
	\begin{equation}
		\frac{m}{\sqrt {-g}}\int d \tau  \delta^4(x^\sigma-\xi^\sigma(\tau)) \nonumber\\
		\times \left(\frac{d u^\mu}{d \tau} +\Gamma^\mu_{\alpha \beta} u^\alpha u^\beta+ \frac{f'_2}{ f_2}  h^{\mu \nu} \nabla_\nu R \right)=0\,,
	\end{equation}
	where $h^{\mu \nu}=g^{\mu \nu}+ u^\mu u^\nu$ is the projection operator. The equation of motion of the point particle is then given by
	\begin{equation}
		\label{eq:nmc_part_acc}
		\mathfrak{a}^\mu=\frac{d u^\mu}{d \tau} +\Gamma^\mu_{\alpha \beta} u^\alpha u^\beta=-
		\frac{f'_2}{ f_2}   h^{\mu \nu} \nabla_\nu R \,,
	\end{equation}
	where
	\begin{equation}
		\label{eq:force}
		\mathfrak{f}^{\nu}=m \mathfrak{a}^\mu=-m \frac{f'_2}{ f_2}   h^{\mu \nu} \nabla_\nu R \,,
	\end{equation}
	is the velocity-dependent 4-force on the particles associated to the NMC to gravity and $\mathfrak{a}^\mu$ is the corresponding 4-acceleration (see \cite{Ayaita2012} for an analogous calculation in the context of growing neutrino models where the neutrino mass is non-minimally coupled to a dark energy scalar field). 
	
	It is important to note that if the particles are part of a fluid, then the 4-acceleration of the individual particles does not, in general, coincide with the 4-acceleration of the fluid element to which they belong, as can be clearly seen by comparing Eqs. \eqref{eq:nmc_fluid_acc} and \eqref{eq:nmc_part_acc} (this point is often overlooked, see \textit{e.g.} \cite{Bertolami2020}). However, in the case of dust $p=0$ and $\mathcal{L}_\text{m}=-\rho$, and the 4-acceleration of a fluid element and of its particles are both given by
	\begin{equation}
		\mathfrak{a}^\mu=-
		\frac{f'_2}{ f_2}   h^{\mu \nu} \nabla_\nu R \,.
	\end{equation}
	
	\subsection{Boltzmann's $\mathcal{H}$-theorem}
	\label{subsec:boltzmann}
	
	The usual collisionless Boltzmann equation given by
	\begin{equation}
		\label{eq:boltzmann1}
		\frac{d \mathcal{F}}{dt}= \frac{\partial  \mathcal{F}}{\partial t}+\nabla_{\vb r} \mathcal{F} \cdot \frac{d\vb r}{dt} + \nabla_{\vb p} \mathcal{F} \cdot {\vb F}=0 \,,
	\end{equation}
	expresses the constancy in time of a six-dimensional phase space volume element ${\mathcal V}_6$ containing a fixed set of particles in the absence of particle collisions. Here, $t$ is the physical time, the six-dimensional phase space is composed of the six positions and momentum coordinates $({\vb r},{\vb p})$ of the particles, ${\vb F}=d{\vb p}/dt$ is the 3-force on the particles (assumed to be independent of ${\vb p}$), and $\mathcal{F}(t,{\vb r},{\vb p}) {\mathcal V}_6 $ is the number of particles in the six-dimensional infinitesimal phase space volume element ${\mathcal V}_6 = d^3 r \, d^3 p$. However, in the presence of NMC to gravity ${\vb F}$ may depend on $\vb p$, and this volume is in general not conserved. In this case, phase-space continuity, expressing particle number conservation in six-dimensional phase space in the absence of collisions,
	\begin{equation}
		\label{eq:phase-space-cont}
		\frac{\partial \mathcal{F}}{\partial t} + \nabla_{\vb r}\cdot\left(\mathcal{F}\frac{d \vb{r}}{dt}\right)
		+\nabla_{\vb p}\cdot\left(\mathcal{F}\vb{F}\right)=0 \,,
	\end{equation}
	should be used rather than Eq. \eqref{eq:boltzmann1}. Here, $\vb{r}$ and $\vb{p}$ are independent variables, thus implying $\nabla_{\vb r}\cdot \vb{p}=0$. Note that no assumption has been made regarding the relativistic or non-relativistic nature of the particles (Eq. \eqref{eq:phase-space-cont} is valid in both regimes).
	
	In a flat homogeneous and isotropic universe, described by the Friedmann-Lemaître-Robertson-Walker metric, the line element is given by
	\begin{equation}
		ds^2=-dt^2+d{\vb r} \cdot d {\vb r}= -dt^2 + a^2(t) d{\vb q} \cdot d{\vb q}\,,
	\end{equation}
	where $a(t)$ is the scale factor and ${\vb q}$ are comoving Cartesian coordinates. In this case, the Ricci scalar is a function of cosmic time alone [$R=R(t)$] and the $i0$ components of the projection operator may be written as  $h^{i0}=\gamma^2 v^i$, where $\gamma=u^0=dt/d\tau$ and $v^i=u^i/\gamma$ are the components of the 3-velocity. Therefore, Eq.~\eqref{eq:force} implies that the 3-force on the particles is given by
	\begin{eqnarray}
		\label{eq:3-force}
		F^i=\frac{d {p}^i}{dt}&=& \frac{\mathfrak{f}^{i}}{\gamma} -\frac{d\ln a}{dt}p^i= -\left(\frac{d\ln a}{dt}+\frac{f'_2}{ f_2}   \frac{d R} {dt} \right)p^i \nonumber \\
		&=&- \left(\frac{d \ln a}{dt}+ \frac{d \ln f_2}{dt}   \right) p^i  \nonumber\\
		&=&  -\frac{d \ln \left(a f_2 \right)}{dt} p^i \,,
	\end{eqnarray}
	This in turn implies that $p^i \propto (f_2 a)^{-1}$, so that 
	\begin{equation}
		\label{V6}
		{\mathcal V}_6 = d^3 r \, d^3 p \propto {f_2}^{-3}\,,
	\end{equation}
	where we have taken into account that $d^3 r= a^3 d^3 q$. Eq. \eqref{V6} explicitly shows that in the presence of a NMC to gravity the phase-space volume is, in general, no longer incompressible.
	
	In a homogeneous and isotropic universe
	\begin{align}
		\label{eq:drdt}
		\frac{d\vb{r}}{dt} &= \frac{da}{dt}\vb{q}+a\frac{d\vb{q}}{dt} = \frac{d \ln a}{dt}\vb{r} + \vb{v} \nonumber\\
		&= \frac{d \ln a}{dt} \vb{r} + \frac{\vb{p}}{\left(m^2+p^2\right)^{1/2}} \,,
	\end{align}
	where $m$ is the rest mass of the particles, thus implying that 
	\begin{equation}
		\label{eq:nabla-r}
		\nabla_{\vb r}\cdot\left(\frac{d\vb{r}}{dt}\right)=3\frac{d \ln a}{dt}\,.
	\end{equation}
	
	Substituting Eqs. \eqref{eq:3-force} and \eqref{eq:nabla-r} into the phase-space continuity equation — note that Eq. \eqref{eq:phase-space-cont} remains valid in a FLRW background — and taking into account that in a homogeneous universe $\mathcal{F}$ is independent of $\vb{r}$ [$\mathcal{F}=\mathcal{F}(t,\vb{p})$], one obtains
	\begin{eqnarray}
		\label{eq:boltzmann2}
		0 &=& \frac{\partial  \mathcal{F}}{\partial t} + \mathcal{F}\nabla_{\vb r}\cdot\left(\frac{d\vb{r}}{dt}\right) + \vb{F}\cdot\nabla_{\vb p}\mathcal{F} +  \mathcal{F} \nabla_{\vb p} \cdot {\vb F}\nonumber \\ 
		&=&\frac{\partial  \mathcal{F}}{\partial t} - \frac{\partial  \mathcal{F}}{\partial p^i}\frac{d \ln \left(a f_2\right)}{dt} p^i -3 \mathcal{F} \frac{d \ln f_2}{dt}  \,.
	\end{eqnarray}
	Note that Eq. \eqref{eq:boltzmann2} does not include collision terms and, therefore, it only applies in the case of collisionless fluids. For example, after neutrino decoupling non-gravitational neutrino interactions may, in general, be neglected and, consequently, Eq. \eqref{eq:boltzmann2} may be used to determine the evolution of the neutrino phase-space distribution function for as long as the Universe remains approximately homogeneous and isotropic (the same applying to photons after recombination, although to a lesser extent). We shall defer to the following subsection a discussion of the impact of collisions in situations where they might be relevant.
	
	Let us start by explicitly verifying the conservation of the number of particles $N$ inside a constant comoving spatial volume $V_q$ defined by $\int d^3 r =a^3 \int d^3 q =a^3 V_q$. Since $N=\int d^3 r \, d^3 p \, \mathcal{F}=a^3 V_q \int \mathcal{F} d^3 p$,
	\begin{align}
		\label{eq:dotN}
		\frac{d N}{dt}&= 3 \frac{d\ln a}{dt} N+a^3 V_q \int  d^3 p \frac{\partial \mathcal{F}}{\partial t}  \nonumber\\
		&= 3 \frac{d\ln (a f_2)}{dt} N + a^3 V_q \int  d^3 p \frac{\partial  \mathcal{F}}{\partial p^i}\frac{d \ln \left(a f_2\right)}{dt} p^i  = 0\,.
	\end{align}
	Here, we have used Eq. ~\eqref{eq:boltzmann2} in order to evaluate $\partial \mathcal{F}/ \partial t$ and performed the momentum integral by parts.
	
	Let us now consider Boltzmann's $\mathcal{H}$ defined by
	\begin{equation}
		\mathcal{H}= \int d^3r \, d^3 p \mathcal{F} \ln \mathcal{F} = a^3 V_q \int d^3 p \mathcal{F} \ln \mathcal{F}  \,,
	\end{equation}
	Taking the derivative of $\mathcal{H}$ with respect to the physical time and using ~\eqref{eq:dotN} one obtains
	\begin{eqnarray}
		\label{eq:dotH}
		\frac{d \mathcal{H}}{dt}&=& 3 \frac{d\ln a}{dt} \mathcal{H}+a^3 V_q \int  \, d^3 p  (1+\ln \mathcal{F})  \frac{\partial \mathcal{F}}{\partial t} \nonumber\\
		&=& 3 \frac{d\ln a}{dt} (\mathcal{H}-N)+ a^3 V_q \int  d^3 p \frac{\partial \mathcal{F}}{\partial t}  \ln \mathcal{F} \,,
	\end{eqnarray}
	where again $\int d^3 r =a^3 \int d^3 q =a^3 V_q$ and $N$ is the number of particles inside $V_q$. Using Eq.~\eqref{eq:boltzmann2}, the integral which appears in the last term of Eq.~\eqref{eq:dotH} may be written as
	\begin{eqnarray}
		I  &=& \int  d^3 p  \frac{\partial \mathcal{F}}{\partial t} \ln \mathcal{F}=3 \frac{d \ln f_2}{dt} \int d^3 p\mathcal{F} \ln \mathcal{F}   \nonumber\\
		&+& \int  d^3 p  \left(\frac{\partial  \mathcal{F}}{\partial p^i}\frac{d \ln \left(af_2\right)}{dt}p^i\right) \ln \mathcal{F} \nonumber \\
		&=& I_1 + I_2 \,,
	\end{eqnarray}
	where
	\begin{eqnarray}
		I_1&=& 3 (a^3 V_q)^{-1} \frac{d \ln f_2}{dt} \mathcal{H}\\
		I_2&=& \int  d^3 p  \left(\frac{\partial  \mathcal{F}}{\partial p^i}\frac{d \ln \left(af_2\right)}{dt}p^i\right) \ln \mathcal{F} \,.
	\end{eqnarray}
	Integrating $I_2$ by parts one obtains
	\begin{eqnarray}
		\label{eq:integralI}
		I_2 &=& - \int  d^3 p  \mathcal{F}\frac{\partial }{\partial p^i} \left[\ln \mathcal{F}\frac{d \ln \left(a f_2\right)}{dt}p^i\right]\nonumber\\
		&=&-3 (a^3 V_q)^{-1} \frac{d \ln \left(a f_2\right)}{dt} \mathcal{H} - \int  d^3 p  \frac{\partial \mathcal{F}}{\partial p^i} \frac{d \ln \left(a f_2\right)}{dt}p^i \nonumber \\
		&=& 3 (a^3 V_q)^{-1} \frac{d \ln \left(a f_2\right)}{dt} (N-\mathcal{H}) \,.
	\end{eqnarray}
	Summing the various contributions, Eq.~\eqref{eq:dotH} finally becomes 
	\begin{equation}
		\label{eq:dotH1}
		\frac{d \mathcal{H}}{dt} = 3 \frac{d\ln f_2}{dt} N \,.
	\end{equation}
	
	In general relativity $f_2$ is equal to unity and, therefore, Boltzmann's $\mathcal{H}$ is a constant in the absence of particle collisions. However, Eq. ~\eqref{eq:dotH1} implies that this is no longer true in the context of NMC theories of gravity. In this case, the evolution of Boltzmann's $\mathcal{H}$ is directly coupled to the evolution of the universe. Boltzmann's $\mathcal{H}$ may either grow or decay, depending on whether $f_2$ is a growing or a decaying function of time, respectively. This provides an explicit demonstration that Boltzmann's $\mathcal{H}$ theorem --- which states that $d\mathcal{H}/dt \le 0$ --- may not hold in the context of NMC theories of gravity.
	
	\subsubsection*{An alternative derivation}
	
	Consider two instants of time $t_\text{A}$ and $t_\text{B}$, with $a_\text{A}=1$. According to Eq. \eqref{eq:3-force}, in the absence of collisions, $\vb{p}\propto(af_2)^{-1}$. Therefore, assuming that the number of particles is conserved, Eq. \eqref{V6} implies that
	\begin{equation}
		\frac{\mathcal{F}_\text{B}}{\mathcal{F}_\text{A}}\equiv\frac{\mathcal{F}(t_\text{B},f_{2,\text{A}}\vb{p}/(a_\text{B}f_{2,\text{B}}))}{\mathcal{F}(t_\text{A},\vb{p})} =\frac{{\mathcal V}_{6,\text{A}}}{{\mathcal V}_{6,\text{B}}}=\left(\frac{f_{2,\text{B}}}{f_{2,\text{A}}}\right)^3 \,.
	\end{equation}
	Hence,
	\begin{align}
		\label{eq:alt-H}
		\mathcal{H}_\text{B}&=a^3_\text{B}V_q\int d^3p_\text{B}\mathcal{F}_\text{B}\ln \mathcal{F}_\text{B} \nonumber\\
		&= a^3_\text{B}V_q\int d^3p_\text{A} \left(\frac{f_{2,\text{A}}}{a_\text{B}f_{2,\text{B}}}\right)^3  \mathcal{F}_\text{B}\ln \mathcal{F}_\text{B} \nonumber \\
		&= V_q\int d^3p_\text{A}  \mathcal{F}_\text{A} \ln\left[\mathcal{F}_\text{A}\left(\frac{f_{2,\text{B}}}{f_{2,\text{A}}}\right)^3 \right]\nonumber \\
		&= V_q\int d^3p_\text{A}  \mathcal{F}_\text{A} \ln\mathcal{F}_\text{A} + 3\ln\left(\frac{f_{2,\text{B}}}{f_{2,\text{A}}}\right)V_q\int d^3p_\text{A}  \mathcal{F}_\text{A} \nonumber \\
		&=\mathcal{H}_\text{A}+3\left(\ln f_{2,\text{B}}-\ln f_{2,\text{A}} \right)N \,.
	\end{align}
	If $t_\text{A}$ and $t_\text{B}$ are sufficiently close, one can write $t_\text{B}-t_\text{A}=dt$, $\mathcal{H}_\text{B}-\mathcal{H}_\text{A}=d\mathcal{H}$, and $\ln f_{2,\text{B}}-\ln f_{2,\text{A}}=d\ln f_2$. Then, dividing Eq. \eqref{eq:alt-H} by $dt$ one obtains Eq. \eqref{eq:dotH1}. This alternative derivation shows, perhaps even more explicitly, how the growth or decay of the magnitude of the linear momentum of the particles associated to the NMC to gravity may contribute, respectively, to a decrease or an increase of Boltzmann's $\mathcal{H}$.
	
	\subsection{Entropy}
	\label{subsec:entropy}
	
	Consider a fluid of $N$ point particles with Gibbs' and Boltzmann's entropies given respectively by
	\begin{eqnarray}
		S_{G}&=&-\int P_{N} \ln P_{N} d^{3} r_{1} d^{3} p_{1} \cdots d^{3} r_{N} d^{3} p_{N}\, \\
		S_{B}&=&-N \int P \ln P d^{3} r d^{3} p\,.
	\end{eqnarray}
	where $P_{N}\left({\vb r}_1, {\vb p}_1, \ldots, {\vb r}_N, {\vb p}_N,t\right)$ and $P\left({\vb r}, {\vb p},t\right)$ are, respectively, the $N$-particle probability density function in $6N$-dimensional phase space and the single particle probability in 6-dimensional phase space. $P$ and $P_N$ are related by
	\begin{equation}
		P\left({\vb r}, {\vb p},t\right)=\int P_N d^{3} r_{2} d^{3} p_{2} \cdots d^{3} r_{N} d^{3} p_{N}
	\end{equation}
	These two definitions of the entropy have been shown to coincide only if
	\begin{equation}
		P_{N}\left({\vb r}_1, {\vb p}_1, \ldots, {\vb r}_N, {\vb p}_N,t\right)= \prod_{i=1}^N P\left({\vb r}_i, {\vb p}_i,t\right)\,,
	\end{equation}
	or, equivalently, if particle correlations can be neglected, as happens for an ideal gas \cite{Jaynes1965}. In the remainder of this section we shall assume that this is the case, so that $S=S_B=S_G$ (otherwise $S_G<S_B$ \cite{Jaynes1965}). We shall also consider a fixed comoving volume $V_q$. 
	
	Close to equilibrium $\mathcal{F}\left({\vb r}, {\vb p},t\right)=NP\left({\vb r}, {\vb p},t\right)$  holds to an excellent approximation and, therefore
	\begin{equation}
		\mathcal{H}=-S + N \ln N\,.
	\end{equation}
	Again, assuming that the particle number $N$ is fixed, Eq. ~\eqref{eq:dotH1} implies that
	\begin{equation}
		\label{eq:dotS}
		\frac{d S}{dt}=-\frac{d \mathcal{H}}{dt}  = - 3 \frac{d \ln f_2}{dt} N\,.
	\end{equation}
	Hence, the entropy $S$ in a homogeneous and isotropic universe may decrease with cosmic time, as long as ${f_2}$ grows with time. This once again shows that the second law of thermodynamics does not generally hold in the context of modified theories of gravity with an NMC between the gravitational and the matter fields.
	
	\subsubsection*{The collision term}
	
	Under the assumption of molecular chaos, \textit{i.e.} that the velocities of colliding particles are uncorrelated and independent of position, adding a two-particle elastic scattering term to Eq. \eqref{eq:phase-space-cont} results in a non-negative contribution to the entropy increase with cosmic time ---  this contribution vanishes for systems in thermodynamic equilibrium. This result holds independently of the NMC coupling to gravity, as acknowledged in \cite{Bertolami2020} where the standard calculation of the impact of the collision term has been performed without taking into account the momentum-dependent forces on the particles due to the NMC to gravity. However, as demonstrated in this section, these momentum-dependent forces may be associated with a further decrease of the magnitude of the linear momentum of the particles (if $f_2$ grows with time) contributing to the growth of Boltzmann’s $\mathcal{H}$ (or, equivalently, to a decrease of the entropy). The existence of particle collisions, although extremely relevant in most cases, does not change this conclusion.
	
	If the particles are non-relativistic, and assuming thermodynamic equilibrium, $\mathcal{F}(\vb p,t)$ follows a Maxwell-Boltzmann distribution. In an FLRW homogeneous and isotropic universe with an NMC to gravity the non-relativistic equilibrium distribution is maintained even if particle collisions are switched off at some later time, since the velocity of the individual particles would simply evolve as ${\vb v} \propto (a f_2)^{-1}$ in the absence of collisions (see Eq.~\eqref{eq:3-force}) --- the temperature, in the case of non-relativistic particles, would evolve as $\mathcal{T}\propto v^2 \propto (af_2)^{-2}$.
	
	If the fluid is an ideal gas of relativistic particles (with $p =\rho/3$), each satisfying equation of motion for a point particle derived in Eq. \eqref{eq:nmc_part_acc} (except, eventually, at quasi-instantaneous scattering events), then its on-shell Lagrangian vanishes ($\mathcal{L}_\text{m,[fluid]}=T=-\rho+3p =0$), and we recover the results found in Section \ref{sec.seclaw}, namely that the entropy density $s$ evolves as $n(\mathcal{T}) \propto s(\mathcal{T}) \propto \mathcal{T}^3\propto a^{-3}f_2^{-3/4}$. This implies that both the number of particles $N$ and the entropy $S$ in a fixed comoving volume are not conserved — they evolve as $N\propto S\propto na^3\propto f_2^{-3/4}$.
	
	Unless $f_2$ is a constant, the equilibrium distribution of the photons cannot be maintained after the Universe becomes transparent at a redshift $z\sim10^3$, given that the number of photons of the cosmic background radiation is essentially conserved after that. Hence, direct identification of Boltzmann’s $\mathcal{H}$ with the entropy should not be made in this case. The requirement that the resulting spectral distortions be compatible with observations has been used to put stringent limits on the evolution of $f_2$ after recombination \cite{Avelino2018}, and we will show in the next chapter.
	
	\subsubsection*{The strength of gravity}
	\label{subsubsec:gstrength}
	
	Existing cosmic microwave background and primordial nucleosynthesis constraints restrict the NMC theory of gravity studied in the present work (or its most obvious generalization) to be very close to General Relativity ($f_2=1$) at late times \cite{Avelino2018,Azevedo2018a}. Before big bang nucleosynthesis, the dynamics of $f_2$ is much less constrained on observational grounds, but it is reasonable to expect that the cosmological principle and the existence of stable particles — assumed throughout this work — would still hold (at least after primordial inflation). This requires the avoidance of pathological instabilities, such as the Dolgov-Kawasaki instability, \textit{i.e.} $\kappa f''_1+f''_2\mathcal{L}_\text{m}\geq0$.
	
	Consider a scenario, free from pathological instabilities, in which the function $f_2$ was much larger at early times than at late times (here, early and late refer to times much before and after primordial nucleosynthesis, respectively). In this scenario, the present value of Newton’s gravitational constant is the result of a dynamical process associated with the decrease of $f_2$, perhaps by many orders of magnitude, from early to late times. More importantly, the high entropy of the Universe and the weakness of gravity would be interrelated in this scenario.
 

\chapter{Constraints on Nonminimally Coupled $f(R)$ Gravity} 
\label{chapter_constr} 

We have shown in Chapter \ref{chapter_nmc} that theories featuring a nonminimal coupling (NMC) between gravity and the matter fields can feature a non-conservation of energy-momentum, which has significant thermodynamic consequences. This non-conservation, along with changes to the equations of motion, also leads to different predictions for both cosmological phenomena and solar system dynamics, and one can use observational data to constrain specific NMC models.

In this chapter we will provide an overview of these constraints, summarising existing local constraints in the literature (Section \ref{sec.solar-const}), and detailing the new cosmological constraints obtained in this work from observations, namely the cosmic microwave background (CMB) (Section \ref{sec.cmb}) \cite{Avelino2018}, big-bang nucleosynthesis (BBN) (Section \ref{sec.bbn}) \cite{Azevedo2018a}, type Ia supernovae (SnIa) and baryon acoustic oscillations (BAO) (Section \ref{sec.ddr}) \cite{Azevedo2021} observations.

Throughout this chapter we will consider the action
\begin{equation}\label{eq:actionf1f2_cons}
	S = \int d^4x \sqrt{-g} \left[\kappa f_1(R)+f_2(R)\mathcal{L}_\text{m} \right] \,,
\end{equation}
where $f_1(R)$ and $f_2(R)$ are arbitrary functions of the Ricci scalar $R$, $\mathcal{L}_\text{m}$ is the Lagrangian of the matter fields and $\kappa = c^4/(16\pi G)$. We will also consider a flat FLRW metric with line element
\begin{equation}
	\label{eq:line-nmc-flat}
	ds^2=-dt^2+a^2(t)\left[dr^2 +r^2 d\theta^2 +r^2\sin^2\theta d\phi^2\right]\,,
\end{equation}
where $a(t)$ is the scale factor (we set $a(t)=1$ at the present time), $t$ is the cosmic time, and $r$, $\theta$ and $\phi$ are polar comoving coordinates, filled by a collection of perfect fluids, with energy-momentum tensor (EMT) of the form
\begin{equation}\label{eq:pf_emt_nmc-cons}
	T^{\mu\nu}=(\rho+p)U^\mu U^\nu + p g^{\mu\nu}\,,
\end{equation}
where $\rho$ and $p$ are respectively the proper density and pressure of the fluid, and $U^\mu$ is the 4-velocity of a fluid element, satisfying $U_\mu U^\mu = -1$. Alternatively, we can also write the line element for this metric as
\begin{equation}
	\label{eq:line-nmc-flat-cart}
	ds^2=-dt^2+a^2(t)d\vb{q}\cdot d\vb{q}\,,
\end{equation}
where $\vb{q}$ are the comoving Cartesian coordinates.

The modified field equations can be obtained from the action \eqref{eq:actionf1f2_cons}
\begin{equation}\label{eq:fieldNMC_cons}
	FG_{\mu\nu}=\frac{1}{2}f_2 T_{\mu\nu} + \Delta_{\mu\nu}F+\frac{1}{2}g_{\mu\nu}\kappa f_1 - \frac{1 }{2}g_{\mu\nu} RF \,,
\end{equation}
where
\begin{equation}
	\label{eq:F_cons}
	F=\kappa f'_1+f'_2\mathcal{L}_\text{m}\,,
\end{equation}
the primes denote a differentiation with respect to the Ricci scalar, $G_{\mu\nu}$ is the Einstein tensor, $\Delta_{\mu\nu}\equiv\nabla_\mu \nabla_\nu-g_{\mu\nu}\square$, with $\square=\nabla_\mu \nabla^\mu$ being the D'Alembertian operator, and $\mathcal{L}_\text{m}$ is the on-shell Lagrangian of the matter fields, which in the case of a perfect fluid composed of point particles is given by the trace of the EMT
\begin{equation}
	\label{eq:Lag_cons}
	\mathcal{L}_\text{m}= T^{\mu\nu}g_{\mu\nu} = T= 3p-\rho \,.
\end{equation}
The modified Friedmann equation (MFE) is
\begin{equation}\label{eq:fried-f1f2-1-cons}
	H^2=\frac{1}{6F}\left[FR- \kappa f_1+f_2\rho-6H\dot{F}\right]\,,
\end{equation}
and the modified Raychaudhuri equation (MRE) is
\begin{equation}\label{eq:ray-f1f2-1-cons}
	2\dot{H}+3H^2=\frac{1}{2F}\left[FR-\kappa f_1-f_2 p-4H\dot{F}-2\ddot{F}\right] \,.
\end{equation}
\section{Solar system constraints}
\label{sec.solar-const}

Much like in the context of $f(R)$ theories in Chapter \ref{chapter_modgrav}, one can also perform a weak-field expansion around a spherical body to derive constraints on NMC gravity \cite{Bertolami2013,Castel-Branco2014,March2017,March2019}. The process is largely the same, and the same conditions apply, extended to the NMC function:

\textit{Condition 1:} $f_1(R)$ and $f_2(R)$ are analytical at the background curvature $R=R_0$.

\textit{Condition 2:} The pressure of the local star-like object is approximately null, $p\simeq 0$. This implies that the trace of the energy-momentum tensor is simply $T\simeq -\rho$.

Likewise, the weak-field expansion feature Yukawa terms, which can be avoided if one adds a third condition:

\textit{Condition 3:} $|m|r\ll1$, where $m$ is the effective mass of the scalar degree of freedom of the theory (defined in Eq. \eqref{eq:NMC-mass-param}) and $r$ is the distance to the local star-like object.

The use of the appropriate Lagrangian is critical to the derivation of constraints. In \cite{Bertolami2013,Castel-Branco2014,March2017,March2019} the Lagrangian used for the perfect fluid is $\mathcal{L}_\text{m}=-\rho$, rather than $\mathcal{L}_\text{m}=T$, as we have previously determined. However, the following analysis considers only dust contributions on both the cosmological and local levels, so the general perfect fluid Lagrangian does indeed reduce to $\mathcal{L}_\text{m}=T=-\rho$. Under this assumption, the results presented in the literature for local constraints on NMC gravity therefore should remain valid in light of this thesis.

\subsection[Post-Newtonian expansion]{Post-Newtonian expansion}
Once again one assumes that the scalar curvature can be expressed as the sum of two components
\begin{equation}
	R(r,t)\equiv R_0(t)+R_1(r) \,,
\end{equation}
where $R_0(t)$ is the background spatially homogeneous scalar curvature, and $R_1(r)$ is a time-independent perturbation to the background curvature. Since the timescales usually considered in Solar System dynamics are much shorter than cosmological ones, we can usually take the background curvature to be constant, i.e. $R_0= \text{const.}$. We can therefore separate the source for the Ricci scalar into two different components, one cosmological and another local. So the trace of the field equations Eq. \eqref{eq:fieldNMC_cons} reads
\begin{align}\label{eq:NMC-trace-dec}
	&\left[\kappa f'_1+f'_2\left(\mathcal{L}_\text{m}^\text{cos}+\mathcal{L}_\text{m}^\text{s}\right)\right]R-\kappa f_1  \nonumber\\
	&\qquad\qquad+3\square\left[\kappa f'_1+f'_2\left(\mathcal{L}_\text{m}^\text{cos}+\mathcal{L}_\text{m}^\text{s}\right)\right]=\frac{1}{2}f_2\left(T^\text{cos}+T^\text{s}\right)\,,
\end{align}
where $\mathcal{L}_\text{m}^\text{cos}=T^\text{cos}=-\rho^\text{cos}$ and $\mathcal{L}_\text{m}^\text{s}=T^\text{s}=-\rho^\text{s}$ are the cosmological and local matter contributions, respectively. If one takes into account that $R_1\ll R_0$ and that $R_0$ solves the terms of the expansion of Eq. \eqref{eq:NMC-trace-dec} that are independent of $R_1$, one can write it as
\begin{align}
	\label{eq:NMC-trace-exp}
	&6\nabla^2\left[\left(\kappa f''_{1,0}+f''_{2,0}\mathcal{L}_\text{m}\right)R_1\right] + \left(-2\kappa f'_{1,0}+f'_{2,0}\mathcal{L}_\text{m}\right)R_1 \nonumber \\
	&+2\left(\kappa f''_{1,0}+f''_{2,0}\mathcal{L}_\text{m}\right)R_0 R_1+6\left[\square\left(\kappa f''_{1,0}-f''_{2,0}\rho^\text{cos}\right)-\rho_\text{s}\square f''_{2,0}\right]R_1 \nonumber \\
	&\qquad=-(1+f_{2,0})\rho_\text{s}+2f'_{2,0}R_0\rho_\text{s}+6\rho_\text{s}\square f'_{2,0}+6f'_{2,0}\nabla^2\rho^\text{s} \,,
\end{align}
where $f_{i,0}\equiv f_i(R_0)$. From here one can define a potential
\begin{equation}
	\label{eq:NMC-potential-ppn}
	U(r)=2\left[\kappa f''_{1,0}+f''_{2,0}\mathcal{L}_\text{m}(r)\right]R_1(r) \,,
\end{equation}
and a mass parameter $m$ as
\begin{equation}
	\label{eq:NMC-mass-param}
	m^2 \equiv \frac{1}{3} \left[\frac{2 \kappa f'_{1,0}-f'_{2,0}\mathcal{L}_\text{m}}{2\left(\kappa f''_{1,0}+f''_{2,0}\mathcal{L}_\text{m}\right)}-R_0-\frac{3\square\left(\kappa f''_{1,0}-f''_{2,0}\rho^\text{cos}\right)-3\rho^\text{s}\square f''_{2,0}}{\kappa f''_{1,0}+f''_{2,0}\mathcal{L}_\text{m}}\right] \,.
\end{equation}
If $|m|r\ll1$, using Eqs. \eqref{eq:NMC-potential-ppn} and \eqref{eq:NMC-mass-param} in Eq. \eqref{eq:NMC-trace-exp} we obtain
\begin{equation}
	\label{eq:trace-exp-pot-mass}
	\nabla^2 U - m^2 U = \frac{1}{3}f_{2,0}\rho^\text{s}+\frac{2}{3}f'_{2,0}\rho^\text{s}R_0 +2\rho^\text{s}\square f'_{2,0} +2f'_{2.0}\nabla^2 \rho^\text{s} \,,
\end{equation}
which outside of the star can be solved for $R_1$
\begin{equation}
	\label{eq:NMC_R1_sol}
	R_1 = \frac{\chi}{8\pi\left(f''_{2,0}\rho^\text{cos}-f''_{1,0}\right)}\frac{M}{r}\,,
\end{equation}
where $M$ is the mass of the star and 
\begin{equation}
	\label{eq:NMC-ppn-eta}
	\chi=-\frac{1}{3}f_{2,0}+\frac{2}{3}f'_{2,0}R_0+2\square f'_{2,0} \,.
\end{equation}

By considering a flat FLRW metric with a spherically symmetric perturbation
\begin{equation}
	\label{eq:pert-NMC-flrw}
	ds^2=-\left[1+2\Psi(r)\right] dt^2 + a^2(t)\left\{\left[1+2\Phi(r)\right]dr^2 +r^2 d\theta^2 +r^2\sin^2\theta d\phi^2\right\} \,,
\end{equation}
and solving the linearized field equations for $\Psi(r)$ and $\Phi(r)$ with the solution obtained for $R_1$ \eqref{eq:NMC_R1_sol}, one obtains \cite{Bertolami2013}
\begin{equation}
	\label{eq:nmc-ppn-psi}
	\Psi=-\frac{f_{2,0}+f'_{2,0}R_0}{12\pi\left(f'_{1,0}-f'_{2,0}\rho^\text{cos}\right)}\frac{M}{r}\,,
\end{equation}
\begin{equation}
	\label{eq:nmc-ppn-phi}
	\Phi=\frac{f_{2,0}+4f'_{2,0}R_0+6\square f'_{2,0}}{24\pi\left(f'_{1,0}-f'_{2,0}\rho^\text{cos}\right)}\frac{M}{r}.
\end{equation}
Comparing Eqs. \eqref{eq:nmc-ppn-psi} and \eqref{eq:nmc-ppn-phi} with the equivalent PPN metric
\begin{equation}
	\label{eq:metric-ppn-nmc}
	ds^2=-\left(1-\frac{2GM}{r}\right) dt^2 + \left(1+\gamma\frac{2GM}{r}\right)\left(dr^2 +r^2 d\theta^2 +r^2\sin^2\theta d\phi^2\right) \,,
\end{equation}
where $\gamma$ is a PPN parameter, one can see that in NMC gravity 
\begin{equation}
	\label{eq:nmc-gamma}
	\gamma=\frac{1}{2}\left[\frac{f_{2,0}+4f'_{2,0}R_0+6\square f'_{2,0}}{f_{2,0}+f'_{2,0}R_0}\right]\,.
\end{equation}
If $f_2=1$, then $\gamma=1/2$, like in $f(R)$ gravity. The tightest bound on this parameter comes from the tracking of the Cassini probe, where it was determined that $\gamma-1= (2.1\pm2.3)\times 10^{-5}$. This in turn can be used to constrain particular NMC models, provided that the linearized limit is valid and that $|m| r\ll 1$ (see \cite{Bertolami2013} for a more in-depth look at this constraint).

\subsection{``Post-Yukawa'' expansion}

Similarly to $f(R)$ theories, the authors in \cite{Castel-Branco2014} have expanded the previous analysis to include the consideration of the Yukawa terms in the expansion, rather than disregard them entirely by imposing a limit to the mass of the scalar degree of freedom. We will briefly summarize the resulting constraints from the leading order of the expansion, and point the reader to \cite{Castel-Branco2014,March2017,March2019} for a more detailed analysis.

The used metric for spacetime around a spherical central object (star) with mass $M$ and density $\rho$ in this scenario mirrors the $f(R)$ case, in that it is also given by a small perturbation on an asymptotically  flat Minkowski metric, and is given by the line element
\begin{equation}
	\label{eq:pert-NMC-mink}
	ds^2=-\left[1+2\Psi(r)\right] dt^2 +\left\{\left[1+2\Phi(r)\right]dr^2 +r^2 d\theta^2 +r^2\sin^2\theta d\phi^2\right\} \,.
\end{equation}
This metric is very similar to the one used for the post-Newtonian expansion \eqref{eq:pert-NMC-flrw}, with the notable exception that one ignores the effects of the background cosmological curvature. In this case, one assumes that the $f(R)$ function admit a Taylor expansion around $R=0$ of the form
\begin{equation}
	\label{eq:yuk-f1}
	f_1(R)=R+\frac{R^2}{6\mathcal{M}^2}+ \mathcal{O}(R^3) \,,
\end{equation}
\begin{equation}
	\label{eq:yuk-f2}
	f_2(R)=1+2\xi \frac{R}{\mathcal{M}^2}+\mathcal{O}(R^2) \,,
\end{equation}
where $\mathcal{M}$ is a characteristic mass scale and $\xi$ is a dimensionless parameter. Solving the trace of the field equations under these assumptions, and ignores terms of order $\mathcal{O}(c^{-3})$ or smaller, one obtains
\begin{equation}
	\label{eq:trace-nmc-1}
	\nabla^2 R - \mathcal{M}^2R=\frac{8\pi G}{c^2}\mathcal{M}^2\left[\rho-6\frac{2\xi}{\mathcal{M}^2}\nabla^2\rho\right]\,.
\end{equation}
The solution for the curvature outside of the star is \cite{Castel-Branco2014}
\begin{equation}
	\label{eq:R-sol-nmc-yuk}
	R(r)=\frac{2GM}{c^2 r}\mathcal{M}^2 (1-12\xi) A(\mathcal{M},r_s)e^{-mr}\,,
\end{equation}
where $r_s$ is the mass of the star and $A(\mathcal{M},r_s)$ is a form factor given by
\begin{equation}
	\label{eq:yuk-form}
	A(\mathcal{M},r_s) = \frac{4\pi}{\mathcal{M}M}\int_{0}^{r_s} \sinh(\mathcal{M}r)\rho(r)r \, dr \,.
\end{equation}

One can now determine the solutions for $\Psi$ and $\Phi$ outside the star, obtaining 
\begin{equation}
	\label{eq:psi-nmc-yuk}
	\Psi = -\frac{GM}{c^2 r}\left[1+\left(\frac{1}{3}-4\xi\right)A(\mathcal{M},r_s)e^{-\mathcal{M}r}\right] \,,
\end{equation}
\begin{equation}
	\label{eq:phi-nmc- yuk}
	\Phi = \frac{GM}{c^2 r}\left[1-\left(\frac{1}{3}-4\xi\right)A(\mathcal{M},r_s)e^{-\mathcal{M}r}(1+\mathcal{M}r)\right] \,.
\end{equation}
Inserting these perturbations back into the metric \eqref{eq:pert-NMC-mink} one can immediately identify an additional Yukawa term to the usual Newtonian potential, which reads
\begin{equation}
	U(r)=-\frac{GM}{r}\left[1+\alpha A(\mathcal{M},r_s)e^{-r/\lambda}\right]\,,
\end{equation}
where $\alpha= 1/3-4\xi$ and $\lambda = 1/\mathcal{M}$ are the strength and characteristic length of the Yukawa potential. One can then use current constraints \cite{Adelberger2003} on a fifth Yukawa-type force to constrain the parameter pair ($\alpha,\lambda$) or, equivalently, ($\xi,\mathcal{M}$) (see Fig. \ref{fig:yuk}).
\begin{figure}[!h]
	\centering
	\subfloat{\includegraphics[width=0.49\textwidth]{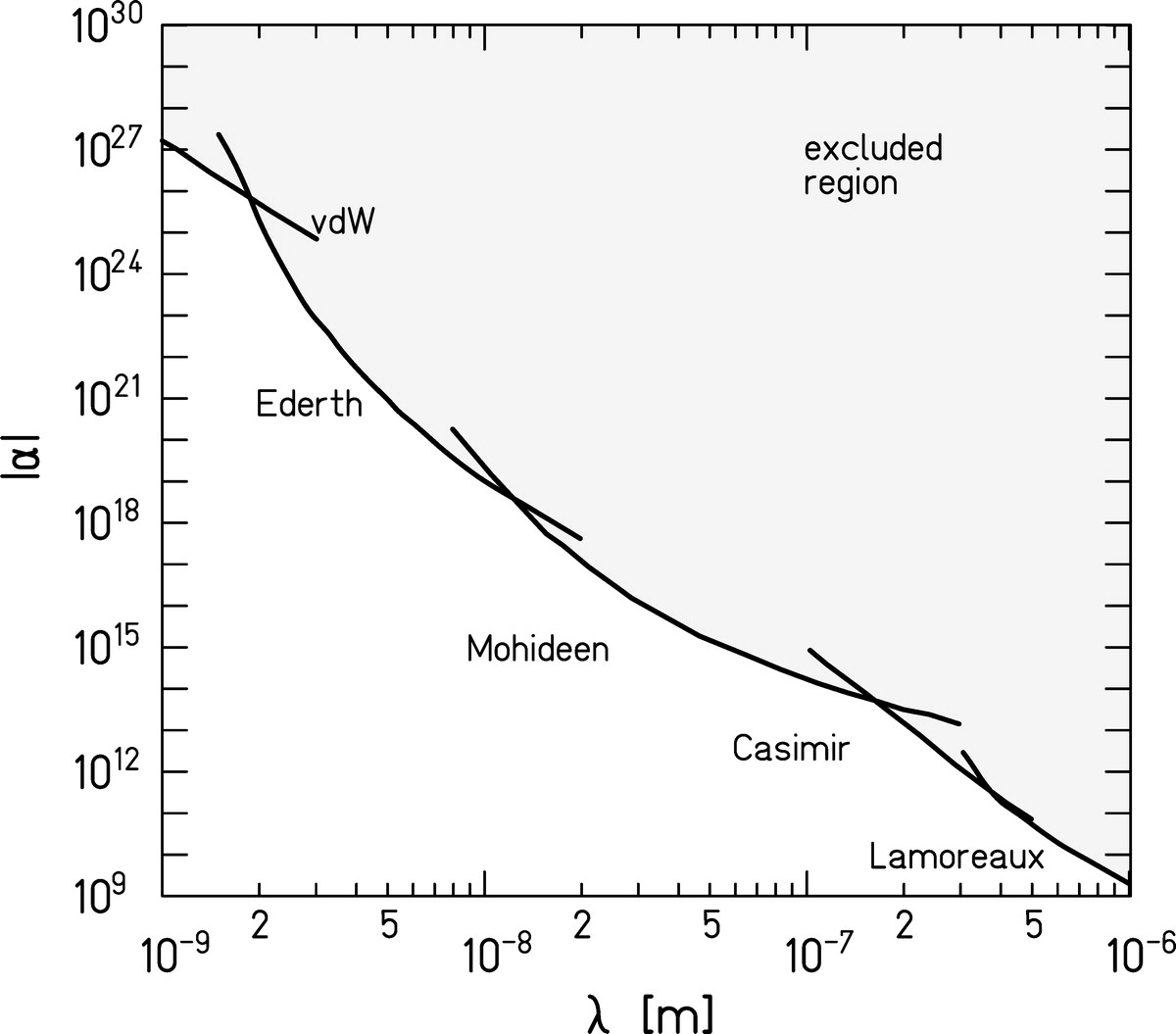}}\hfill
	\subfloat{\includegraphics[width=0.49\textwidth]{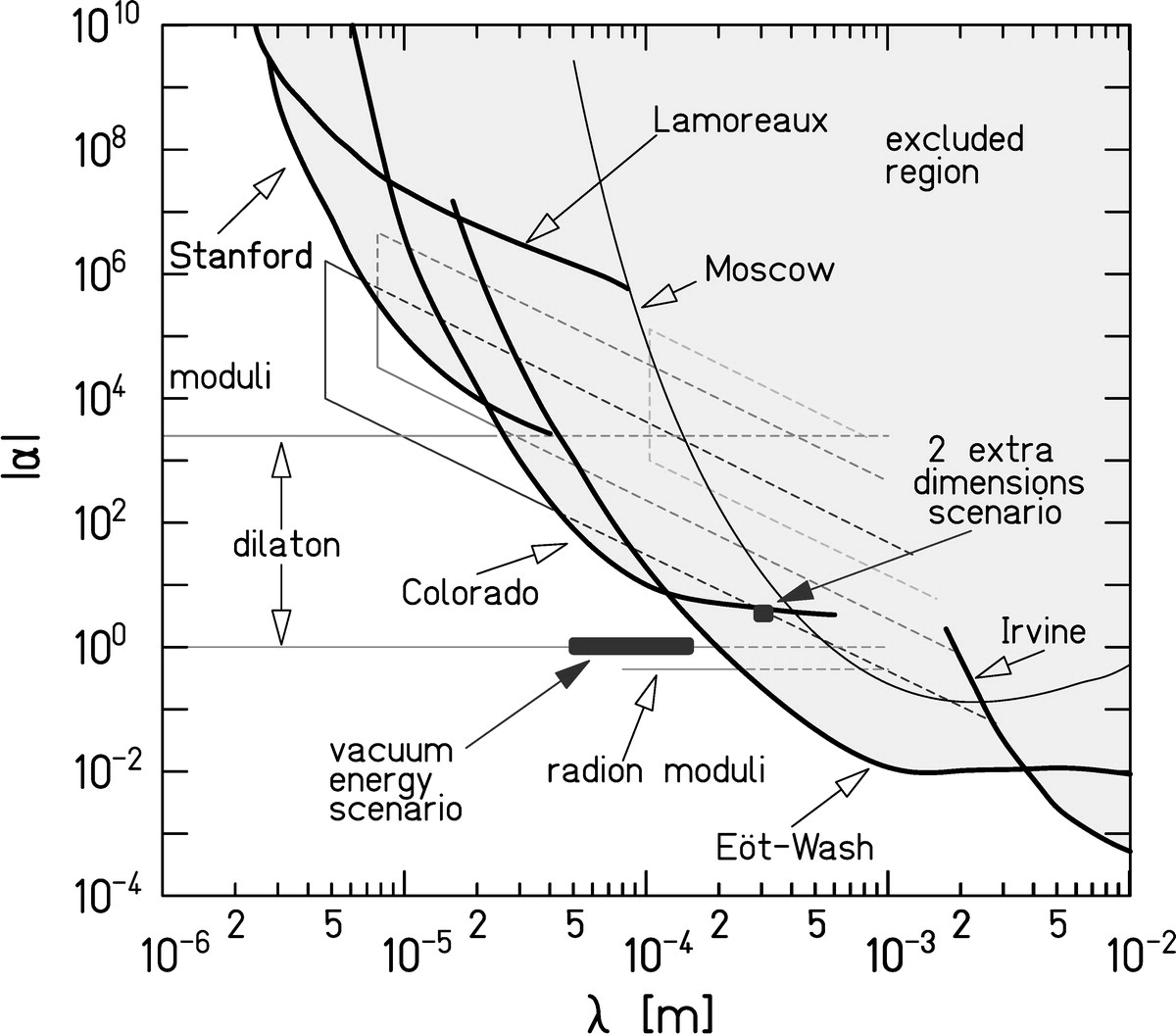}}\par
	\subfloat{\includegraphics[width=0.49\textwidth]{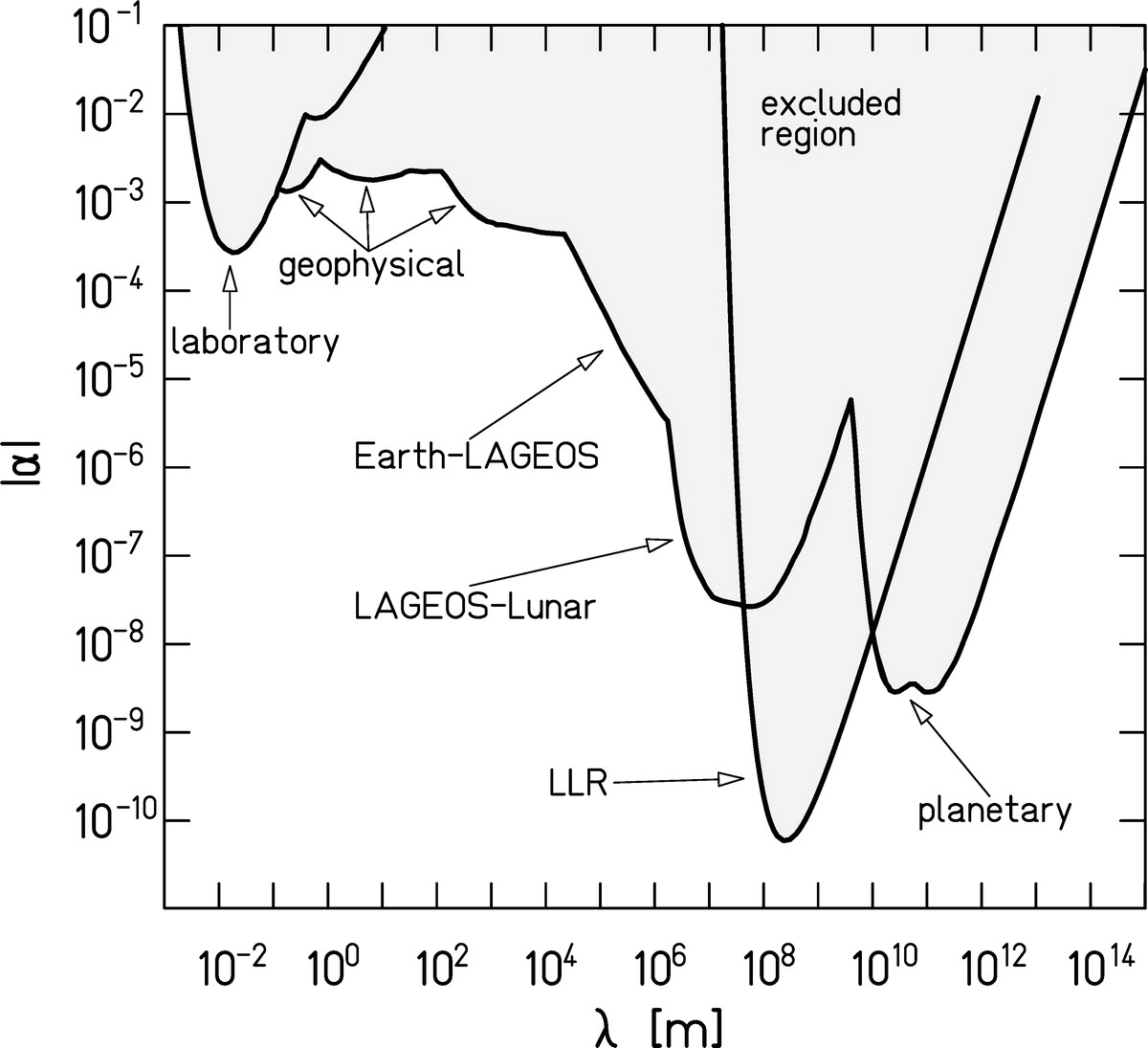}}
	\caption[Constraints on the Yukawa force up to the nano scale]{Constraints on the Yukawa force strength $\alpha$ and range $\lambda$ parameters \cite{Adelberger2003}.}
	\label{fig:yuk}
\end{figure}

This analysis can of course be extended to a higher-order expansion of the functions $f_1$ and $f_2$. Though the calculations are much more involved, constraints on the expansion parameters have been obtained from the precession of Mercury's orbit \cite{March2017} and from ocean experiments \cite{March2019}.

\section{Spectral distortions in the cosmic microwave background}
\label{sec.cmb}

The cosmic microwave background (CMB) has been the subject of many studies and observational missions ever since it was discovered. Most of this effort concerns the study of the small temperature variations across the CMB, usually categorized as multiple types of spectral distortions, some of which can be gravitational in origin. Naturally, one would expect that a modified theory of gravity could have an impact on these distortions, and this is indeed the case in NMC theories of gravity. Since photons couple differently to the background curvature in NMC theories (when compared to GR), one can derive a new type of spectral distortions, which we dub $n$-type spectral distortions, affecting the normalization of the spectral energy density \cite{Avelino2018}.

Let us start by assuming that the CMB has a perfect black body spectrum with temperature $T_{\rm dec}$ at the time of decoupling between baryons and photons (neglecting tiny temperature fluctuations of $1$ part in $10^5$). The spectral energy density and number density of a perfect black body are given by
\begin{equation}
	\label{eq:black-bod-cons}
	u(\nu)=\frac{8 \pi h \nu^3}{e^{h \nu/(k_B T)}-1}\,, \qquad n(\nu) = \frac{u(\nu)}{h \nu}\,,
\end{equation}
respectively, where $h$ is Planck's constant and $k_B$ is Boltzmann's constant, $T$ is the temperature and $E_\gamma = h \nu$ is the energy of a photon of frequency $\nu$.  In the standard scenario, assuming that the universe becomes transparent for $T < T_{\rm dec}$, the CMB radiation retains a black body spectrum after decoupling. This happens because the photon number density evolves as $n_\gamma \propto a^{-3}$ (assuming that the number of photons is conserved) while their frequency is inversely proportional to the scale factor $a$, so that $\nu \propto a^{-1}$. In the case studied in the present paper, the number of CMB photons is still assumed to be conserved (so that $n_\gamma \propto a^{-3}$)

However, recall that the momentum of particles in NMC gravity evolves as determined in Eq. \eqref{eq:momev}, and therefore the energy and frequency of photons also follows 
\begin{equation}
	\label{eq:nu-ev}
	E_\gamma \propto \nu \propto (a f_2)^{-1}\,.
\end{equation}
Alternatively, one can arrive at the same result by recalling that the equation of motion of a point particle in NMC gravity is given by
\begin{equation}
	\label{eq:geodesic}
	\frac{du^{\mu} }{ ds}+\Gamma^\mu_{\alpha\beta} u^\alpha u^\beta = \mathfrak{a}^\mu \,,
\end{equation}
where $\mathfrak{f}^\mu = m\mathfrak{a}^\mu$ is a momentum-dependent four-force, given by
\begin{equation}
	\label{eq:extraforce}
	\mathfrak{f}^\mu=-m\frac{f'_2}{ f_2}h^{\mu\nu}\nabla_\nu R \, ,
\end{equation}
and $h^{\mu\nu}=g^{\mu\nu}+u^{\mu}u^{\nu}$ is the projection operator. Solving Eqs. \eqref{eq:geodesic} and \eqref{eq:extraforce} with the metric \eqref{eq:line-nmc-flat} one finds that the components of the three-force on particles $\vb{\mathfrak{f}}=d\vb{p}/dt$ are given by \cite{Avelino2020} (in Cartesian coordinates)
\begin{equation}
	\label{eq:3force}
	\mathfrak{f}^i = -\frac{d\ln(af_2)}{dt}p^i \,,
\end{equation}
where $p^i$ are the components of the particle's three-momentum $\vb{p}$, which therefore evolve as
\begin{equation}
	p^i \propto (a f_2)^{-1}\,. \label{eq:momev-cons}
\end{equation}Hence, taking into account that $n_\gamma \propto a^{-3} \propto (f_2)^3 \times (af_2)^{-3}$, the spectral energy density at a given  redshift $z$ after decoupling may be written as
\begin{equation}
	u(\nu)_{[z]} =  \frac{(f_{2[z]})^3}{(f_{2[z_{\rm dec}]})^3} \frac{8 \pi h \nu^3}{e^{h \nu/(k_B T_{[z]})}-1}\,, 
\end{equation}
where 
\begin{equation}
	T_{[z]} \equiv T_{ [z_{\rm dec}]} \frac{(1+z) f_{2 [z_{dec}]}}{(1+z_{dec})f_{2 [z]}}\,.
\end{equation}

The evolution of this spectral density is similar to that of a perfect black body (see Eqs. \eqref{eq:black_spectral_red} and \eqref{eq:black_temp_red}), except for the different normalization. Also note that a small fractional variation $\Delta f_2/f_2$ on the value of $f_2$ produces a fractional change in the normalization of the spectral density equal to $3 \Delta f_2/f_2$.  

The FIRAS (Far InfraRed Absolute Spectrophotometer) instrument onboard COBE (COsmic Background Explorer) measured the spectral energy of the nearly perfect CMB black body spectrum \cite{Fixsen1996,Fixsen2009}. The weighted root-mean-square deviation between the observed CMB spectral radiance and the blackbody spectrum fit was found to be less than $5$ parts in $10^5$ of the peak brightness. Hence, we estimate that $f_2$ can vary by at most by a few parts in $10^5$ from the time of decoupling up to the present time. This provides a stringent constraint on NMC theories of gravity, independently of any further considerations about the impact of such theories on the background evolution of the Universe.

Let us define
\begin{equation}
	\Delta f_2^ {i \to f} \equiv \left|f_2(z_f) - f_2(z_i)\right| \,, 
\end{equation}
which is assumed to be much smaller than unity. Here, $z_i$ and $z_f$ are given initial and final redshifts with $z_i > z_f$. Consider a power law model for $f_2$ defined by
\begin{equation}
	\label{eq:model}
	f_1(R) \sim R\,, \qquad f_2(R) \propto R^n\,,
\end{equation}
in the redshift range $[z_f,z_i]$, where $n$ is a real number with $|n| \ll 1$ and $f_2 \sim 1$ at all times. In this case
\begin{equation}
	\label{eq:f2var}
	\Delta f_2^ {i \to f}  \sim   \left| \left(\frac{R(z_f)}{R(z_i)}\right)^n -1\right| 
	\sim  3 \left| n \right| \ln \left( \frac{1+z_f}{1+z_i}\right)\,.
\end{equation}
Therefore, we have 
\begin{equation}
	\left| n \right| \lesssim \frac{\Delta f_2}{3}^{CMB \to 0} \left[ \ln \left(1+z_\text{CMB}\right) \right]^{-1}\,,
	\label{eq:nconstraint3}
\end{equation}
and it is simple to show that this translates into $|n|\lesssim {\rm few} \times 10^{-6}$.

\section{Big-bang nucleosynthesis }
\label{sec.bbn}

Primordial nucleosynthesis may be described by the evolution of a set of differential equations, namely the Friedmann equation, the evolution of baryon and entropy densities, and the Boltzmann equations describing the evolution of the average density of each nuclide and neutrino species. As one could expect, even taking into account experimental values for the reaction cross-sections instead of theoretical derivations from particle physics, the accurate computation of element abundances cannot be done without resorting to numerical algorithms \cite{Wagoner1973,Kawano1992,Smith:1992yy,Pisanti2008,Consiglio2017}. The prediction of these quantities in the context of NMC theories is beyond the scope of this thesis, but it is worthy of note that these codes require one particular parameter to be set a priori: the baryon-to-photon ratio $\eta$.

While in GR the baryon-to-photon ratio is fixed around nucleosynthesis, in general the same does not  happen in the context of NMC theories. To show this, recall that the evolution of the density of photons and baryons (which are always non-relativistic from the primordial nucleosynthesis epoch up to the present era) is given by Eq. \eqref{eq:densityevo} as
\begin{equation}
	\label{eq:matradevo}
	\rho_\gamma = \rho_{\gamma,0} a^{-4}f_2^{-1}\,, \qquad \rho_\text{b} =\rho_{\text{b},0} a^{-3}\,.
\end{equation}
While the baryon number in a fixed comoving volume is conserved ($n_\text{b}\propto a^{-3}$, where $n_\text{b}$ is the baryon number density), before recombination photons are in thermal equilibrium, so the photon number density is directly related to the temperature by
\begin{equation}
	\label{eq:photntemp}
	n_\gamma={2\zeta(3)\over \pi^2}T^3\,.
\end{equation}
Since the photon energy density also relates to the temperature as
\begin{equation}
	\label{eq:photenergdens}
	\rho_\gamma={\pi^2\over 15}T^4\,,
\end{equation}
combining Eqs. \eqref{eq:matradevo}, \eqref{eq:photntemp} and \eqref{eq:photenergdens} one obtains that that the baryon-to-photon ratio $\eta$ between BBN (at a redshift $z_\text{BBN} \sim 10^9$) and photon decoupling (at a redshift $z_\text{CMB} \sim 10^3$) evolves as
\begin{equation}
	\label{eq:eta}
	\eta\equiv {n_\text{b}\over n_\gamma}\propto f_2^{3/4}\,,
\end{equation}
as opposed to the GR result, $\eta=\text{const}$. After photon decoupling the baryon-to-photon ratio is conserved in NMC theories, with the energy of individual photons evolving as $E_\gamma \propto (a f_2)^{-1}$ (as opposed to the standard $E_\gamma \propto a^{-1}$ result).

We will assume that the modifications to the dynamics of the universe with respect to GR are small, in particular to the evolution of $R$ and $H$ with the redshift $z$. This can be seen as a ``best-case'' scenario for the theory, as any significant changes to $R(z)$ and $H(z)$ are expected to worsen the compatibility between the model's predictions and observational data. Hence, in the following, we shall assume that 
\begin{eqnarray}
	R &=& 3 H_0^2 \left[ \Omega_{\text{m},0} (1+z)^{3} + 4 \Omega_{\Lambda,0} \right] \nonumber  \\
	&\sim&  3 H_0^2 \Omega_{\text{m},0} (1+z)^{3} \propto (1+z)^{3}\,,
\end{eqnarray}
where $\Omega_{\text{m},0} \equiv (\rho_{\text{m},0})/(6 \kappa H_0^2)$ and $\Omega_{\Lambda,0} \equiv (\rho_{\Lambda,0})/(6 \kappa H_0^2)$ are the matter and dark energy density parameters (here dark energy is modelled as a cosmological constant $\Lambda$), and the approximation is valid all times, except very close to the present time.

Let us define
\begin{equation}
	\frac{\Delta \eta}{\eta}^{i \to f} \equiv \frac{\left|\eta(z_f) - \eta(z_i)\right|}{\eta(z_i)} \,,
\end{equation}
which is assumed to be much smaller than unity. Eq. \eqref{eq:f2var} then implies that
\begin{equation}
	\left| n \right| \lesssim \frac49 \frac{\Delta \eta}{\eta}^{i \to f} \left[ \ln \left( \frac{1+z_f}{1+z_i}\right) \right]^{-1}\,,
	\label{eq:nconstraint}
\end{equation}
assuming a small relative variation of $\eta$ satisfying Eq. \eqref{eq:eta}
\begin{equation}
	\label{eq:etavar}
	\Delta f_2^ {i \to f}  \sim \frac43 \frac{\Delta \eta}{\eta}^{i \to f}\,.
\end{equation}

There are two main ways of estimating the value of $\eta$ at different stages of cosmological evolution. On one hand, one may combine the observational constraints on the light element abundances with numerical simulations of primordial BBN nucleosynthesis to infer the allowed range of $\eta$. This is the method used in \cite{Iocco2009}, among others, leading to
\begin{equation}
	\label{eq:etabbnconst}
	\eta_\text{BBN}=(5.7\pm0.6)\times 10^{-10}
\end{equation}
at 95\% credibility interval (CI) just after nucleosynthesis (at a redshift $z_\text{BBN} \sim10^9$). More recently, an updated version of the program {\fontfamily{qcr}\selectfont PArthENoPE} ({\fontfamily{qcr}\selectfont PArthENoPE} 2.0), which computes the abundances of light elements produced during BBN, was used to obtain new limits on the baryon-to-photon ratio, at $2\sigma$ \cite{Consiglio2017}
\begin{equation}
	\label{eq:etabbn2const}
	\eta_\text{BBN}=(6.23^{+0.24}_{-0.28})\times 10^{-10}\,.
\end{equation}

There is some variation of $\eta$ during nucleosynthesis due to the entropy transfer to photons associated with the $e^\pm$ annihilation. The ratio between the values of $\eta$ at the beginning and the end of BBN is given approximately by a factor of $2.73$ \cite{Serpico2004}. Although the NMC will lead to further changes ratio, we will not consider this effect, since it will be subdominant for $|n| \ll 1$. We will therefore use the above standard values obtained for $\eta_\text{BBN}$ immediately after nucleosynthesis to constrain NMC gravity.

The neutron-to-photon ratio also affects the acoustic peaks observed in the CMB, generated at a redshift $z_\text{CMB} \sim10^3$. The full-mission Planck analysis \cite{Ade2016} constrains the baryon density $\omega_\text{b} = \Omega_\text{b} (H_0/[100 \text{ km s}^{-1}\text{ Mpc}^{-1}] )$ with the inclusion of  BAO, at 95\% CI,
\begin{equation}
	\label{eq:omegacmb-cons}
	\omega_\text{b}=0.02229^{+0.00029}_{-0.00027}\, .
\end{equation}
This quantity is related to the baryon-to-photon ratio via $\eta = 273.7\times10^{-10} \omega_\text{b}$, leading to
\begin{equation}
	\label{eq:etacmb-cons}
	\eta_\text{CMB}= 6.101^{+0.079}_{-0.074}\times 10^{-10}\,.
\end{equation}
Here, we implicitly assume that no significant change to $\eta$ occurs after $z_\text{CMB} \sim10^3$, as shown in Ref. \cite{Avelino2018}.

Taking these results into consideration, we shall determine conservative constraints on $n$ using the maximum allowed variation of $\eta$ from $z_\text{BBN} \sim10^9$ to $z_\text{CMB} \sim10^3$, using the appropriate lower and upper limits given by Eqs. \eqref{eq:etabbnconst}, \eqref{eq:etabbn2const} and \eqref{eq:etacmb-cons}. Combining Eqs. \eqref{eq:eta} and \eqref{eq:model} to obtain
\begin{equation}
	\label{eq:etamodel}
	\eta\propto R^{3n/4}\,,
\end{equation}
it is easy to see that the sign of $n$ will affect whether $\eta$ is decreasing or increasing throughout the history of the universe, and thus, since $R$ monotonically decreases towards the future, a positive (negative) $n$ will imply a decreasing (increasing) $\eta$. This being the case, for the allowed range in Eq. \eqref{eq:etabbnconst}, we have for positive $n$
\begin{equation}
	\label{eq:npos}
	\frac{\Delta \eta}{\eta} = \frac{\left|(6.101 - 0.074) - (5.7 + 0.6)\right|}{5.7 + 0.6} \simeq 0.04 \,,
\end{equation}
and for negative $n$
\begin{equation}
	\label{eq:nneg}
	\frac{\Delta \eta}{\eta} = \frac{\left|(6.101 + 0.079) - (5.7 - 0.6)\right|}{5.7 - 0.6} \simeq 0.21 \,,
\end{equation}
Therefore we find
\begin{equation}
	\label{eq:nconstr1}
	-0.007<n<0.002\, ,
\end{equation}
and using the limits given in Eq. \eqref{eq:etabbn2const} \cite{Consiglio2017},
\begin{equation}
	\label{eq:nconstr2}
	-0.002<n<0.003\, .
\end{equation}

In the previous section the NMC has been shown to lead to $n$-type spectral distortions in the CMB, affecting the normalization of the spectral energy density, and for a power-law $f_2\propto R^n$ we obtained $|n|\lesssim {\rm few} \times 10^{-6}$, which is roughly $3$ orders of magnitude stronger than the constraint coming from the baryon-to-photon ratio. Still, this constraint and those given by Eqs. \eqref{eq:nconstr1} and \eqref{eq:nconstr2} are associated with cosmological observations which probe different epochs and, as such, can be considered complementary: while the former limits an effective value of the power-law index $n$ in the redshift range $[0,10^3]$, the later is sensitive to its value at higher redshifts in the range $[10^3,10^9]$.

Furthermore, NMC theories with a power-law coupling $f_2(R)$ have been considered as a substitute for dark matter in previous works \cite{Bertolami2012,Silva2018}. There it has been shown that $n$ would have to be in the range $-1\leq n \leq -1/7$ to explain the observed galactic rotation curves. However, such values of $n$ are excluded by the present study.

\section{Distance duality relation}
\label{sec.ddr}

Etherington's relation, also known as the distance-duality relation (DDR), directly relates the luminosity distance $d_\text{L}$ and angular-diameter distance $d_\text{A}$ in GR, where they differ only by a specific function of the redshift
\begin{equation}
	\label{eq:DDR_cons}
	\frac{d_\text{L}}{d_\text{A}}=(1+z)^2\,.
\end{equation}
Naturally, if a modified gravity theory features different expressions for the luminosity or angular-diameter distances, this relationship may also change. The DDR has therefore recently come into focus given the possibility of performing more accurate tests of Etherington's relation  with new cosmological surveys of type Ia supernovae (SnIa) and baryon acoustic oscillations (BAO)\cite{Bassett2004,Ruan2018,Xu2020,Martinelli2020,Lin2021,Zhou2021}, as well as observations from Euclid \cite{Laureijs2011,Astier2014} and gravitational wave observatories \cite{Cai2017}. In this section, we derive the impact of NMC theories on the DDR and use the most recent data available to impose constraints on a broad class of NMC models.

\subsection{The distance-duality relation in NMC theories}

The luminosity distance $d_\text{L}$ of an astronomical object relates its absolute luminosity $L$, i.e. its radiated energy per unit time, and its energy flux at the detector $l$, so that they maintain the usual Euclidian relation
\begin{equation}
	\label{eq:apparent luminosity-cons}
	l = \frac{L}{4\pi d_\text{L}^2}\,,
\end{equation}
or in terms of the luminosity distance
\begin{equation}
	\label{eq:lum_dist_1_const}
	d_\text{L} = \sqrt{\frac{L}{4\pi l}}\,.
\end{equation}
Over a small emission time $\Delta t_\text{em}$ the absolute luminosity can be written as
\begin{equation}
	\label{eq:abs_lum_1_const}
	L = \frac{N_{\gamma,\text{em}} E_\text{em}}{\Delta t_\text{em}}\,,
\end{equation}
where $N_{\gamma,\text{em}}$ is the number of emitted photons and $E_\text{em}$ is the average photon energy. An observer at a coordinate distance $r$ from the source will, however, observe an energy flux given by
\begin{equation}
	\label{eq:app_lum_const}
	l = \frac{N_{\gamma,\text{obs}}E_\text{obs}}{\Delta t_\text{obs} 4\pi r^2}
\end{equation}
where $N_{\gamma,\text{obs}}$ is the number of observed photons and $E_\text{obs}$ is their average energy.

Recall that the energy of a photon in a flat FLRW metric in NMC gravity is given by
\begin{equation}
	\label{eq:photon_freq}
	E\propto\nu\propto \frac{1}{af_2} = \frac{1+z}{f_2} \, .
\end{equation}
Note that while the number of photons is conserved, $N_{\gamma,\text{obs}} = N_{\gamma,\text{em}}$, the time that it takes to receive the photons is increased by a factor of $1+z$, $t_\text{obs}= (1+z)t_\text{em}$, and as per Eq. \eqref{eq:photon_freq}, their energy is reduced as 
\begin{equation}
	\label{eq:energy_obs_const}
	E_\text{obs} = \frac{E_\text{em}}{1+z}\frac{f_2(z)}{f_2(0)} \,,
\end{equation}
where $f_2(z)=f_2[R(z)]$ and $f_2(0)=f_2[R(0)]$ are respectively the values of the function $f_2$ at emission and at the present time. The distance $r$ can be calculated by just integrating over a null geodesic, that is
\begin{align}
	\label{eq:null geodesic_const}
	ds^2 &= - dt^2 + a(t)^2dr^2 = 0\nonumber \\
	\Rightarrow dr &= -\frac{dt}{a(t)} \nonumber \\
	\Rightarrow  r &= \int_{t_\text{em}}^{t_\text{obs}} \frac{dt}{a(t)}= \int_0^z \frac{dz'}{H(z')} \,,
\end{align}
Using Eqs. \eqref{eq:abs_lum_1_const}, \eqref{eq:app_lum_const}, \eqref{eq:energy_obs_const} and \eqref{eq:null geodesic_const} in Eq. \eqref{eq:lum_dist_1_const}, we finally obtain
\begin{equation}
	\label{eq:lum_dist_cons}
	d_\text{L} = (1+z)\sqrt{\frac{f_2(0)}{f_2(z)}}\int_0^z \frac{dz'}{H(z')} \, .
\end{equation}
In the GR limit $f_2=\text{const.}$, and we recover the standard result.

The angular-diameter distance $d_\text{A}$, on the other hand, is defined so that the angular diameter $\theta$ of a source that extends over a proper distance $s$ perpendicularly to the line of sight is given by the usual Euclidean relation
\begin{equation}
	\label{eq:angle-cons}
	\theta=\frac{s}{d_\text{A}}\,.
\end{equation}
In a FLRW universe, the proper distance $s$ corresponding to an angle $\theta$ is simply
\begin{equation}
	\label{eq:prop_dist-cons}
	s= a(t)r\theta = \frac{r\theta}{1+z}\,,
\end{equation}
where the scale factor has been set to unity at the present time. So the angular-diameter distance is just
\begin{equation}
	\label{eq:ang_dist_cons}
	d_\text{A}=\frac{1}{1+z}\int_0^z \frac{dz'}{H(z')}\,.
\end{equation}
Comparing Eqs. \eqref{eq:lum_dist_cons} and \eqref{eq:ang_dist_cons} one finds
\begin{equation}
	\label{eq:mod_DDR}
	\frac{d_\text{L}}{d_\text{A}}=(1+z)^2 \sqrt{\frac{f_2(0)}{f_2(z)}}\,.
\end{equation}

Deviations from the standard DDR are usually parametrized by the factor $\upeta$ as 
\begin{equation}
	\label{eq:par_DDR}
	\frac{d_\text{L}}{d_\text{A}}=(1+z)^2 \upeta\,.
\end{equation}
Constraints on the value of $\upeta$ are derived from observational data for both $d_\text{A}$ and $d_\text{L}$.
Comparing Eqs. \eqref{eq:mod_DDR} and \eqref{eq:par_DDR} one immediately obtains
\begin{equation}
	\label{eq:eta_ddr}
	\upeta(z) = \sqrt{\frac{f_2(0)}{f_2(z)}}\,,
\end{equation}
(see also \cite{Minazzoli2014,Hees2014} for a derivation of this result in theories with an NMC between the matter fields and a scalar field). 

If $f_1=R$, like in GR, any choice of the NMC function apart from $f_2 = 1$ would lead to a deviation from the standard $\Lambda$CDM background cosmology and would therefore require a computation of the modified $H(z)$ and $R(z)$ for every different $f_2$ that is probed. However, it is possible to choose a function $f_1$ such that the cosmological background evolution remains the same as in $\Lambda$CDM. In this case, the Hubble factor is simply 
\begin{equation}
	\label{eq:Hubble}
	H(z)=H_0\left[\Omega_{\text{r},0}(1+z)^4+\Omega_{\text{m},0}(1+z)^3+\Omega_{\Lambda,0}\right]^{1/2} \,,
\end{equation}
and the scalar curvature $R$ is given by
\begin{equation}
	\label{eq:scalar_curv}
	R(z)=3H_0^2\left[\Omega_{\text{m},0}(1+z)^3+4\Omega_{\Lambda,0}\right] \,,
\end{equation}
where $H_0$ is the Hubble constant, $\Omega_{\text{r},0}$, $\Omega_{\text{m},0}$ and $\Omega_{\Lambda,0}$ are the radiation, matter and cosmological constant density parameters at present time. The calculation of the appropriate function $f_1$ must be done numerically, by integrating either the MFE \eqref{eq:fried-f1f2-1-cons} or the MRE \eqref{eq:ray-f1f2-1-cons} for $f_1$ with the appropriate initial conditions at $z=0$ (when integrating the MRE, the MFE serves as an additional constraint). Considering that GR is strongly constrained at the present time, the natural choice of initial conditions is $f_1(0)=R(0)$ and $\dot{f}_1(0)=\dot{R}(0)$.

Nevertheless, in this section we will consider that significant deviations of $f_2$ from unity are allowed only at relatively low redshift, since CMB and BBN constraints on NMC theories have already constrained $f_2$ to be very close to unity at large redshifts \cite{Avelino2018,Azevedo2018a}. We have verified that the function $f_1(z)$ required for Eqs. \eqref{eq:Hubble} and \eqref{eq:scalar_curv} to be satisfied deviates no more than $3\%$ from the GR prediction $f_1=R$ for $z\lesssim1.5$, for the models investigated in this paper (using the best-fit parameters in Tables \ref{tab:results_full_power} and \ref{tab:results_full_exp}).

\subsection{Methodology and results}\label{sec:results}

In \cite{Martinelli2020}, the authors used Pantheon and BAO data to constrain a parametrization of the DDR deviation of the type 
\begin{equation}
	\label{eq:par_eta_eps-cons}
	\upeta(z) = (1+z)^{\epsilon}\,.
\end{equation}
and obtained, for a constant $\epsilon$, $\epsilon=0.013\pm0.029$ at the 68\% credible interval (CI). Here, we use the same datasets and a similar methodology to derive constraints for specific NMC models. We present a brief description of the methodology for completeness, but refer the reader to \cite{Martinelli2020} for a more detailed discussion. 

In general, BAO data provides measurements of the ratio $d_z$ (see, for example, \cite{Beutler2011}), defined as
\begin{equation}
	d_z\equiv \frac{r_\text{s}(z_\text{d})}{D_V(z)} \,,
\end{equation}
where $D_V(z)$ is the volume-averaged distance \cite{Eisenstein2005}
\begin{equation}
	D_V(z)=\left[(1+z)^2 d_\text{A}^2(z) \frac{c z}{H(z)}\right]^{1/3} \,,
\end{equation}
and $r_\text{s}(z_\text{d})$ is the comoving sound horizon at the drag epoch. Assuming that the evolution of the Universe is close to $\Lambda$CDM, $r_\text{s}(z_\text{d})$ can be approximated as \cite{Eisenstein1998}
\begin{equation}
	\label{eq:sound_hor}
	r_\text{s}(z_\text{d}) \simeq \frac{44.5 \ln \left(\frac{9.83}{\Omega_{\text{m},0}h^2}\right)} {\sqrt{1+10(\Omega_{\text{b},0}h^2)^{3/4}}} \,,
\end{equation}
where $\Omega_{\text{b},0}$ is the baryon density parameter and $h$ is the dimensionless Hubble constant. Here we shall assume that $\Omega_{\text{b},0}h^2=0.02225$ in agreement with the latest \textit{Planck} release \cite{Aghanim2020}. Notice that the BAO observations are used to estimate $d_\text{A}$, which remains unchanged in NMC theories provided that the evolution of $H(z)$ and $R(z)$ is unchanged with respect to the $\Lambda$CDM model. Thus, BAO data will ultimately provide us with constraints on $H_0$ and $\Omega_{\text{m},0}$. The original datasets that we shall consider in the present paper come from the surveys 6dFGS \cite{Beutler2011}, SDDS \cite{Anderson2014}, BOSS CMASS \cite{Xu2012}, WiggleZ \cite{Blake2012}, MGS \cite{Ross2015}, BOSS DR12 \cite{Gil-Marin2016}, DES \cite{Abbott2019}, Ly-$\alpha$ observations \cite{Blomqvist2019}, SDSS DR14 LRG \cite{Bautista2018} and quasar observations \cite{Ata2018}, but the relevant data is conveniently compiled and combined in Appendix A of \cite{Martinelli2020}.

Likewise, the luminosity distance can be constrained using SnIa data, via measurements of their apparent magnitude
\begin{equation}
	\label{eq:appar_mag}
	m(z)=M_0 +5 \log_{10}\left[\frac{d_\text{L}(z)}{\text{Mpc}}\right]+25 \,,
\end{equation}
or, equivalently,
\begin{equation}
	\label{eq:appar_mag_explicit}
	m(z) = M_0 - 5\log_{10}(H_0)+ 5\log_{10}\left[\upeta(z)\hat{d}_\text{L}(z)\right] +25 \,,
\end{equation}
where $M_0$ is the intrinsic magnitude of the supernova and $\hat{d}_\text{L}(z)$ is the GR Hubble-constant-free luminosity distance. Note, that the intrinsic magnitude $M_0$ is completely degenerate with the Hubble constant $H_0$, and thus simultaneous constraints on both quantities cannot be derived from SnIa data alone. As per \cite{Martinelli2020}, we use the marginalized likelihood expression from Appendix C in \cite{Conley2011}, which takes into account the marginalization of both $M_0$ and $H_0$, whenever possible.  Likewise, we use the full 1048 point Pantheon compilation from \cite{Scolnic2018}.

For simplicity, we shall consider two NMC models with a single free parameter (the NMC parameter) which is assumed to be a constant in the relevant redshift range ($0<z<1.5$), and assume a flat Universe evolving essentially as $\Lambda$CDM. Since the contribution of radiation to the overall energy density is very small at low redshift we ignore its contribution, and therefore $\Omega_{\Lambda,0}=1-\Omega_{\text{m},0}$.

We use the Markov chain Monte Carlo (MCMC) sampler in the publicly available Python package \texttt{emcee} \cite{Foreman-Mackey2013} to build the posterior likelihoods for the cosmological parameters, $H_0$ and $\Omega_{\text{m},0}$, as well as the NMC parameter, assuming flat priors for all of them. The MCMC chains are then analyzed using the Python package \texttt{GetDist} \cite{Lewis2019}, to calculate the marginalized means and CIs, as well as plots of the 2D contours of the resulting distributions.

\subsubsection*{Power Law}
\label{subsubsec:powerlaw}

Consider a power law NMC function of the type
\begin{equation}
	\label{eq:power_law}
	f_2\propto R^n \,,
\end{equation}
where $n$ is the NMC parameter (GR is recovered when $n=0$). Using Eqs. \eqref{eq:scalar_curv} and \eqref{eq:power_law} in Eq. \eqref{eq:eta_ddr}, one obtains
\begin{equation}
	\label{eq:eta_power}
	\upeta(z;n,\Omega_{\text{m},0}) = \left[\frac{\Omega_{\text{m},0}+4(1-\Omega_{\text{m},0})}{\Omega_{\text{m},0}(1+z)^3+4(1-\Omega_{\text{m},0})}\right]^{n/2} \,.
\end{equation}

The marginalized 68\% CI results can be found in Table \ref{tab:results_full_power}, and the 2D distributions for $n$ and $\Omega_{\text{m},0}$ are displayed in Fig.~\ref{fig:n_Om} (see also Fig. \ref{fig:tri_n_Tot} for the remaining distribution plots). A reconstruction of Eq. \eqref{eq:eta_power} is also shown in Fig.~\ref{fig:eta_power_law}. Note that Eq. \eqref{eq:eta_power} implies that in the power-law case $\upeta(z)$ only depends on the parameters $n$ and $\Omega_{\text{m},0}$, which are not completely degenerate. Therefore,  SnIa data alone is able to constrain both of these parameters. However, since BAO data constrains both $H_0$ and $\Omega_{\text{m},0}$, we are able to combine the two datasets to significantly improve the constraints on $n$ and $\Omega_{\text{m},0}$.

\begin{table}
	\centering
	\caption[Best-fit values and marginalized means and limits on cosmological and power-law NMC parameters from BAO and SnIa]{Best-fit values and marginalized means, 68\%, 95\% and 99\% CI limits obtained from currently available data on the cosmological parameters $\Omega_{\text{m},0}$ and $H_0$ (in units of km s$^{-1}$ Mpc$^{−1}$) and on the NMC parameter $n$ (dimensionless).}
	\label{tab:results_full_power}
	\bgroup
	\def\arraystretch{1.2}
	\begin{tabular}{lc|ccccc}
		\hline\hline
		Parameter                        & Probe    & Best fit & Mean    & 68\%                 & 95\%                 & 99\%                 \\ \hline
		& BAO      & $66.4$   & $66.8$  & $^{+1.2}_{-1.4}$     & $^{+2.7}_{-2.5}$     & $^{+3.7}_{-3.2}$     \\
		{\boldmath$H_0$}                 & SnIa     & \multicolumn{5}{c}{unconstrained}                                                      \\
		& SnIa+BAO & $66.0$   & $66.1$  & $\pm 1.2$            & $^{+2.5}_{-2.3}$     & $^{+3.4}_{-3.0}$     \\ \hline
		& BAO      & $0.291$  & $0.300$ & $^{+0.027}_{-0.035}$ & $^{+0.064}_{-0.060}$ & $^{+0.094}_{-0.071}$ \\
		{\boldmath$\Omega_{\text{\bf m},0}$} & SnIa     & $0.181$  & $0.191$ & $^{+0.037}_{-0.061}$ & $^{+0.11}_{-0.094}$  & $^{+0.18}_{-0.10}$   \\
		& SnIa+BAO & $0.276$  & $0.279$ & $^{+0.024}_{-0.030}$ & $^{+0.054}_{-0.052}$ & $^{+0.079}_{-0.062}$ \\ \hline
		& BAO      & \multicolumn{5}{c}{unconstrained}                                                      \\
		{\boldmath$n$}                   & SnIa     & $0.178$  & $0.184$ & $^{+0.092}_{-0.15}$  & $^{+0.26}_{-0.24}$   & $^{+0.45}_{-0.26}$   \\
		& SnIa+BAO & $0.014$  & $0.013$ & $\pm 0.035$          & $^{+0.071}_{-0.066}$ & $^{+0.097}_{-0.085}$\\
		\hline\hline
	\end{tabular}
	\egroup
\end{table}
\begin{figure}
	\centering
	\includegraphics[width=0.85\textwidth]{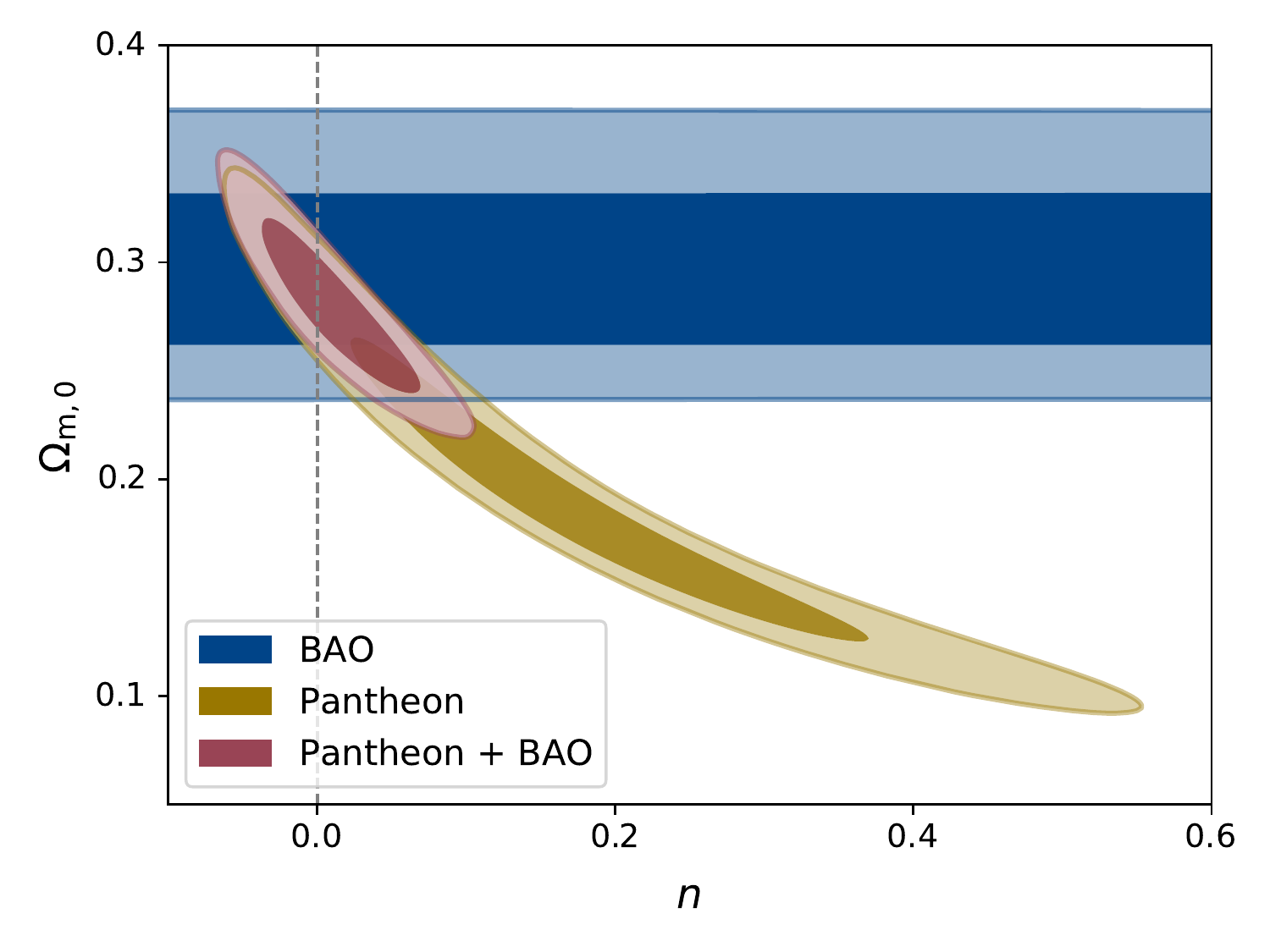}
	\caption[Constraints on the power law parameter $n$ and $\Omega_{\text{m},0}$]{2D contours on the power law parameter $n$ and $\Omega_{\text{m},0}$ using data from BAO (blue), SnIa (yellow) and the combination of the two (red). The darker and lighter concentric regions represent the 68\% and 95\% credible intervals, respectively. Colour-blind friendly colour-scheme from \cite{Tol}.}
	\label{fig:n_Om}
\end{figure}
\begin{figure}
	\centering
	\includegraphics[width=0.85\textwidth]{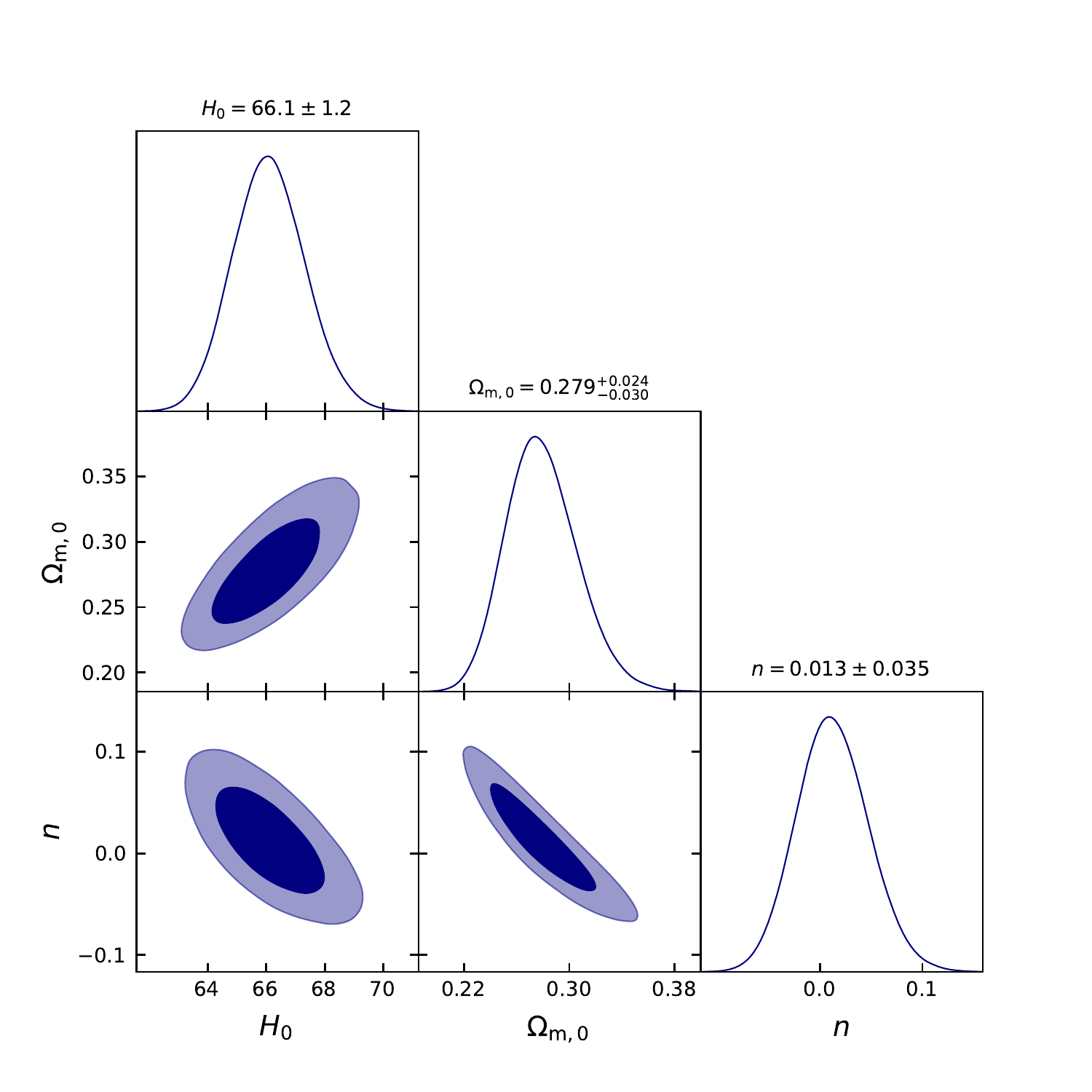}
	\caption[Constraints on the power law parameter $n$, $H_0$ and $\Omega_{\text{m},0}$]{Constraints on the power law parameter $n$, $H_0$ and $\Omega_{\text{m},0}$ using combined data from BAO and SnIa. The darker and lighter regions represent the 68\% and 95\% credible intervals, respectively.}
	\label{fig:tri_n_Tot}
\end{figure}
\begin{figure}
	\centering
	\includegraphics[width=0.85\textwidth]{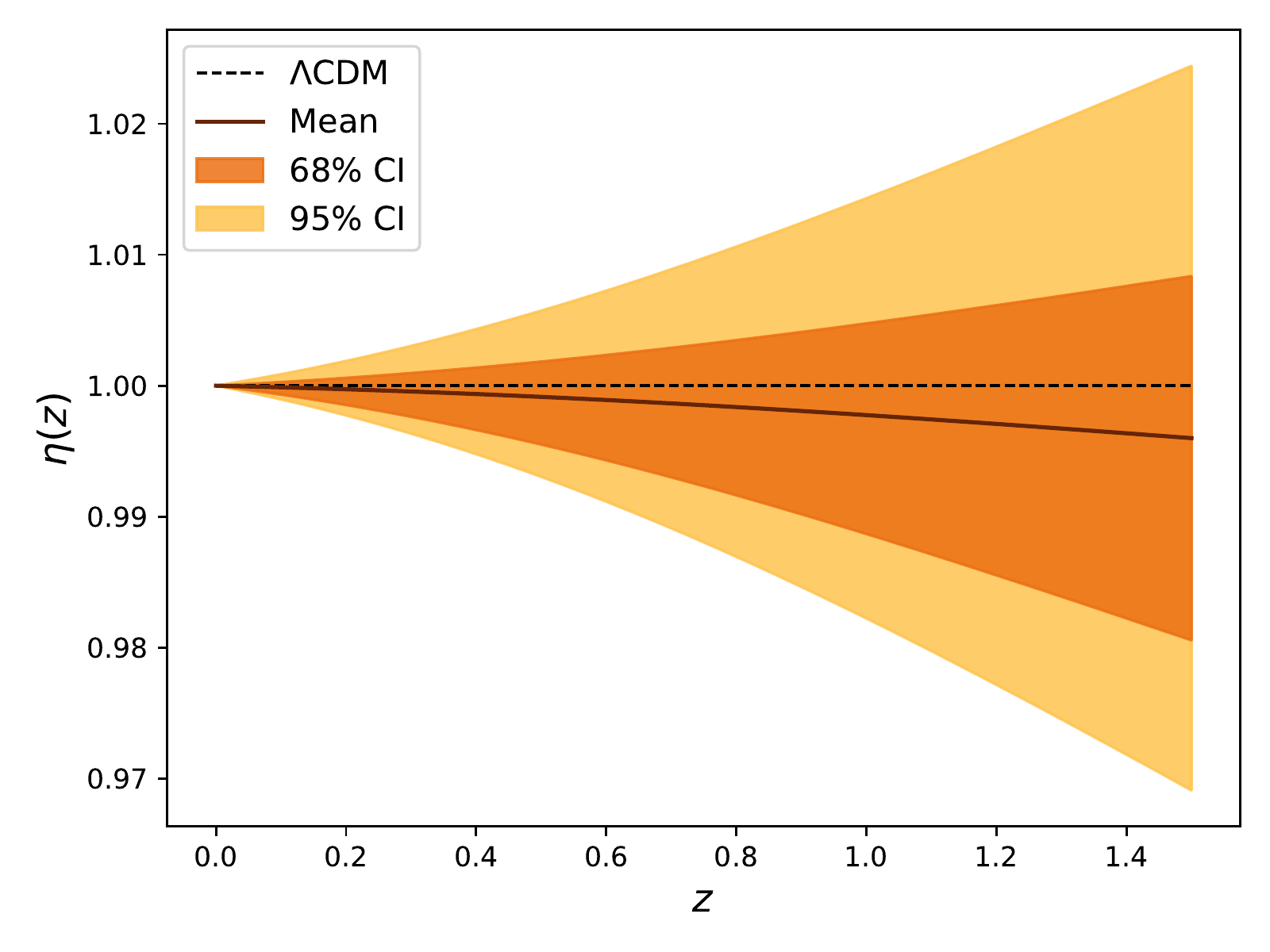}
	\caption[Reconstruction of $\upeta(z)$ for the power law NMC model]{Reconstruction of $\upeta(z)$ for the power law NMC model from combined BAO and SnIa data. The dashed line represents the GR prediction $\upeta=1$, while the solid red line represents the mean value of $\upeta$ at every redshift. The orange (darker) and yellow (lighter) contours represent the 68\% and 95\% credible intervals, respectively.}
	\label{fig:eta_power_law}
\end{figure}

The combined SnIa and BAO datasets constrain the NMC parameter to $n=0.013\pm 0.035$ (68\% CI). While this constraint falls short of the ones previously obtained from the black-body spectrum of the CMB, $|n|\lesssim \text{few}\times 10^{-6}$, or from BBN, $-0.002<n<0.003$, the present results are again complementary as they more directly constrain the value of $n$ at much smaller redshifts. Note that this still rules out NMC models designed to mimic dark matter, as these would require a power law with exponent in the range $-1\leq n \leq -1/7$ \cite{Bertolami2012,Silva2018}.

\subsubsection*{Exponential}
\label{subsubsec:exp}

Consider now an exponential NMC function,
\begin{equation}
	\label{eq:exp_f2}
	f_2\propto e^{\beta R} \,,
\end{equation}
where $\beta$ is the NMC parameter, with dimensions of $R^{-1}$ (GR is recovered when $\beta=0$). Using Eqs. \eqref{eq:scalar_curv} and \eqref{eq:exp_f2} in Eq. \eqref{eq:eta_ddr}, one obtains
\begin{equation}
	\label{eq:eta_exp}
	\upeta(z;\beta,\Omega_{\text{m},0}, H_0) = \exp\left[\frac{3}{2}\beta H_0^2 \Omega_{\text{m},0} (1-(1+z)^3)\right] \,.
\end{equation}
Note that $\upeta$ now depends on all three free parameters, $\beta$, $\Omega_{\text{m},0}$ and $H_0$. Furthermore, since $H_0$ is now also degenerate with $\beta$ and $\Omega_{\text{m},0}$, we can no longer analytically marginalize over $H_0$, and SnIa data alone cannot be used to derive useful constraints on any of these parameters. By combining the BAO and SnIa datasets, however, one is able to break this degeneracy, and derive constraints on all three parameters. The marginalized results can be found in Table \ref{tab:results_full_exp}, and the 2D distributions for $\beta$ and $\Omega_{\text{m},0}$ can be found in Fig.~\ref{fig:beta_Om} (see also Fig.~\ref{fig:tri_beta_Tot} for the remaining distribution plots). A reconstruction of Eq. \eqref{eq:eta_exp} is also shown in Fig.~\ref{fig:eta_exp}.

\begin{table}
	\centering
	\caption[Best-fit values and marginalized means and limits on cosmological and exponential NMC parameters from BAO and SnIa]{Best-fit values and marginalized means, 68\%, 95\% and 99\% CI limits obtained from currently available data on the cosmological parameters $\Omega_{\text{m},0}$ and $H_0$ (in units of km s$^{-1}$ Mpc$^{−1}$) and on the NMC parameter $\beta$ (in units of km$^{-2}$ s$^2$ Mpc$^2$).}
	\label{tab:results_full_exp}
	\bgroup
	\def\arraystretch{1.2}
	\begin{tabular}{lc|ccccc}
		\hline\hline
		Parameter                        & Probe    & Best fit & Mean    & 68\%                 & 95\%                 & 99\%                 \\ \hline
		& BAO      & $66.4$   & $66.8$  & $^{+1.2}_{-1.4}$     & $^{+2.7}_{-2.5}$     & $^{+3.7}_{-3.2}$     \\
		{\boldmath$H_0$}                 & SnIa     & \multicolumn{5}{c}{unconstrained}                                                       \\
		& SnIa+BAO & $65.7$   & $65.7$  & $\pm 1.0$            & $^{+2.1}_{-2.0}$     & $^{+3.4}_{-3.0}$     \\ \hline
		& BAO      & $0.291$  & $0.300$ & $^{+0.027}_{-0.035}$ & $^{+0.064}_{-0.060}$ & $^{+0.094}_{-0.071}$ \\
		{\boldmath$\Omega_{\text{\bf m},0}$} & SnIa     & \multicolumn{5}{c}{unconstrained}                                                       \\
		& SnIa+BAO & $0.268$  & $0.268$ & $\pm0.019$           & $^{+0.038}_{-0.036}$ & $^{+0.052}_{-0.046}$ \\ \hline
		& BAO      & \multicolumn{5}{c}{unconstrained}                                                       \\
		{\boldmath$\beta\cdot10^6$}      & SnIa     & \multicolumn{5}{c}{unconstrained}                                                       \\
		& SnIa+BAO & $1.18$   & $1.24$  & $^{+0.97}_{-1.2}$    & $^{+2.2}_{-2.1}$     & $^{+3.3}_{-2.5}$     \\ \hline\hline
	\end{tabular}
	\egroup
\end{table}
\begin{figure}
	\centering
	\includegraphics[width=0.85\textwidth]{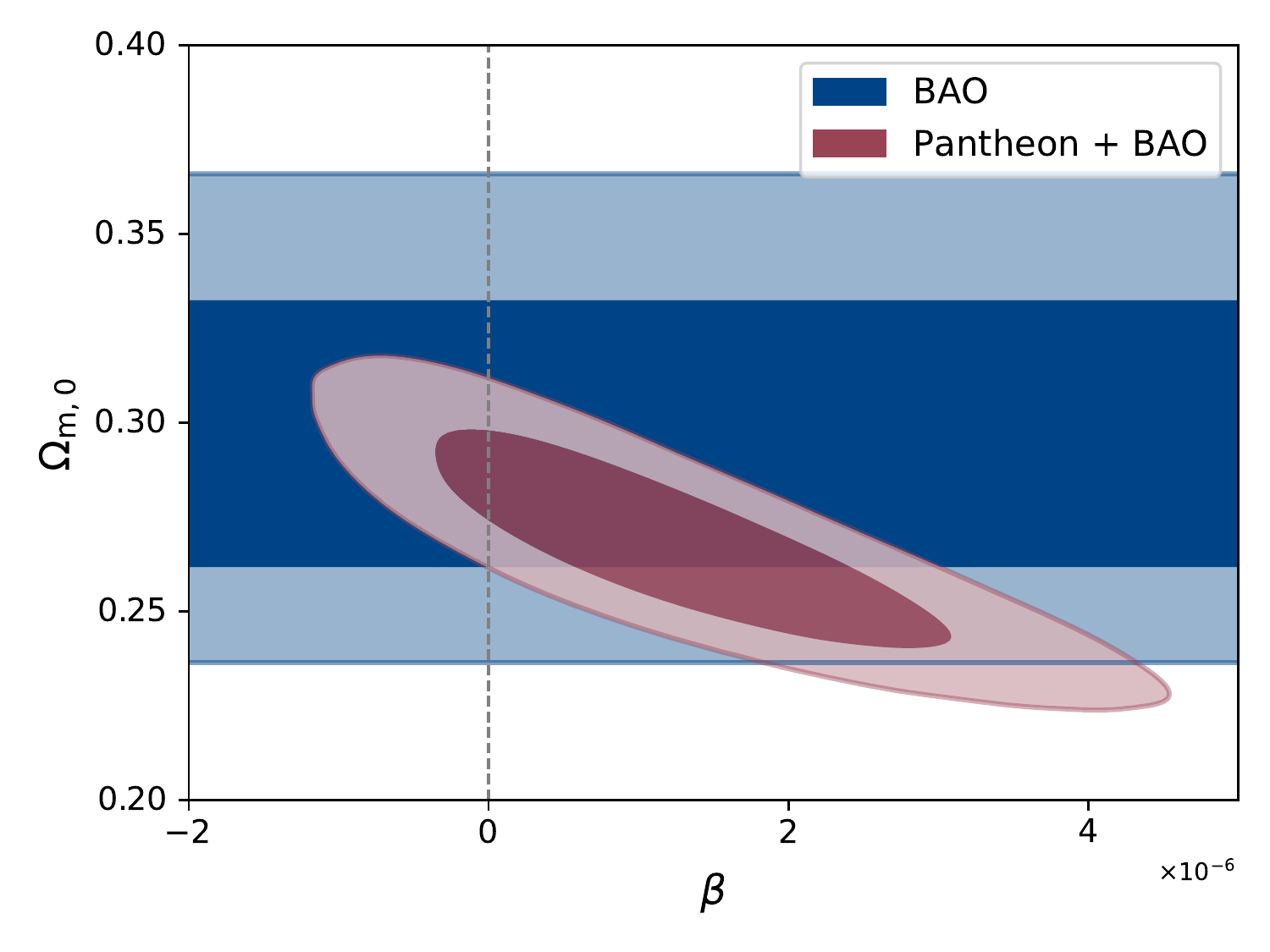}
	\caption[Constraints on the exponential parameter $\beta$ and $\Omega_{\text{m},0}$]{2D contours on the exponential parameter $\beta$ and $\Omega_{\text{m},0}$ using data from BAO (blue) and the combination of the SnIa and BAO (red). The darker and lighter concentric regions represent the 68\% and 95\% credible intervals, respectively.}
	\label{fig:beta_Om}
\end{figure}
\begin{figure}
	\centering
	\includegraphics[width=0.85\textwidth]{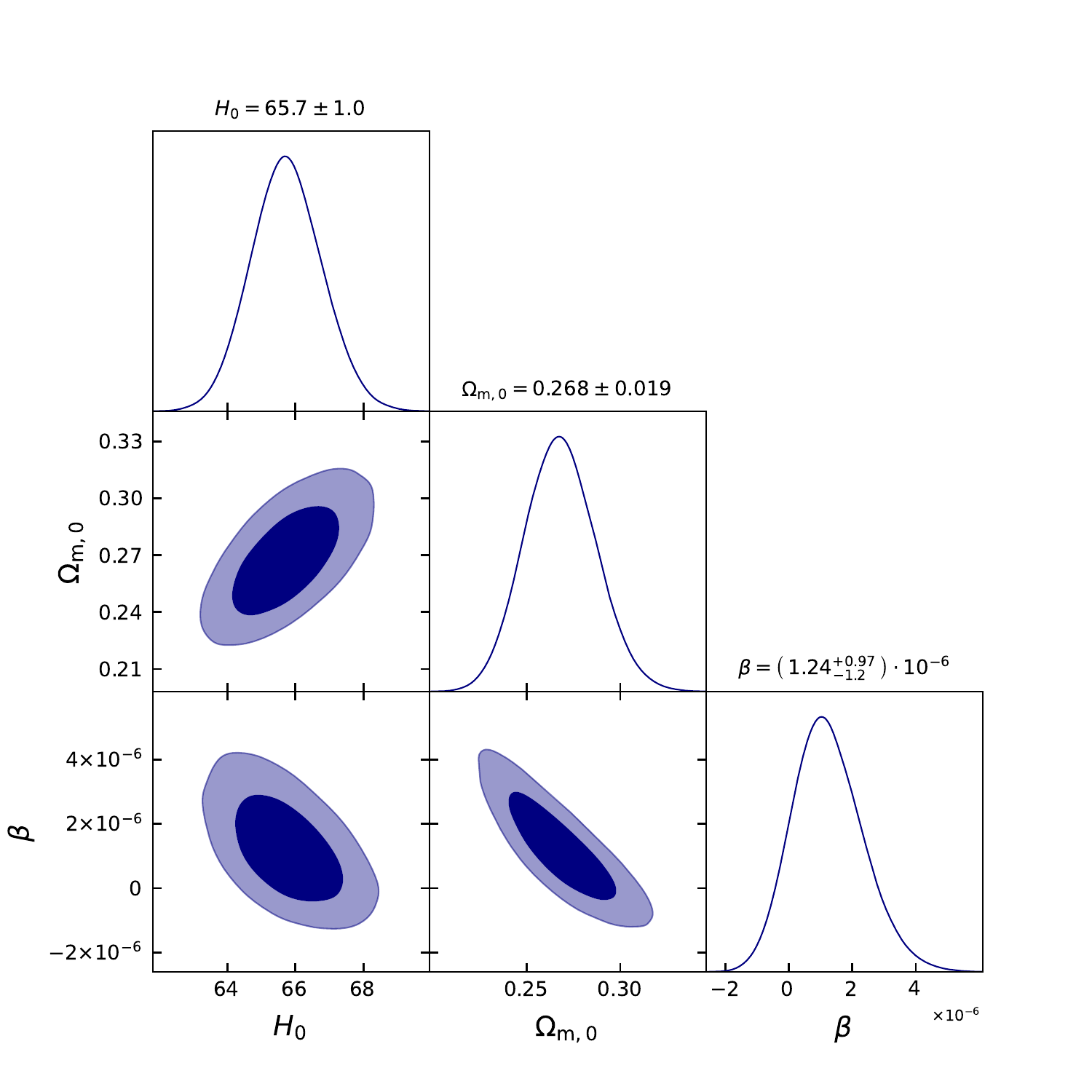}
	\caption[Constraints on the exponential parameter $\beta$, $H_0$ and $\Omega_{\text{m},0}$]{Constraints on the exponential parameter $\beta$, $H_0$ and $\Omega_{\text{m},0}$ using combined data from BAO and SnIa. The darker and lighter regions represent the 68\% and 95\% credible intervals, respectively.}
	\label{fig:tri_beta_Tot}
\end{figure}
\begin{figure}
	\centering
	\includegraphics[width=0.85\textwidth]{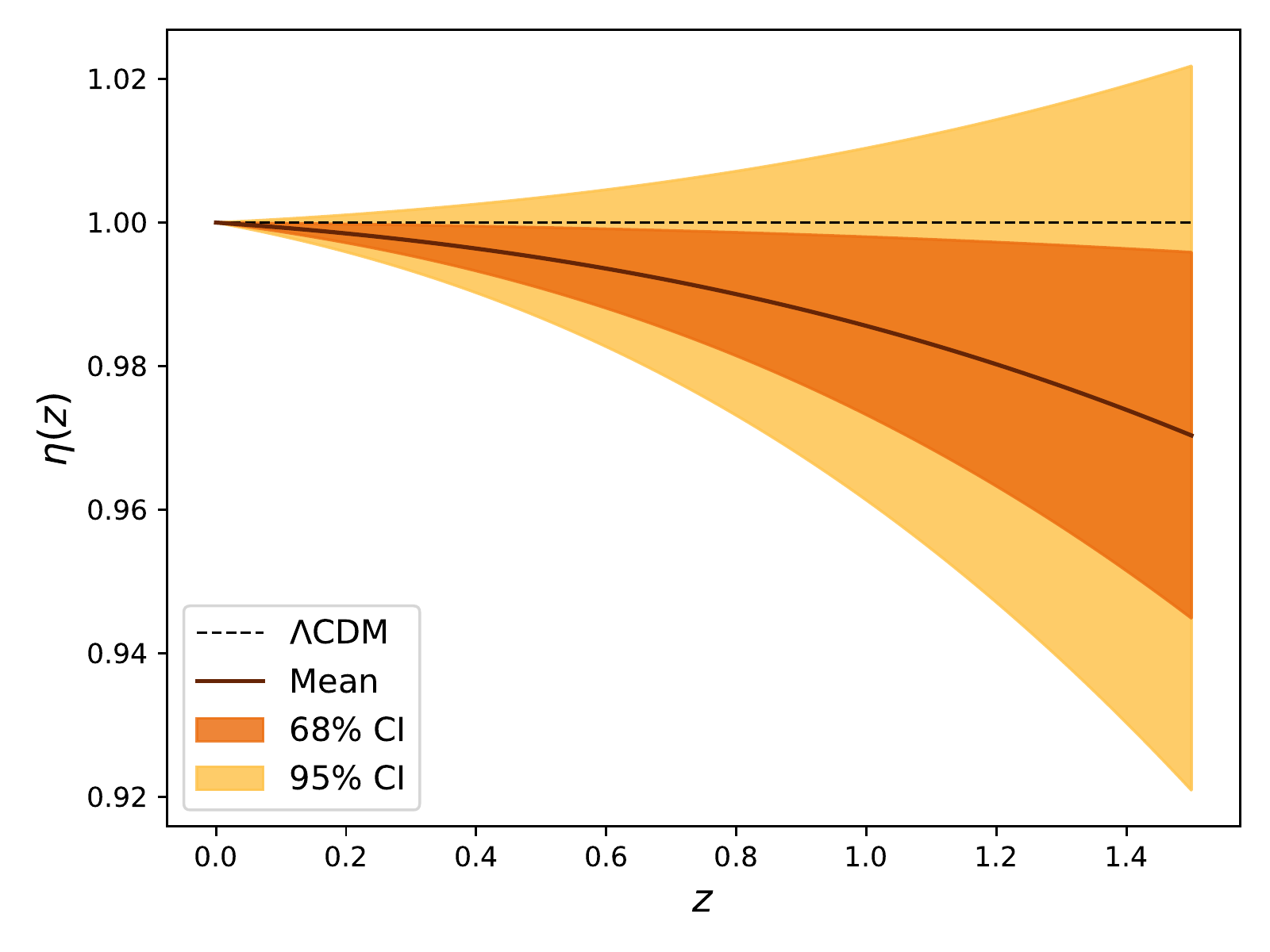}
	\caption[Reconstruction of $\upeta(z)$ for the exponential NMC model]{Reconstruction of $\upeta(z)$ for the exponential NMC model from combined BAO and SnIa data. The dashed line represents the GR prediction $\upeta=1$, while the solid red line represents the mean value of $\upeta$ at every redshift. The orange (darker) and yellow (lighter) contours represent the 68\% and 95\% credible intervals, respectively.}
	\label{fig:eta_exp}
\end{figure}

The combined SnIa and BAO datasets constrain the NMC parameter to $\beta=\left(1.24^{+0.97}_{-1.2}\right)\cdot10^{-6}$ (68\% CI), in units of km$^{-2}$ s$^2$ Mpc$^2$. Once again this result complements the one found for the same function using the method presented in \cite{Azevedo2018a} for the variation of the baryon to photon ratio, $|\beta|\lesssim10^{-28}$, as they constrain the same parameter in significantly different redshift ranges. Also notice that while the marginalized results do not contain the GR limit $\beta=0$ at the 68\% CI, that limit is contained in both the marginalized 95\% CI, $\beta=\left(1.2^{+2.2}_{-2.1}\right)\cdot10^{-6}$, and the 2D 68\% credible region.

Future observations from LSST and the Euclid DESIRE survey \cite{Laureijs2011,Astier2014} will provide more data points in the range $z\in[0.1, 1.6]$. For the $\epsilon$ parametrization used in \cite{Martinelli2020}, this will result in an improvement on the constraint of about one order of magnitude. If this is the case, one could expect a corresponding improvement on the NMC parameter constraints. Third-generation gravitational-wave observatories will also be able to provide data points at even higher redshift (up to $z\sim5$), which will serve as independent and complementary data \cite{Cai2017}.

\chapter{Conclusions} 

\label{chapter_conc} 

In this thesis, we have studied a class of $f(R)$ gravity theories that feature a nonminimal coupling (NMC) between gravity and the matter fields. These NMC theories feature two functions of the Ricci scalar, $f_1(R)$ and $f_2(R)$. While the former takes the place of the Ricci scalar in the Einstein-Hilbert action, the latter couples directly with the Lagrangian of the matter fields. As a consequence, the on-shell Lagrangian of the matter fields appears explicitly in the equations of motion, in contrast with minimally coupled $f(R)$ gravity. Hence, its correct determination is crucial to the derivation of the  dynamics of the gravitational and matter fields. Several different on-shell Lagrangians have been used to describe perfect fluids in the literature, such as $\mathcal{L}_\text{m}^\text{on}= -\rho$ or $\mathcal{L}_\text{m}^\text{on}=p$, but their use in the context of NMC theories of gravity generally  leads to different results, contradicting the claim that they describe the same type of fluid. This suggested that a more in-depth analysis of the on-shell Lagrangian of perfect fluids was required, and that has been the central problem tackled in this thesis.

This final chapter concludes our work by summarising its main contributions and discussing possible future research directions.

\section{Main contributions}

We focused on three key points of research: (1) the characterization of the on-shell Lagrangian of the matter fields; (2) the background-level thermodynamics of a Universe filled with perfect fluids composed of solitonic particles of fixed mass and structure; and (3) the derivation of novel constraints on specific NMC gravity models.

\subsection{The Lagrangian of perfect fluids}

In NMC theories, the minimum action principle results in a set of equations of motion that explicitly feature the on-shell Lagrangian of the matter fields. In the case of NMC gravity, this also means that the motion of particles and fluids, and the evolution of the energy-momentum tensor (EMT) will depend on the form of the on-shell Lagrangian of the matter fields.

In Chapter \ref{chapter_lag}, we have shown that the correct form of the on-shell Lagrangian depends not only on the fluid's EMT but also on the microscopic properties of the fluid itself. This is one of the key contributions of this thesis.

For a perfect fluid that is minimally coupled with gravity and that conserves particle number and entropy, the on-shell Lagrangian can take many different forms that do not change the equations of motion or the EMT. We have shown that this is still the case for a barotropic fluid.

The other major contributions in this area is the demonstration that the correct on-shell Lagrangian for a fluid composed of solitonic particles must be given by the trace of the fluid's EMT, \textit{i.e.} $\mathcal{L}_\text{m}^\text{on}= T=T^{\mu\nu}g_{\mu\nu}$. This result is independent of the coupling of the fluid to gravity or to other fields, but is particularly relevant when this coupling is nonminimal, since the on-shell Lagrangian appears explicitly in the equations of motion.

In Chapter \ref{chapter_nmc}, we also found that consistency between the evolution of the energy and linear momentum in the context of theories of gravity with an NMC coupling to the matter fields requires that the condition $\mathcal L^{\rm on}=T$ is required for consistency, giving no margin to other possibilities for the on-shell Lagrangian of an ideal gas.

\subsection{The nonnminimal coupling and thermodynamics}

In Chapter \ref{chapter_nmc}, we have shown that fluids composed of radiation (such as photons or neutrinos), couple differently to gravity than dust (such as cold dark matter or baryons), since their on-shell Lagrangian density vanishes. This, together with the well-known nonconservation of the EMT present in NMC theories, can lead to unusual thermodynamic behaviour.

In particular, we  have found that the nonconservation of the EMT translates into a non-adiabatic expansion of a homogeneous and isotropic Friedmann-Lemaître-Robertson-Walker universe, in stark contrast with general relativity. This in turn implies that the entropy of such a universe is in general not conserved. Moreover, we have shown that Boltzmann's $\mathcal{H}$-theorem may not hold in the context of NMC theories.

\subsection{Observational constraints}

In Chapter \ref{chapter_constr}, we have constrained $n$-type spectral distortions in the cosmic microwave background (CMB) associated with the NMC to gravity, affecting the normalization of the black-body spectrum. Using data from COBE and WMAP, we constrained the NMC function $f_2$ to vary by only a few parts in $10^5$ from decoupling ($z\sim10^3$) up to the present day. For a power-law NMC of the type $f_2 \propto R^n$, this translate into $|n|\lesssim {\rm few} \times 10^{-6}$.

We have also shown that the baryon-to-photon ratio is no longer a constant in NMC gravity after big bang nucleosynthesis (BBN). Using data from the Planck satellite and from computational codes that simulate the process of BBN, we have obtained constraints on power-law NMC of the type $f_2 \propto R^n$ of $-0.007<n<0.002$ or $-0.002<n<0.003$ (depending on the version of BBN code used), in the redshift range $z\in[10^3,10^9]$.

We have also shown that the NMC causes a violation of Etherington's distance-duality relation, which can be verified using data from Type Ia supernovae (SnIa) and baryon acoustic oscillations (BAO) observations. These observations effectively constrain NMC models in the redshift range $z\in[0,1.5]$. For a power-law NMC of the type $f_2 \propto R^n$, we obtain the marginalized constraint $n=0.013\pm 0.035$, while for an exponential NMC of the type $f_2\propto e^{\beta R}$, we obtain the marginalized constraint $\beta=\left(1.24^{+0.97}_{-1.2}\right)\cdot10^{-6}$.

\section{Discussion and future work}

As with any research, this thesis is not without limitations. While the fluids described in Chapter \ref{chapter_lag} can cover many different cases, they rely on a few assumptions. Namely, the derivation of the on-shell Lagrangian for the solitonic-particle fluid $\mathcal{L}_\text{m}^\text{on}=T$ assumes that the particles have fixed rest mass and structure.

In light of these results, it may be worthy to re-examine previous works in the literature in which other forms of the Lagrangian have been assumed for cosmic fluids that are well described by an ideal gas. While some results are expected to hold, particularly when  $\mathcal{L}_\text{m}^\text{on}=-\rho$ is used to describe dust, this will not be the case in general.

The stability of NMC models is another aspect that was not explored in this thesis. Some models in this class of theories are subject to instabilities, such the well-known Ostrograsdky and Dolgov-Kawasaki instabilities, and therefore be unsuitable or undesirable as mechanisms for the explanation of cosmological phenomena. This was not the focus of this thesis but is a topic that requires more attention.

Another issue that merits more research is the possible nonconservation of the EMT of matter, and its consequences. It is not immediately clear how the difference in how fluids couple to gravity can arise from first principles, nor why some fluids (like dust) maintain the conservation of energy-momentum, while others (like radiation) do not. Moreover, since this non-conservation is tied to an NMC with gravity, the possibility of the existence of a purely gravitational EMT in this context may be worthy of consideration. This idea is still a matter of discussion and debate in GR but has not been explored in the context of NMC theories.

Lastly, it is worthy of note that while we have focused on a particular family of $f(R)$ NMC gravity models, many issues we have covered in this thesis are transversal to many other theories, including those that feature an NMC between other fields (such as dark energy) and the Lagrangian of the matter fields. Two clear examples are already given in Section \ref{sec.nmclagrole}, where the dark energy field is coupled to either neutrinos or the electromagnetic field, but the main results of Section \ref{sec.nmclag} are not limited to any particular coupling or gravitational theory.


\bibliographystyle{naturemag}
\bibliography{Dissertation.bib}

\end{document}